\documentclass[fleqn,10pt]{wlscirep}
\usepackage[utf8]{inputenc}
\usepackage[T1]{fontenc}
\usepackage{cite}
\usepackage{amsmath,amssymb,amsfonts}
\usepackage{algorithmic}
\usepackage{epstopdf}
\usepackage{graphicx}
\usepackage{comment}
\usepackage{multirow}
\usepackage{subcaption}
\usepackage{xcolor}
\usepackage{makecell}
\usepackage{textcomp}
\usepackage{url}
\usepackage{flushend}
\usepackage{graphicx} 
\usepackage[export]{adjustbox}
\usepackage{algorithm}
\usepackage{algorithmic}
\usepackage{authblk}
\usepackage{enumitem}
\usepackage{colortbl}
\usepackage[symbol]{footmisc}
\newcolumntype{P}[1]{>{\centering\arraybackslash}p{#1}}
\makeatletter
\newcommand\notsotiny{\@setfontsize\notsotiny\@vipt\@viipt}
\makeatother

\title{\LARGE Exploring Out-of-distribution Detection for Sparse-view Computed Tomography with Diffusion Models}

\author[1,*]{Ezgi Demircan-Tureyen}
\author[1]{Felix Lucka}
\author[1,2]{Tristan van Leeuwen}
\affil[1]{Computational Imaging, Centrum Wiskunde and Informatica, 1098 XG Amsterdam, Netherlands}
\affil[2]{Mathematical Institute, Utrecht University, 3584 CD Utrecht, Netherlands}

\affil[*]{edt@cwi.nl}

\begin{abstract}
Recent works demonstrate the effectiveness of diffusion models as unsupervised solvers for inverse imaging problems. Sparse-view computed tomography (CT) has greatly benefited from these advancements, achieving improved generalization without reliance on measurement parameters. However, this comes at the cost of potential hallucinations, especially when handling out-of-distribution (OOD) data. To ensure reliability, it is essential to study OOD detection for CT reconstruction across both clinical and industrial applications. This need further extends to enabling the OOD detector to function effectively as an anomaly inspection tool. In this paper, we explore the use of a diffusion model, trained to capture the target distribution for CT reconstruction, as an in-distribution prior. Building on recent research, we employ the model to reconstruct partially diffused input images and assess OOD-ness through multiple reconstruction errors. Adapting this approach for sparse-view CT requires redefining the notions of ``input'' and ``reconstruction error''. Here, we use filtered backprojection (FBP) reconstructions as input and investigate various definitions of reconstruction error. Our proof-of-concept experiments on the MNIST dataset highlight both successes and failures, demonstrating the potential and limitations of integrating such an OOD detector into a CT reconstruction system. Our findings suggest that effective OOD detection can be achieved by comparing measurements with forward-projected reconstructions, provided that reconstructions from noisy FBP inputs are conditioned on the measurements. However, conditioning can sometimes lead the OOD detector to inadvertently reconstruct OOD images well. To counter this, we introduce a weighting approach that improves robustness against highly informative OOD measurements, albeit with a trade-off in performance in certain cases.

\end{abstract}
\begin{document}

\flushbottom
\maketitle
%
%
\thispagestyle{empty}

\section*{Introduction}
Computational imaging is among many application areas that have gained attention with the uptake of diffusion models. CT reconstruction is a typical computational imaging challenge, aiming to reconstruct images from raw measurements (i.e., X-ray projections of object slices acquired from several directions). Given some projection measurements $\mathbf{y}$ and a forward measurement operator $\mathbf{A}$, the goal is to retrieve $\mathbf{x}\in \mathbb{R}^{n}$ from the forward model: 
\begin{equation}
    \mathbf{y} = \mathbf{Ax} + \mathbf{\eta}, \quad \mathbf{y} \in \mathbf{R}^{m}, \quad \mathbf{A} \in \mathbb{R}^{m \times n},
\label{eq:forwardmodel}
\end{equation}
where $\mathbf{\eta} \in \mathbb{R}^{m}$ denotes a noise vector. As high doses of radiation are usually not well-tolerated, achieving to reconstruct high-quality images from significantly reduced number of projections (i.e., $m \ll n$) is crucial. This requires addressing the ill-posed nature of the problem, since without the utilization of prior knowledge on $\mathbf{x}$, exact retrieval of it is not possible ($\mathbf{x} \mapsto \mathbf{y}$ is many-to-one). Some recent works leveraged the diffusion model as prior \cite{song2021solving, chung2022improving, chung2022diffusion, guan2023generative} to tackle general inverse imaging problems from Bayesian viewpoint. They rely on the idea of training a diffusion model on a set $\left\{\mathbf{x}^{(1)}, \mathbf{x}^{(2)}, \cdots, \mathbf{x}^{(N)}\right\} \sim p(\mathbf{x})$ to implicitly learn $p(\mathbf{x})$, so that one can sample from it. They simply incorporate the measurement process to the iterative sampling scheme to approximate conditional samples (i.e., samples from posterior $p(\mathbf{x}|\mathbf{y})$) by maintaining consistency between the prior and the measurement model. This approach suggests a better generalization capability than a conditional generative model, as the model remains unaware of the measurement process, thereby eliminating the need for retraining when the measurement parameters change. Moreover, since the model is not trained on a specific task, it can flexibly be adapt to different tasks. 

\begin{figure*}[t!]
 \centering
  \begin{subfigure}[b]{0.7\textwidth}
  \centering
{\includegraphics[width=\textwidth,keepaspectratio=true]{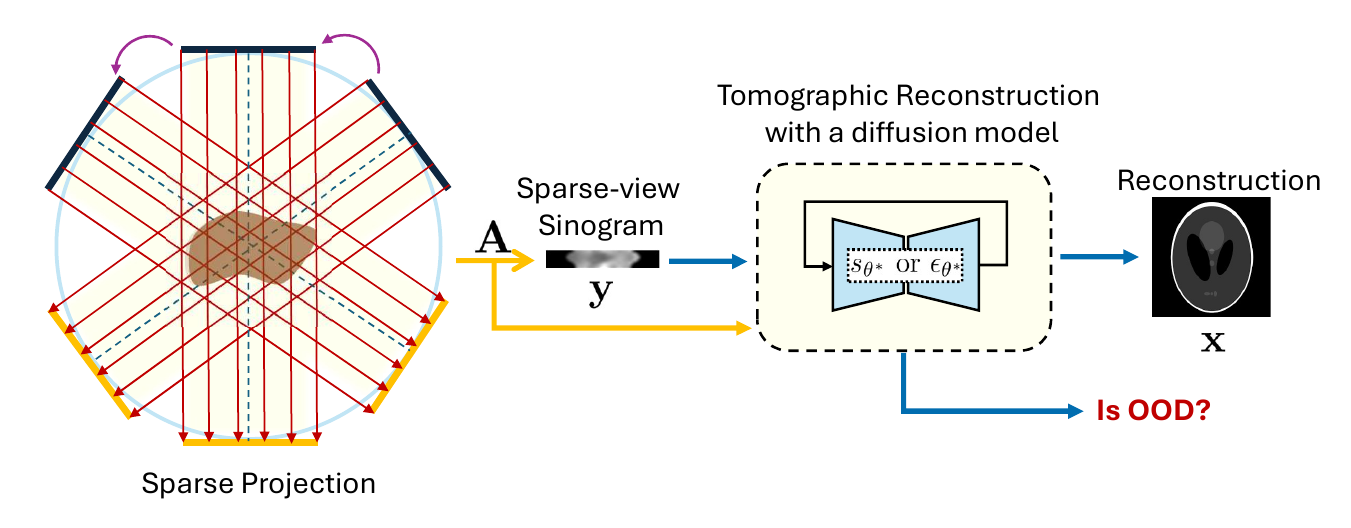}}
 \end{subfigure}
\caption{Schematic of sparse-view imaging setup and reconstruction framework augmented with out-of-distribution detector.}
\label{fig:intro}
\end{figure*}

Another application area for diffusion models is out-of-distribution (OOD) detection. This versatile task serves various purposes, such as ensuring reliability and aiding in inspection. Due to the diverse nature of OOD data in most scenarios, it is impractical to formalize OOD detection as a supervised classification task. Unsupervised OOD detection has garnered significant attention, as it relies solely on ID data for training (i.e., \textit{one-class learning}), mitigating generalization issues stemming from underrepresented OOD data. The literature is dominated by two approaches: \textit{likelihood-based} and \textit{reconstruction-based} OOD detection. The former involves training a density estimation model, allowing detection of OOD samples by assigning them low likelihoods. However, recent studies \cite{choi2018waic, nalisnick2018deep, ren2019likelihood, serra2019input, zhang2021understanding} have shown that these models often assign unexpectedly higher probabilities to the OOD samples from an unrelated distribution (e.g., a model trained on Fashion-MNIST assigns higher likelihoods to the MNIST samples than those from Fashion-MNIST.) The latter, on the other hand, involves reconstructing input data using a generative model and comparing the reconstructed samples with the original ones. This method relies on the assumption that the model struggles to accurately reconstruct OOD samples due to their dissimilarity to the training distribution. Therefore, when presented with OOD samples, the reconstruction errors are expected to be higher compared to in-distribution samples.
Due to the superiority of diffusion models in terms of mode coverage and sample quality compared to other generative models, they become a promising option for reconstruction-based OOD detection \cite{wyatt2022anoddpm, wolleb2022diffusion, graham2023denoising, tur2023exploring, liu2023unsupervised, linmans2024diffusion}. 

An accurate and automated CT reconstruction pipeline is a sought-after target for many clinical and industrial practices. The aforementioned diffusion models appear to be aligned with this objective but necessitate an OOD detection mechanism, either to prevent reconstructions in situations that surpass the model's generalization capacity or to purposely inspect abnormalities. Despite the extensive literature demonstrating the potential of diffusion models as unsupervised inverse problem solvers and OOD detection tools separately, to the best of our knowledge, this will be the first study at their intersection. Moreover, from a broader point of view, there is a gap in the literature regarding the applicability of the reconstruction-based OOD detection paradigm to inverse problems, where the inputs are incomplete in some manner. To our knowledge, the recent study \cite{bhuttouncertainty} stands as the only research employing a reconstruction-based OOD detection approach for image reconstruction, where the models are trained in supervised manner and the OOD detection task is approached from an uncertainty estimation perspective. So, the main focus of this paper is exploring the use of diffusion models for OOD detection in unsupervised sparse-view CT reconstruction. Our goal is to address questions such as: Can a diffusion model trained to capture the prior distribution for a sparse-view CT reconstruction task be also utilized for OOD detection as shown in Fig. \ref{fig:intro}? Given the in-distribution dataset of full-view images, how well can this model generalize to the distribution shift caused by the sparse-view mappings of in-distribution images? Does conditioning on the measurements aid or hinder the OOD detection process? How should we redefine the notion of reconstruction error, given that its usual definition is impractical in the case of sparse-view tomography? We believe that the answers to these questions are crucial for better understanding and may provide insights for future research directions on reliable CT reconstruction pipelines. We conducted all experiments on MNIST images \cite{lecun1998mnist,deng2012mnist} within a one-class experimental setup. In this setup, the diffusion model is trained on one digit, treating all other digits as OOD. 

The reader should be aware that the term ``reconstruction'' throughout the paper may refer to either CT reconstruction or reconstruction for OOD detection. Although these tasks may occasionally coincide, context is crucial for understanding the intended meaning. We will specifically use ``CT reconstruction'' to denote the former. Similarly, the term ``data'' can sometimes be ambiguous. It usually refers to projection data or measurements, as in ``data fidelity'', but when used in phrases like ``data manifold'' and ``training data,'' it refers to images.

\section{Background}
\label{sec:background}
\subsubsection*{A Brief Look at OOD Detection} 

While the notion of ``out-of-distribution (OOD)'' universally refers to any shift from the training distribution, the task of ``OOD detection'' often lack clarity due to the diverse sources of shifts to be detected and the varied motivations behind detection efforts. Distribution shifts may stem from various factors, such as content alterations, domain discrepancies, changes in acquisition setups, instrumental variations, style differences, corruptions, or adversarial attacks. They are usually categorized into two classes: covariate and semantic shift. Given an input space $\mathcal{X}$ and the space of semantic categories $\mathcal{Y}$; covariate shifts manifest themselves by affecting $\mathcal{X}$, while $\mathcal{Y}$ remains unchanged (e.g., clean to noisy image). Namely, these changes pertain to appearance rather than content. Semantic shifts, on the other side, involve alterations within $\mathcal{Y}$, which indirectly lead to shifts in $\mathcal{X}$ (e.g., digit ``7'' to digit ``9'') \cite{yang2021generalized}. 

The motivation behind OOD detection is usually ensuring the reliability of the system, as any kind of shift is possible in post-deployment open-world scenario \cite{drummond2006open}. For instance, a trustworthy medical diagnosis system designed to identify specific diseases based on symptoms should have the capability to detect when provided with symptoms lying beyond its expertise. Such symptoms might be indicating new or rare conditions, thus the system should refrain from providing a potentially incorrect diagnosis instead of blindly classifying them. Or the data could simply be corrupted (without any content deviation) which complicates the model's robustness, as the ML models are vulnerable to covariate shifts \cite{hendrycks2019benchmarking}. In such cases, the decision to detect them as OOD or not largely hinges on the model's ability to generalize across existing covariate shifts \cite{yang2023full}. 

Beyond reliability considerations, the motivation behind OOD detection can be extended to various tasks, such as detecting adverse pathologies in medicine \cite{wyatt2022anoddpm, wolleb2022diffusion, bercea2024towards}, identifying defects in industrial processes \cite{he2022survey, presenti2022automatic, zhang2023unsupervised}, recognizing novel features in planetary observation \cite{kerner2019novelty} or video surveillance \cite{idrees2018enhancing} for further analysis. In \cite{yang2021generalized}, the authors present a taxonomy to eliminate confusion arising from the versatility of OOD detection. Through the lens of that work, we can position ``OOD detection for CT reconstruction'' in anomaly detection (AD) category, since OOD samples are viewed as either a threat to the reliability of the reconstruction system or indicating internally erroneous/malicious subjects, possibly pointing to conditions like diseases or defects that demand attention. Considering these dual motivations, our experimental setup can be interpreted in two distinct ways: Firstly, it can be perceived as facilitating near-OOD \cite{ming2022impact} scenarios, aligning with the objective of ``anomaly detection for reliability''. Secondly, it can also address ``anomaly detection for inspection'', since for instance, digit ``3'' can be considered a defected version of a digit ``8''. Note that, in \cite{yang2021generalized}, anomaly detection is further divided into two: sensory and semantic AD, based on the covariate/semantic shift dichotomy. Our focus in this work is semantic AD, as the inputs inherently exhibit covariate shift, regardless of whether they are anomalous or not, by definition of sparse-view tomography. Here, we consider images reconstructed by applying filtered backprojection (FBP) to sparse view sinograms as inputs.

\subsubsection*{Denoising Diffusion Probabilistic Models (DDPMs)} 
Score matching with Langevin dynamics (SMLD) \cite{song2019generative} and denoising diffusion probabilistic models (DDPMs)\cite{ho2020denoising} represent two blueprints of the modern diffusion models. They both rely on the data generation by denoising principle. Specifically, they both operate by handcrafting a diffusion process that gradually perturb the data, and learning a model to be used while reversing this process. Their differences lie primarily in their training and sampling methodologies. SMLD is trained with score matching to predict \textit{score} (i.e., the gradient of the log likelihood with respect to data) at each noise scale and utilizes Langevin dynamics for sampling, while DDPM is trained with a re-weighted variant of the evidence lower bound (ELBO) to predict noise realization at each noise scale and adopts ancestral sampling. However, it has been shown in \cite{song2020score} that SMLD and DDPM can be unified within a  framework that defines forward and reverse stochastic differential equations (SDEs) respectively for the noising and denoising processes, within a continuous state space context. Under this framework, SMLD and DDPM correspond to the discretizations of two separate SDEs: variance exploding (VE-SDE) and variance preserving (VP-SDE), respectively \cite{song2020score}.

In \cite{song2020score}, the continuous diffusion process for the trajectory of samples $\left\{\mathbf{x}_t \in \mathbb{R}^n\right\}_{t \in[0,T]}$ has the form of 
\begin{equation}
  \mathrm{d} \mathbf{x}_t=f(t) \mathbf{x}_t \mathrm{~d} t+g(t) \mathrm{d} \mathbf{w}_t,
  \label{eq:forwardsde}
\end{equation}
where $\left\{{\mathbf{w}}_t\right\}_{t \in[0,T]}$ is the standard Wiener process. Here, the functions $f(t)$ and $g(t)$ are chosen in such a way that for any time step $t$, the transition density $p_{0t}(\mathbf{x}_t \mid \mathbf{x}_0)$ is a conditional linear Gaussian distribution, and for any sample $\mathbf{x}_0$ drawn from the unknown data distribution $p_0(\mathbf{x}) \equiv p(\mathbf{x})$, one should have $\mathbf{x}_t \sim \mathcal{N}(\mathbf{0}, \mathbf{I})$ at time step $t=T$. Eq. \eqref{eq:forwardsde} is coupled with the reverse SDE \cite{anderson1982reverse}, i.e.,
\begin{equation}
\mathrm{d} \mathbf{x}_t=\left[f(t) \mathbf{x}_t-g(t)^2 \nabla_{\mathbf{x}_t} \log p_t\left(\mathbf{x}_t\right)\right] \mathrm{d} t+g(t) \mathrm{d} \bar{\mathbf{w}}_t,
\label{eq:backwardsde}
\end{equation}
where $\left\{\bar{\mathbf{w}}_t\right\}_{t \in[0,T]}$ is the standard Wiener process in backward direction, and $\nabla_{\mathbf{x}_t} \log p_t\left(\mathbf{x}_t\right)$ is the score of $p_t\left(\mathbf{x}_t\right)$. Using the reparameterization trick \cite{kingma2013auto} and considering only VP-SDE (i.e., DDPM with Score SDE formulation), the discrete forward diffusion is implemented as
\begin{equation}
\mathbf{x}_t=\sqrt{\bar{\alpha}_t} \mathbf{x}_0+\sqrt{1-\bar{\alpha}_t} \mathbf{z}, \quad \mathbf{z} \sim \mathcal{N}(0, \mathbf{I}) 
  \label{eq:forwardsdediscrete}
\end{equation}
where $\bar{\alpha}_t \triangleq \prod_{k=1}^t \alpha_k$ for $\alpha_t \triangleq 1-\beta_t$ and $\beta_t$ defining the noise schedule in monotonically increasing fashion as a function of $t$ \cite{ho2020denoising}, and the reverse diffusion has the form of
\begin{equation}
\mathbf{x}_{t-1}=\frac{1}{\sqrt{\alpha_t}}\left(\mathbf{x}_t+\left(1-\alpha_t\right) \nabla_{\mathbf{x}_t} \log p_t\left(\mathbf{x}_t\right) \right)+\sqrt{\sigma_t} \mathbf{z}, \quad  \mathbf{z} \sim \mathcal{N}(0, \mathbf{I})
  \label{eq:backwardsdediscrete}
\end{equation}
 Therefore, in order to reverse the perturbation process, one must have an initial noise sample $\mathbf{x}_T$ and gradually denoise it to eventually obtain a data sample $\mathbf{x}_0 \sim p(x)$. According to Eq. \eqref{eq:backwardsdediscrete}, denoising requires the score at each intermediate time step, which is estimated by a neural network, i.e., $\mathbf{s}_\theta\left(\mathbf{x}_t, t\right) \approx \nabla_{\mathbf{x}_t} \log p_t\left(\mathbf{x}_t\right)$.  However, it can also be expressed in terms of noise, using a model trained with $\epsilon$-matching to predict noise realisation \cite{ho2020denoising}, based on the relation $\mathbf{\epsilon}_\theta\left(\mathbf{x}_t, t\right) = -\sqrt{1-\bar{\alpha}_t} \mathbf{s}_\theta\left(\mathbf{x}_t, t\right)$. Sticking with the score-based formulation, and given that the true data score is unknown, the neural network is trained with denoising score matching \cite{vincent2011connection}. In this approach, the model (i.e., the score model) parameterizes the score of the noisy data by employing the score of transition density $p_{0t}(\mathbf{x}_t \mid \mathbf{x}_0)=\mathcal{N}\left(\mathbf{x}_t \mid \alpha_t \mathbf{x}_0, \beta^2_t \mathbf{I}\right)$ in the objective, i.e.,
 \begin{equation}
\theta^*=\underset{\theta}{\arg \min } \ \mathbb{E}_{t, \mathbf{x}_t, \mathbf{x}_0} \left[\left\|\mathbf{s}_\theta(\mathbf{x}_t, t)-\nabla_{\mathbf{x}_t} \log p_{0t}(\mathbf{x}_t \mid \mathbf{x}_0)\right\|_2^2\right]
\label{eq:loss}
\end{equation}
Thus, at any intermediate time step $t$, given the estimate $\mathbf{\hat{x}}_t$, the trained model $\mathbf{s}_{\theta^*}\left(\mathbf{\hat{x}}_t, t\right)$, and the noise schedule, one can determine $\mathbf{\hat{x}}_{t-1}$ using Eq. \eqref{eq:backwardsdediscrete}.

\subsubsection*{Diffusion Models in CT reconstruction} 
Supervised learning methods for sparse-view CT reconstruction aim to learn a mapping from sparse measurements to the dense images by training the model on paired data \cite{ArMaOkSc19,SlYeEl23}. This approach limits the model to specific data acquisition parameters, necessitating retraining whenever the measurement parameters change. In contrast, unsupervised methods eliminate the need for retraining, freeing the model from dependency on the measurement process.
This disentanglement is theoretically achievable through Bayesian inference, which breaks down the posterior distribution \( p(\textbf{x}|\textbf{y}) \) into the likelihood \( p(\textbf{y}|\textbf{x}) \) and the prior \( p(\textbf{x}) \). The likelihood represents the knowledge of the measurement process and is already known in non-blind inverse problems. Thus, the learning problem is reduced to the unsupervised task of training a generative model to accurately capture the prior distribution. Finally, the solution to an inverse problem (CT reconstruction in our case) becomes nothing but a sample drawn from the posterior distribution.

Following the demonstration of their effectiveness in image generation \cite{dhariwal2021diffusion}, diffusion models have been employed as implicit or explicit generative priors for inverse imaging problems \cite{song2021solving, chung2022improving, chung2022diffusion, kawar2022denoising, chung2022come, graikos2022diffusion, feng2023score}. The baseline work \cite{song2021solving} suggests to train an unconditional diffusion model to capture the prior distribution of medical images. In time-conditioned setup of diffusion, Bayesian inference demands \( p_t(\textbf{y}|\textbf{x}_t) \) for every time step $t$, which is intractable \cite{chung2022diffusion}. For that reason, \cite{song2021solving} enforces the measurement consistency by the applications of one-step reverse diffusion followed by a projection onto the measurement subspace. This approach can be seen as an attempt to sample from an approximation of true posterior, rather than explicitly approximating \( p_t(\textbf{y}|\textbf{x}_t) \). In \cite{chung2022improving}, the authors improve \cite{song2021solving} by introducing a correction term that ensures the sample remains on the data manifold, which is coined manifold constrained gradient (MCG). In \cite{chung2022diffusion}, diffusion posterior sampling (DPS) is proposed, where the authors achieve measurement consistency by following a gradient toward higher measurement likelihood without performing any projection, thereby relaxing the linearity assumption. In \cite{chung2022come}, the authors observe that starting reverse diffusion from partially diffused images of incomplete observations rather than pure Gaussian noise can accelerate inference time without compromising reconstruction quality. In \cite{graikos2022diffusion}, the authors suggest to use unconditional diffusion models as plug-and-play priors. In \cite{kawar2022denoising}, the authors design a variational inference objective to approximate \( p_t(\textbf{y}|\textbf{x}_t) \), which enables direct sampling from the posterior. In \cite{feng2023score}, the authors benefit the score-based diffusion models' ability to provide explicit priors (i.e., exact probabilities) to determine resulting posterior.

Among the references above, in \cite{song2021solving} and \cite{chung2022improving}, the authors highlight CT reconstruction as one of their case studies, marking the first instances where a generative model achieved state-of-the-art results on clinical CT data. We will refer the baseline work \cite{song2021solving} as Score-SDE, following the convention in \cite{chung2022improving}. Given that the Score-SDE can modify any sampling algorithm to maintain measurement consistency, let's denote a generic sampling procedure as ${h}$, i.e., 
\begin{equation}
\hat{\mathbf{x}}^u_{t-1}={h}\left(\hat{\mathbf{x}}^c_{t}, \mathbf{z}_t, \mathbf{s}_{\mathbf{\theta}^*}\left(\hat{\mathbf{x}}^c_{t}, t\right)\right) \quad \hat{\mathbf{x}}^c_{T}, \mathbf{z}_t \sim \mathcal{N}(\mathbf{0}, \mathbf{I}), \quad t=T, T-1, \cdots, 0
\label{eq:unconditionalupdate}
\end{equation}
Here, at every iteration, ${h}$ processes a noisy sample $\hat{\mathbf{x}}^c_{t}$ and performs one-step noise reduction using the unconditional score model $\mathbf{s}_{\mathbf{\theta}^*}$. Note that the samples with superscripts $\hat{\mathbf{x}}^u$ and $\hat{\mathbf{x}}^c$ denote unconditional and conditional samples, respectively. Using the alternative formulation in \cite{song2021solving}, the measurement operator $\mathbf{A}$ in Eq. \eqref{eq:forwardmodel} can be decomposed as $\mathbf{A}=\mathcal{P}(\Lambda) \mathbf{R}$, where $\mathbf{R}$ corresponds to the discrete Radon Transform, and $\mathcal{P}(\Lambda) \in \{0,1\}^{m \times n}$ subsamples the sinogram into $\textbf{y}$ by removing each $i$th measurement according to the subsampling mask $\operatorname{diag}(\Lambda)$ with $\Lambda_{ii} = 0$. Therefore, as $tr(\Lambda) = m$ decreases, the measurements become sparser. In order to promote the measurements $\mathbf{y}$, Score-SDE implements a conditional update step as proximal mapping, which could later be written in closed form as follows: 

\begin{equation}
\hat{\mathbf{x}}^c_{t-1}=\mathbf{R}^{\dag}\left[\lambda \Lambda \mathcal{P}^{-1}(\Lambda) \hat{\mathbf{y}}_{t}+(1-\lambda) \Lambda \mathbf{R} \hat{\mathbf{x}}^u_{t-1}+(\mathbf{I}-\Lambda) \mathbf{R} \hat{\mathbf{x}}^u_{t-1}\right], \quad \hat{\mathbf{y}}_{t} \sim p_{t}\left(\mathbf{y}_{t} \mid \mathbf{y}\right).
\label{eq:songsconditionalupdate}
\end{equation}
where $\mathbf{R}^{\dag}$ denotes filtered backprojection (FBP). Since for the stochastic process $\left\{\mathbf{y}_t\right\}_{t \in[0,T]}$, $\mathbf{y}_t=\mathbf{A} \mathbf{x}_t+\alpha_t \mathbf{\eta}$; one can tractably generate an intermediate sample $\hat{\mathbf{y}}_t$ as $\hat{\mathbf{y}}_t=\alpha_t \mathbf{y}+\beta_t \mathbf{A} \mathbf{z}$, where $\mathbf{z} \in \mathbb{R}^n \sim \mathcal{N}(\mathbf{0}, \mathbf{I})$. So, Eq. \eqref{eq:songsconditionalupdate} computes a sinogram that promotes both the prior and the data fidelity, subsequently backprojecting it to the image domain, resulting in what is referred to as a conditional sample. An intuitive explanation to how it computes such a sinogram is as follows: After increasing the dimensionality of $\mathbf{y}$ from $m$ to $n$, for the observed parts of that sinogram, the available projections are combined with the forward-projected unconditional sample. The impact of each term is adjusted by the hyperparameter $\lambda \in [0,1]$. For the remaining parts of the sinogram, where no measurement is available, the forward-projected unconditional sample is used directly. This straightforward measurement consistency step corresponds to the projection onto the measurement subspace. 


\subsubsection*{Diffusion Models in OOD Detection} 
Following the demonstration of failure cases when utilizing generative models for likelihood-based OOD detection \cite{choi2018waic, nalisnick2018deep, ren2019likelihood, serra2019input, zhang2021understanding}, reconstruction-based OOD detection emerged as a promising alternative. 
It involves training a reconstruction model, which will ideally demonstrate poor performance when applied to an OOD sample, so that one can utilize the reconstruction error for scoring OOD-ness. This approach asks for a model capable of compressing all relevant information to a degree that ensures effective reconstruction of ID data while faltering on OOD data. This state of balance is governed by the concept of the \textit{information bottleneck}, which can be determined by the size of the latent dimension (as in the case of autoencoders) or the amount of noise (as in the case of denoising diffusion probabilistic models -- DDPMs \cite{ho2020denoising}). In this regard, DDPMs emerge as a favorable reconstruction-based OOD detection tool \cite{wyatt2022anoddpm, wolleb2022diffusion, graham2023denoising, tur2023exploring, liu2023unsupervised, linmans2024diffusion}, as the amount of noise is not a hyperparameter and can be externally tuned during inference. 

The main concept in these studies revolves around using diffusion models not for image generation, but instead for reconstruction. The difference is about where the reverse diffusion begins: pure Gaussian noise or partially diffused input. Given an unconditional diffusion model, the reverse diffusion typically starts with a sample $\mathbf{x}_T \sim \mathcal{N}(\mathbf{0}, \mathbf{I})$ and gradually denoise it until an image $\mathbf{\hat{x}}_0 \sim p(x)$ is eventually generated. But, in the case of OOD detection, the initial sample $\mathbf{x}_0$ is partially diffused up to $t_0$, i.e., $\mathbf{x}_{t_0} \sim p_{0{t_0}}(\mathbf{x}_{t_0} \mid \mathbf{x}_0)$, and the reverse diffusion concludes with $\mathbf{\hat{x}}_0 \sim p(\mathbf{x} \mid \mathbf{x}_0)$. For in-distribution (ID) samples, \(\mathbf{\hat{x}}_0\) should be close to \(\mathbf{x}_0\), whereas for OOD samples, \(\mathbf{\hat{x}}_0\) should be significantly different from \(\mathbf{x}_0\), i.e., ${d}(\mathbf{x}_{\text{in}}, \mathbf{\hat{x}}_{\text{in}}) < {d}(\mathbf{x}_{\text{out}}, \mathbf{\hat{x}}_{\text{out}})$, where $d$ denotes the reconstruction error. Note that, whenever we specify the distribution type as a subscript, it always refers to \(\mathbf{x}_0\) (or \(\mathbf{\hat{x}}_0\)) from that distribution, even though we omit the subscript ``\(0\)'' for readability. Two commonly used metrics for the error are the mean-squared error (MSE) and Learned Perceptual Image Patch Similarity (LPIPS), which computes the distance between features of a neural network extracts from two inputs.

The starting point of reverse diffusion ($t_0$) significantly impacts OOD detection performance. When $t_0$ is close to $N$, the reverse diffusion process begins from a highly noisy state, increasing the likelihood that the trajectory will result in a dissimilar reconstruction, even for an ID image. Conversely, when \( t_0 \) is close to zero, the residual signal already contains fine-grained features, leaving little for the model to fill in. As a result, it can accurately reconstruct both ID and OOD samples. This phenomenon is also known as the \textit{realism-faithfulness trade-off} \cite{meng2021sdedit}, which dictates that the higher $t_0$, the more realistic the sample becomes, but this comes at the cost of losing the fidelity to the input. For both high and low values of $t_0$, ID and OOD reconstruction errors become less distinguishable, resulting in subpar OOD detection performance. Therefore, the choice of $t_0$ needs to be tuned according to the training data as in \cite{wyatt2022anoddpm}. This issue can also be addressed by using multiple reconstructions from a range of noise values, as proposed in \cite{graham2023denoising} and further explored in \cite{liu2023unsupervised, linmans2024diffusion}. We refer to this approach as ``multi-scale scoring,'' as varying noise levels enable conditioning on different scales of image features during reconstruction. In this work, we adopt the approach in \cite{graham2023denoising,graham2023unsupervised}, utilizing their publicly available code.


\section{Problem Statement} 
\label{sec:problem}

\begin{figure*}[t!]
 \centering
  \begin{subfigure}[b]{0.95\textwidth}
  \centering
{\includegraphics[width=\textwidth,keepaspectratio=true]{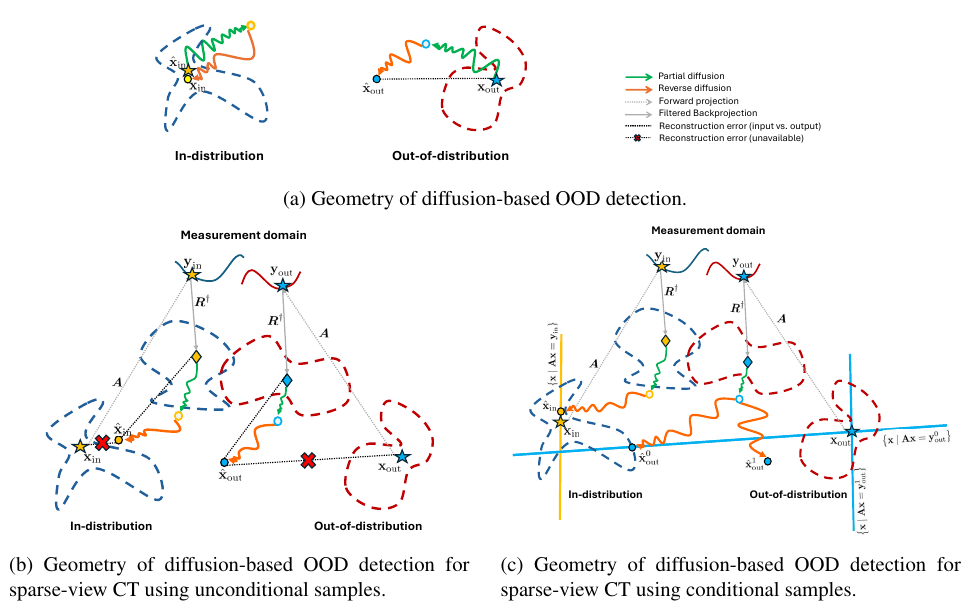}}
 \end{subfigure}
\caption{Diffusion-based OOD detection illustrations from a data manifold perspective. In each figure, input images undergo partial diffusion (green arrows) followed by denoising (orange arrows). Ideally, the reconstruction error is represented by the distance between the input and output, shown as matching-colored stars and circles in (a), and matching-colored circles and rhombuses in (b) and (c). In (a), the reconstruction is directed toward the learned ID data manifold, pulling $\hat{\textbf{x}}_{\text{out}}$ away from $\textbf{x}_{\text{out}}$, resulting in a larger reconstruction error. In (b) and (c), the inputs are FBP mappings of measurements (sinograms), which may lie on their own manifolds. The reconstruction errors between the ground-truth images and the reconstructions are not available due to the absence of ground-truths (stars). The subfigures (b) and (c) compare unconditional and conditional sampling for OOD detection, respectively. In (c), three different measurement subspaces are displayed. Regardless of the solver used, the measurement subspace always acts as an attractor and may confuse the OOD detector by diminishing the influence of the ID data manifold, as $\left\{ \textbf{x} \mid \textbf{A}\textbf{x} = \textbf{y}^1_{\text{out}}\right\}$ could potentially does.}
\label{fig:illustration_ood_vs_oodct}
\end{figure*}

Let us denote the entire partial diffusion and denoising machinery as $f_{\theta^*}$, regardless of whether we train the model using score matching or $\epsilon$-matching. When applied, it performs: $\mathbf{\hat{x}}_{0}^{u \mid t_0} \leftarrow f_{\theta^*}(\mathbf{{x}}_0, t_0)$, meaning that some Gaussian noise is added to the input $\mathbf{{x}}_0$ in $t_0$ steps, with step sizes controlled by the noise schedule, and subsequently reconstructed through reverse diffusion. The superscript \( u \) indicates that the computation of this sample solely depends on the prior, without involving any knowledge about the physical model. As mentioned earlier, reconstruction-based OOD detection links the ability of $f_\theta$ to reconstruct a test image $\mathbf{{x}}_0$ to its distribution (ID or OOD) by typically assessing the reconstruction error $d(\mathbf{{x}}_0, \mathbf{\hat{x}}_{0}^{u \mid t_0})$. Since the model is trained on the in-distribution images, it attempts to map any noisy test image towards in-distribution data manifold. This phenomenon is illustrated in Fig. \ref{fig:illustration_ood_vs_oodct}a, from a geometrical viewpoint of the data manifolds. Given a noisy (partially diffused) input image, the diffusion model maps it towards in-distribution data manifold, which will cause a larger reconstruction error when provided with an OOD image. In this study, the downstream task for the OOD detector is CT reconstruction, which amounts to recovering a hidden image $\mathbf{{x}}$ from its measurements $\mathbf{y}$, assuming that the physical model of the measurement process Eq. \eqref{eq:forwardmodel} is known. Given that the problem is ill-posed, regularization is required. A diffusion model capturing the target distribution serves as the learned prior, enabling us to sample from it. Therefore, we already have a pre-trained diffusion model for the CT reconstruction task, which may also represent the in-distribution from the perspective of the OOD detector. 

In CT reconstruction, at test time, all that we have access to are the sparsely acquired measurement data. By the definition of the problem, the ground-truth is unavailable. It could be possible to utilize the FBP reconstructions as input, i.e., $\mathbf{x}_{0} = \mathbf{R}^\dag {\mathbf{y}}$, since the noisy data manifolds of FBP images from sparse projections and the noisy data manifolds of ground-truth images are likely to overlap. Our intuition builds on research demonstrating that diffusion models can guide off-manifold images back onto the data manifold, supporting the learned distribution as described by the manifold hypothesis \cite{wenliang2022score}. This generative bias has been widely utilized in OOD detection, as shown in \cite{liu2023unsupervised}, where reverse diffusion was applied to masked images deliberately lifted from their original manifolds. The model successfully reconstructed in-distribution images by leveraging the retained structural cues. While FBP artifacts differ from simple masked regions and may introduce structures absent from the training distribution, the reconstructions may still preserve meaningful structural information that aligns with the learned data manifold.  Consequently, one may expect the model to map in-distribution FBP reconstructions of sparse projections to the in-distribution data manifold. A geometric interpretation of this approach is illustrated in Fig. \ref{fig:illustration_ood_vs_oodct}b. Nonetheless, in this scenario, the input $\mathbf{R}^\dag {\mathbf{y}}$ and the output $\mathbf{\hat{x}}_{0}^{u \mid t_0}$ are not directly comparable. Namely, the error $d(\mathbf{R}^\dag {\mathbf{y}}, \mathbf{\hat{x}}_{0}^{u \mid t_0})$ no longer reflects the intended reconstruction error and as sparsity increases, its effectiveness drops even more, since the in-distribution FBP samples diverge further from their hidden ground-truth. Considering these facts, one question that we will address is how to define the notion of reconstruction error in the case of sparse-view CT reconstruction.

As described in Section \ref{sec:background}, diffusion-based CT reconstruction methods iteratively sample in a way that ensures consistency with both the prior and observed measurements. Consequently, the final reconstruction can be viewed as an approximate conditional sample from the posterior, i.e., $\mathbf{\hat{x}}_{0}^{c} \sim p(\mathbf{x}|\mathbf{y})$. Let's denote a conditional sample obtained in the same manner, but by reversing a partially diffused input over $t_0$ steps, as $\mathbf{\hat{x}}_{0}^{c \mid t_0}$. By again using the FBP reconstructions as input, we can measure the reconstruction error employing these conditional samples. Fig. \ref{fig:illustration_ood_vs_oodct}c illustrates how the incorporation of the measurements changes the reverse diffusion output for OOD detection. Although the precise output depends on the solver (e.g., Score-SDE \cite{song2021solving}, MCG \cite{chung2022improving}, DPS \cite{chung2022diffusion}), the measurement subspace consistently acts as an attractor. The reconstruction is likely to compromise between the in-distribution data manifold and the measurement subspace. When an ID image’s reconstruction is drawn toward the measurement subspace, it typically yields a clearer ID signal. However, a strong attraction for an OOD image’s reconstruction can reduce the effectiveness of the reconstruction error as an OOD score. Building on these insights, we will further investigate if and when the common practice of conditioning in sparse-view CT reconstruction complicates OOD detection.

\begin{figure*}[t!]
 \centering
  \begin{subfigure}[b]{0.9\textwidth}
  \centering
  {\includegraphics[width=\textwidth,keepaspectratio=true]{ 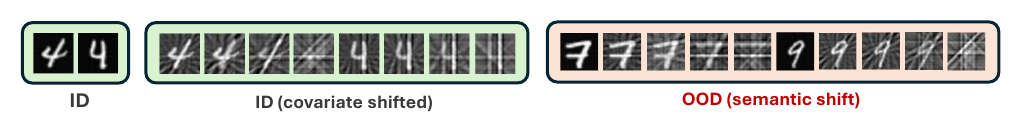}}
 \end{subfigure}
\caption{Example thumbnails from a scenario in which the diffusion model is trained on handwritten ``4''s from the MNIST dataset \cite{lecun1998mnist,deng2012mnist} to serve as a prior for the sparse-view CT reconstruction task. Performing OOD detection under these conditions requires distinguishing the FBP reconstructions of partially measured ``4''s from those of other digits. Similar to semantic anomaly detection, the detector should treat covariate-shifted images as ID while identifying semantic shifts as OOD.}
\label{fig:mnist_id_vs_ood}
\end{figure*}

In summary, we have two tasks that involve reconstruction: CT reconstruction and OOD detection. When OOD detection adopts the conditioning from CT reconstruction, the two tasks are likely to operate in tandem. However, an image could fall outside the in-distribution, yet still be reconstructed with near-ID quality. The dilemma is whether the OOD detector should assign a high or low OOD score to such an image. Fig. \ref{fig:mnist_id_vs_ood} provides example thumbnails to illustrate our criteria for classifying images as ID or OOD. Our approach is closely related to Semantic AD, where research aims to define OOD scores that capture the semantic essence of images (e.g., \cite{peng2019domain, lin2021domain, zhou2022domain, yang2023full}). In this context, FBP reconstructions from sparse measurements of ID samples can be interpreted as covariate shifts. Consequently, we may expect our OOD detector to effectively discriminate semantic shifts, while remaining robust against covariate shifts. Using \(\mathbf{\hat{x}}_{0}^{c \mid t_0}\) alone appears to limit effective discrimination of semantic shifts. On the other hand, relying solely on the generative bias towards reconstructing ID training samples by using \(\mathbf{\hat{x}}_{0}^{u \mid t_0}\) to compute the reconstruction error may lack resilience to covariate shifts. This insight leads us to propose an OOD score that incorporates both conditional and unconditional versions of the reconstruction error, effectively balancing them.

\section{Methodology}
\label{sec:methodology}
In this work, we adopt the multiple reconstructions strategy utilized in \cite{graham2023denoising, liu2023unsupervised, linmans2024diffusion}. Recalling that $d$ measures the reconstruction error, an OOD score based on multiple reconstructions of the inputs noised to different extents can be generically defined as follows:
 \begin{equation}
    \mathcal{S} = \frac{1}{|\mathcal{T}|} \sum_{t_0 \in \mathcal{T}} F\big( d(\mathbf{z}_{\text{inp}}, \mathbf{z}^{t_0}_\text{outp}) \big)
\label{eq:genericOODscore}
\end{equation}
where \(\mathbf{z}_\text{inp}\) and \(\mathbf{z}^{t_0}_\text{outp}\) are substitutes for any input and any output, respectively. The function \(F\) measures how the reconstruction error compares to the expected reconstruction errors for in-distribution images. For that purpose, \cite{graham2023denoising} suggests to use the Z-score, representing the deviation of the reconstruction error from the mean of the errors obtained from a validation set. We follow the same approach by using a validation set composed of full-view in-distribution images. Thus, similar to \cite{graham2023denoising}, the resulting score \(\mathcal{S}\) is the average Z-score across the individual Z-scores, each of which is computed with respect to different \(t_0\) values in order to maintain the realism-faithfulness trade-off.

For the diffusion model, we used the DDPM as provided in the adopted code from \cite{graham2023denoising}. This model is a time-conditioned U-Net \cite{ronneberger2015u} with an architecture identical to that described in \cite{rombach2022high}. It is trained with $\epsilon$-matching to predict the added noise at step $t$. During training, we set \( T = 1000 \) and employed a linear noise schedule with \(\beta_t\) ranging from 0.0015 to 0.0195. Number of sampling steps to use with the PLMS sampler\cite{liu2022pseudo} is set to 100, with equally spaced $t_0$ values from $10$ to $1000$. This leads to a tenfold speed-up. For multiple reconstructions, we skipped some of the sampling points by factor $7$, resulting in $t_0$ values of $\mathcal{T} = \{10, 80, 150, ..., 920, 990\}$ to start reconstruction. We later chose \( t_0 = 150 \) as minimum $t_0$ and excluded \( t_0 = 10 \) and \( t_0 = 80 \) based on analyses indicating that excessive reliance on the FBP input adversely impacts results, especially in projection domain comparisons. Supplementary Figure S1 online provides some findings from these analyses. Additionally, \( t_0 = 990\) was excluded as it is too close to \( t_0 = 1000 \), which is reserved specifically for CT reconstruction. Consequently, we perform $|\mathcal{T}| = 12$ reconstructions leading to a 12-dimensional reconstruction error, spanning from the most faithful reconstruction to the filtered-backprojected observation ($t_0 = 150$), to the most realistic one ($t_0 = 920$). Note that, this setup requires 642 model evaluations.

For CT reconstruction, we employ MCG \cite{chung2022improving} to perform posterior sampling given the measurements. It improves the reconstruction quality achieved by the Score-SDE through the introduction of an additional constraint. It is implemented using a gradient step between Score-SDE's unconditional and conditional updates, with the gradient \(\nabla_{\mathbf{x}_t}\left\|\mathbf{y}-\mathbf{A} \bar{\mathbf{x}}_0(\mathbf{x_t})\right\|_2^2 \). Here, $\bar{\mathbf{x}}_0(\mathbf{x_t})$ computes a posterior expectation $\mathbb{E}\left[\mathbf{x}_0 \mid \mathbf{x}_t\right]$ using Tweedie’s Bayes optimal denoising step \cite{robbins1992empirical}, i.e., $\bar{\mathbf{x}}_0(\mathbf{x_t}) = \left(\mathbf{x}_t+\beta_t^2 \mathbf{s}_{\theta^*}\left(\mathbf{{x}}_t, t\right)\right) / a_t$ given that the forward diffusion is modelled as $\mathbf{x}_t \sim \mathcal{N}\left(a_t \mathbf{x}_0, \beta_t^2 I\right)$. Since $\bar{\mathbf{x}}_0(\mathbf{x_t})$ is intended to lie on the data manifold, the MCG step mitigates the risk of the reverse diffusion trajectory straying from the generative manifolds. In other words, this step ensures that the unconditional sample drawn using Eq. \eqref{eq:unconditionalupdate} remains on the plane tangent to the data manifold, thus preventing significant accumulation of inference errors during the subsequent measurement consistency step in Eq. \eqref{eq:songsconditionalupdate}. Note that, our CT reconstruction uses the predictor-corrector (PC) sampler of VP-SDE from \cite{song2020score}. Following \cite{chung2022improving}, we apply the MCG step after completing one cycle of the corrector-predictor update.

{
\setlength{\parindent}{0mm}
\begin{minipage}{0.65\textwidth}
We confine our experimental design to MNIST images \cite{lecun1998mnist,deng2012mnist} of size $28 \times 28$. We conduct experiments using a one-class setup, where the diffusion model is trained on a single digit. The training digits are cherry-picked based on their likelihood of being confused with other digits after performing OOD detection as described in \cite{graham2023denoising}. Accordingly, the digits 4, 7, and 8 have become the training digits. Moreover, we select three OOD digits for each training digit, according to their visual resemblance, in order to accommodate near-OOD challenges in the context of CT. The resulting experimental setup is shown in Table \ref{tab:mnistIDandOODdigits}. From now on, we will use ``MNISTx'' to refer to a dataset consisting of handwritten images of a digit \(x\). Additionally, in the experiments, we consider two types of sparse-view test sets representing moderate and high sparsity scenarios, unless otherwise stated. The moderate sparsity set includes ID and OOD images with \{18, 12, 9\} projection angles, while the high sparsity set includes \{6, 5, 4\} projection angles. For each projection count, we have 200 ID and 200 OOD images, resulting in a total of 1200 images per test set. 
\end{minipage}%
\hfill%
\begin{minipage}{0.3\textwidth}
\centering
\captionof{table}{ID and corresponding OOD digits, whose MNIST images \cite{lecun1998mnist,deng2012mnist} are used in the experiments.}
\begin{tabular}{c|c}
\small{ID Digit} & \small{OOD Digits} \\
\hline
\hline
4            & 6, 7, 9    \\
\hline
7            & 1, 4, 9    \\
\hline
8            & 3, 5, 9   
\end{tabular}
\label{tab:mnistIDandOODdigits}
\end{minipage}
}

\subsubsection*{RQ-1: How should the OOD reconstruction error be defined in sparse-view tomography?}
\begin{figure*}[t!]
 \centering
 \begin{subfigure}[b]{0.95\textwidth}
 \centering
 {\includegraphics[width=\textwidth,keepaspectratio=true]{ 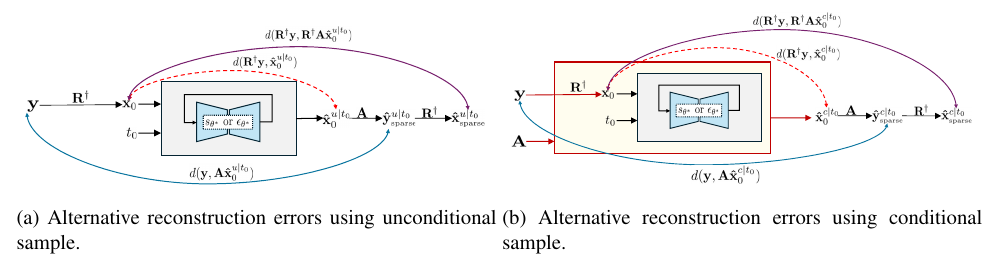}}
 \end{subfigure}
\caption{Block diagrams illustrating alternative components across various spaces for reconstruction error. The black-bordered box encapsulates the diffusion model, which, given $t_0$ and the input $\mathbf{x}_0$ (the filtered-backprojected $y$), performs reconstruction starting from corrupted $\mathbf{x}_0$. Curved double arrows indicate potential comparisons used to measure the reconstruction error. The red dashed arrow indicates the classical comparison between the input and the output of the diffusion model, which does not seem valid in our case. The blue arrow extends the comparison to the projection domain by forward projecting the reconstruction. In the case of the purple arrow, the output is subjected to both forward projection and FBP, to further map the sparse-view reconstruction back into the image domain. In (b), the same comparisons are shown but this time by considering the conditional sample obtained through the CT reconstruction from partially diffused FBP images, represented by the red-bordered box surrounding the black-bordered one.}
 \label{fig:blockdiagrams}
 \end{figure*}

Given Eq. \eqref{eq:genericOODscore}, the question now is how to define \(\mathbf{z}_\text{inp}\) and \(\mathbf{z}^{t_0}_\text{outp}\), so that the disparity between them gives rise to a valid and a distinctive reconstruction error. In Fig. \ref{fig:blockdiagrams}, some potential pairs are illustrated. The variable \(\mathbf{z}_\text{inp}\) can either be the measurement $\mathbf{y}$ fed into the entire CT reconstruction system, or the FBP reconstruction $\mathbf{R}^{\dagger} \mathbf{y}$, which is the main subject of the partial diffusion. The variable \(\mathbf{z}^{t_0}_\text{outp}\) can be derived from the unconditional reconstruction $\mathbf{\hat{x}}_{0}^{u \mid t_0}$ from the noisy input $(\mathbf{R}^{\dagger} \mathbf{y})_{t_0}$. Recall that the model is trained on the full-view images. Since the training data manifold differs from the manifold on which sparse-view FBP images lie, we cannot expect the reconstruction to closely approximate $\mathbf{x}_0 = \mathbf{R}^{\dagger} \mathbf{y}$, as was the case in Fig. \ref{fig:illustration_ood_vs_oodct}a. Namely, the intention is not to reconstruct something close to \(\mathbf{x}_0\), but rather to reconstruct a full-view image by remaining faithful to the information available in a corrupted version of \(\mathbf{x}_0\). This argument detains $d(\mathbf{R}^{\dagger} \mathbf{y}, \mathbf{\hat{x}}_{0}^{u \mid t_0})$ from being a valid error. It is also not so distinctive, as increasing sparsity causes the input manifold of ID images to shift as far away as the OOD data manifolds (See Fig. \ref{fig:illustration_ood_vs_oodct}b for a visualization of almost indistinguishable disparities.)

For a valid comparison, one can consider forward projecting the reconstruction and comparing it with the raw measurements, i.e., \(d(\mathbf{y}, \mathbf{A} \mathbf{\hat{x}}_{0}^{u \mid t_0})\). The idea is straightforward: because we cannot compare the reconstruction with the ground-truth directly due to its unavailability in the downstream task, we can instead compare the sparse projections of them. However, disparities in spatial domain may get diminished when propagating to the projection domain. This means that comparisons in the projection domain may not accurately reflect the magnitude of structural differences in the spatial domain, as even slight changes in a few projections can result from significant differences in the image. Therefore, we can also consider calculating the reconstruction error in the spatial domain by further performing FBP, i.e., \(d(\mathbf{R}^{\dagger} \mathbf{y}, \mathbf{R}^{\dagger} \mathbf{A} \mathbf{\hat{x}}_{0}^{u \mid t_0})\). This error measure may be influenced by the fact that FBP can introduce additional artifacts \cite{frikel2013characterization}, which might skew the comparison, and smooth out the differences between images. Nevertheless, it still retains the potential to act as a surrogate for reconstruction error in some cases, as empirically demonstrated in the following section. 

On the surface, the posterior distribution $p(\mathbf{x}|\mathbf{y})$ may appear irrelevant to OOD detection approach we adopted from \cite{graham2023denoising}. The diffusion model already captures the in-distribution $p(\mathbf{x})$, and the role of the OOD detector is to return distinctively high reconstruction errors for the images that do not belong to that distribution. Distinctively high errors occur when OOD images are poorly but ID images are well reconstructed, and the quality depends on the amount of residual signals remaining after noise (i.e., information bottlenecks). In our case, those residual signals originate not from the original image we aim to reconstruct, but from the FBP reconstructions $\mathbf{R}^{\dagger} \mathbf{y}$, whose accuracy depends on the physical model. Therefore, the information bottlenecks may not only be less representative but also potentially misleading, increasing the likelihood of misclassifying ID images as OOD. This situation calls for an alternative source of data fidelity, prompting the use of $\mathbf{\hat{x}}_{0}^{c \mid t_0}$ to calculate the reconstruction error, either in the projection domain, i.e., \(d(\mathbf{y}, \mathbf{A} \mathbf{\hat{x}}_{0}^{c \mid t_0})\), or by applying further FBP, i.e., \(d(\mathbf{R}^{\dagger} \mathbf{y}, \mathbf{R}^{\dagger} \mathbf{A} \mathbf{\hat{x}}_{0}^{c \mid t_0})\). As a result, we have a total of six comparison schemes to evaluate, as depicted previously in Fig. \ref{fig:blockdiagrams}.

\subsubsection*{RQ-2: What are the failure scenarios associated with conditional sampling in OOD detection?}
When the CT reconstruction is constrained on the measurement $\mathbf{y}$ originating from an OOD image, it is likely to fail, especially if the measurement lacks sufficient information (e.g., in sparse-view CT). However, there are scenarios where it can still generalize to OOD images: (1) when the ID and OOD distributions are very close or overlapping, and (2) when the distributions are separate enough, yet the measurements are sufficiently informative to counteract the misleading prior. These scenarios may also apply when we conditionally reconstruct from partially diffused FBP reconstructions, leading to challenges in accurate OOD detection. Namely, while the use of the unconditional samples $\mathbf{\hat{x}}_{0}^{u \mid t_0}$ clearly blur the ID signal, the conditional samples $\mathbf{\hat{x}}_{0}^{c \mid t_0}$ may, in some cases, distort the OOD signal. Both result in a reduced contrast between ID and OOD scores. In this regard, we design experiments to highlight the cases where conditional sampling complicates the OOD detection by yielding reconstruction errors for OOD images that are as low as those for ID images.
 
\subsubsection*{RQ-3: Can the contrast between ID and OOD scores be enhanced by weighting conditional and unconditional reconstruction errors?}
When the OOD detector's inability to avoid accurately reconstructing OOD images is due to data fidelity (i.e., conditioning on the measurements), benefiting from unconditional samples appears to be reasonable. The question is, can we disentangle such cases from the others? In typical reconstruction-based OOD detection, it is usually reasonable to assume that the further an image is from the in-distribution, the higher the OOD scores it is likely to produce. We may expect to observe this behaviour when using $\mathbf{\hat{x}}_{0}^{u \mid t_0}$. However, conditioning alters this relation by enabling better reconstruction even for the distant images. Referring back to Fig. \ref{fig:illustration_ood_vs_oodct}c, one can speculate that for a distant image, the farther the measurement subspace is from the in-distribution manifold, the more likely an OOD reconstruction is to escape the influence of the prior distribution. In other words, the more the OOD measurement deviates from ID measurements, the more likely it is to result in a quality reconstruction. Accordingly, we propose adjusting the conditional score based on the distance of the measurement subspace from the in-distribution mean by incorporating errors derived from unconditional samples, i.e.,
\begin{equation}
    \mathcal{S} = \frac{1}{|\mathcal{T}|} \sum_{t_0 \in \mathcal{T}} F\Big(\big(1-{w}\big) d\big({\Gamma} \mathbf{y}, {\Gamma} \mathbf{A} \mathbf{\hat{x}}_{0}^{c \mid t_0}\big)+ w d\big({\Gamma} \mathbf{y}, {\Gamma} \mathbf{A} \mathbf{\hat{x}}_{0}^{u \mid t_0}\big)\Big), \quad \quad \quad w = \frac{\left\|(\mathbf{y} - \mathbf{A} \mu_\theta)\right\|^2}{\left\|\mathbf{y}\right\|^2+\left\|\mathbf{A} \mu_\theta \right\|^2} 
    \label{eq:weighting}
\end{equation}
where ${\Gamma}$ is either $\mathbf{I}$ or $\mathbf{R}^{\dagger}$, and $w$ computes the distance in the projection domain as a normalized $L^2$ distance between the measurement and and the forward-projected mean of the training set, $\mu_\theta$. In Eq. \eqref{eq:weighting}, when $w$ is large, the OOD detector increasingly relies on the prior. In contrast, when $w$ is small, the OOD detector places greater emphasis on reconstruction errors derived from conditional samples. Note that, using the mean to describe in-distribution in Eq. \eqref{eq:weighting} overlooks the possibility of elongated data manifolds, where the measurement subspaces associated with ID samples can also lead to large $w$ values. This may promote unconditional sampling within the validation set, leading to a reference distribution with high variance, which subsequently reduces the ability to detect OOD images due to a collapse in Z-scores. Therefore, this approach can be seen as a trade-off between the detector's ability to maintain performance across different conditions and its overall effectiveness. In other words, it helps prevent the detector from being completely misled by conditioning, albeit at the expense of some performance loss, particularly in scenarios with skewed in-distribution.


\section*{Experimental Results}

\subsubsection*{RQ-1: How should reconstruction error be defined in sparse-view tomography?}
\begin{figure*}[t!] 
 \centering
 \begin{subfigure}[b]{0.95\textwidth}
 \centering
 {\includegraphics[width=\textwidth,keepaspectratio=true]{ 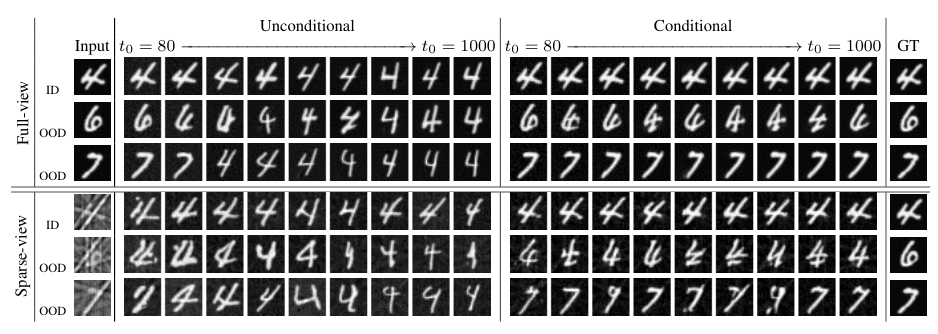}}
 \end{subfigure}
\caption{Example unconditional and conditional reconstructions for nine different $t_0$'s, from the model trained on MNIST4. Rows 1-3 present reconstructions from full-view input images from MNIST4, MNIST6, and MNIST7, respectively, whereas rows 4-6 display reconstructions from FBP mappings of sparse measurements with 5 projection angles for the same images. The images in the final column of the conditional reconstruction section correspond to the estimated CT reconstructions, which can be compared to the ground-truths (GT) shown in the last column of the plot.}
\label{fig:examplereconstructionsmnist}
\end{figure*}

Fig. \ref{fig:examplereconstructionsmnist} presents visual examples that offer insight into how reconstructions from partial diffusions evolve with respect to $t_0$, in both conditional and unconditional cases. The first quarter of the figure aligns with the approach proposed in \cite{graham2023denoising}, as the images are full-view and the reconstructions rely solely on the DDPM model trained on MNIST4, representing unconditional samples. It can be observed that the OOD reconstructions start to diverge from the ground-truth (also input) images sooner than the ID reconstructions, ensuring an overall higher OOD score. However, in sparse-view tomography using the same model, it is less obvious how the ID and OOD reconstructions differ in their deviation from the ground truth. Although semantically OOD samples still diverge early in the chain, error metrics may not effectively capture this difference, especially since the comparisons are limited to the sparse domain. Consequently, even ID images exhibit early divergence, reducing the contrast between ID and OOD errors. On the other hand, conditional reconstructions appear to significantly strengthen the ID signal. Across various $t_0$ values, the reconstructions for the digit ``4''  closely resemble the ground truth in both full- and sparse-view settings (with 5 projection angles). In contrast, the OOD digit ``6'' diverges early, achieving a clear distinction. This outcome aligns with our expectations. However, the situation of OOD digit ``7'' is ambiguous. Its CT reconstruction (reconstruction from $t_0 = 1000$) closely resembles the ground truth, suggesting it does not exhibit the typical failures expected from OOD images where CT reconstruction would generally falter. Despite this, its status as a semantic shift still requires it to be identified as OOD. Although visual inspection suggests of earlier reconstructions in the $t_0$ chain hint at its OOD-ness, this may not be adequately captured by the reconstruction errors. This discrepancy likely stems from the error being evaluated in a sparse domain, where only partial signals are compared. As a result, our OOD detector may incorrectly classify this signal as ID. Before presenting quantitative and broader analyses, we refer the reader to Supplementary Figure S2 online, which provides an ablation study on the effect of using noisy FBP reconstructions as input, considering the same ID and OOD dataset pair as in Fig. \ref{fig:examplereconstructionsmnist}. This analysis compares OOD detection performance using area under the curve (AUC) scores, evaluating multi-scale reconstructions from noisy FBP inputs against multiple instances of unconditional and conditional generations. In the latter case, forward and reverse diffusions run fully (i.e., $t_0 = 1000$), with conditional generation corresponding to CT reconstruction.
 
\begin{figure*}[t!]
 \centering
  \begin{subfigure}[b]{\textwidth}
  \centering
{\includegraphics[width=\textwidth,keepaspectratio=true]{ 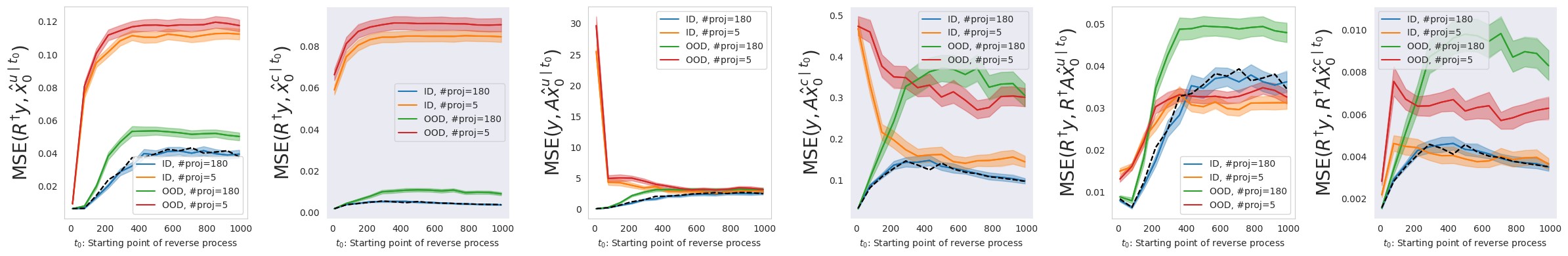}}
 \end{subfigure}
 \caption{Mean-squared errors as a function of $t_0$ based on the model trained on MNIST4 with MNIST7 treated as OOD dataset. The analysis considers full-view and sparse-view cases (with 5 projections) for both ID and OOD datasets. Each plot shows the comparison in a different domain, using either unconditional (light background) or conditional (dark background) reconstructions. Black dashed line shows how the respective error evolved for the validation set. The closer the errors are to this line, the lower the OOD score. This statistical analysis is performed over 200 samples per case, and the shaded areas around the colored lines indicate standard deviation.}
 \label{fig:rq1msewrtt0}
\end{figure*}

In Fig. \ref{fig:rq1msewrtt0}, each plot depicts how the reconstruction error (MSE) varies along the $t_0$ chain for ID and OOD images, considering both full-view and sparse-view (with 5 projection angles) cases, and using the surrogate reconstruction errors illustrated in Fig. \ref{fig:blockdiagrams}. Ideally, OOD errors (green and red lines) should deviate as much as possible from the validation set errors (black dashed line), while ID errors (blue and orange lines) should remain as close as possible. The first two plots highlight the impact of direct comparison between the input and the output. Since the reference input image varies with the number of projections, it’s not possible to accurately classify an image as ID or OOD without taking the number of projections into account. In other words, it is not possible to set a single threshold to separate the sparsely acquired ID signals from the full-view (or less sparse) OOD signals (See the orange and the green lines), because the way they differentiate from each other varies with the number of projections. Moreover, even when adjusting our analysis to factor in the number of projections, the differentiation between errors in both unconditional and conditional cases (See the red and the orange lines) remains unclear. Although Fig. \ref{fig:rq1msewrtt0} specifically uses the model trained on MNIST4 with MNIST7 as the OOD dataset, this behavior is consistent across the other ID vs. OOD cases we examined. The next pair of plots (3rd and 4th) is concerned with the distance between sparse sinograms. In the unconditional case, as $t_0$ increases, the errors in full-view ID reconstructions (blue line) grow, since the process shifts towards image generation driven by the prior, rather than reconstruction. For the conditional reconstructions, the error initially rises gradually with increasing $t_0$, until reaching a point where the prior and data fidelity are sufficiently balanced to keep the reconstruction on the data manifold. After this point, the error smoothly declines to a value slightly higher than the initial error. For the full-view OOD reconstructions (green line), the trend of rapid initial growth followed by stabilization is observed in both the unconditional and conditional cases. The distinction between full-view ID and OOD sinograms appears clearer when conditional reconstructions are used. When it comes to the sparse-view case, the pattern is somewhat reversed. We observe gradual decline, followed by a slight rise at the end of $t_0$ chain. The sparse-view ID errors (orange line) initially exceed the full-view OOD errors (green line), and this remains true until around $t_0 \sim 400$ in the unconditional case. Afterward, while the average errors for both full-view and sparse-view OOD images rise above those for ID images, there is significant overlap, leading to insufficient separation between ID and OOD errors. Fortunately, with the use of conditional reconstructions, full-view OOD errors begin to surpass sparse-view ID errors much earlier, around $t_0 \sim 200$, with a substantial margin. For the specific use case depicted in Fig. \ref{fig:rq1msewrtt0}, employing the sinogram of the conditional reconstruction for comparison appears to provide clear separation between ID and OOD errors, and one can even further enhance the OOD detection performance by excluding $t_0$ values below 200. The 5th and the 6th plots examine the distances in the FBP domain, where the filtered-backprojected sinograms from sparse measurements present. In the conditional case, the overall pattern is similar to what we observed in the sinogram domain, except that the errors from sparse-view measurements are proportionally lower in the FBP domain. This is likely because the FBP process averages out small differences across individual projections, causing them to have a reduced impact on the final error. On the other hand, in the unconditional case, both full-view and sparse-view reconstructions result in rapidly increasing errors that eventually stabilize. This behavior contrasts with the sinogram domain comparisons for sparse-view reconstructions, where errors decrease as $t_0$ increases. The difference arises because, early in the $t_0$ chain, the reconstructions closely align with the input FBP mapping, which we also use as the reference image for comparisons. As $t_0$ increases, the prior progressively dominates, leading to larger reconstruction errors unless the MSE's pixel-level assessment happens to capture semantically matching ID images. Nevertheless, the application of FBP seems to reduce the visibility of errors in considerably sparse signals, making them less pronounced compared to those observed in the sinogram domain.

\definecolor{verylightgray}{rgb}{0.92, 0.92, 0.92}
\definecolor{verylightyellow}{rgb}{0.98, 0.98, 0.80}
\newcommand{\blue}[1]{\textcolor{blue}{#1}}
\newcommand{\red}[1]{\textcolor{red}{#1}}

\begin{table}[t!]
\centering
\caption{AUC scores for three models trained on full-view MNIST4, MNIST7, and MNIST8 images. Red text indicates the highest value in each column separately for two cases: Case-1: Moderate Sparsity, where the number of projection angles (\#proj) is \{18, 12, 9\}, and Case-2: High Sparsity, where \#proj = \{6, 5, 4\}. We highlight the columns with unsatisfactory performance. Confidence bounds are reported for all cases using 1000-fold bootstrapping.}
\resizebox{\textwidth}{!}{%
\arrayrulecolor{black}
\begin{tabular}{p{0.3cm}|p{2.7cm}|P{5.2em}>{\columncolor{verylightyellow}}P{5.2em}>{\columncolor{verylightyellow}}P{5.2em}|>{\columncolor{verylightyellow}}P{5.2em}P{5.2em}P{5.2em}|P{5.2em}P{5.2em}P{5.2em}}
& \multicolumn{1}{l|}{ID Dataset:} & \multicolumn{3}{c|}{MNIST4} & \multicolumn{3}{c|}{MNIST7} & \multicolumn{3}{c}{MNIST8} \\
\hline\hline
&\multicolumn{1}{l|}{OOD Dataset:} & \small MNIST6 & \small MNIST7 & \small MNIST9 & \small MNIST1 & \small MNIST4 & \small MNIST9 & \small MNIST3 & \small MNIST5 & \small MNIST9 \\
\hline
\multirow{6}{*}{\rotatebox[origin=c]{90}{\parbox{3cm}{\centering {Moderate Sparsity}}}} & MSE(${R}^{\dagger} {y}, {\hat{x}}_{0}^{u \mid t_0}$)       & 77.01 
\notsotiny  [74.3, 79.6]    & 63.26 
\notsotiny  [59.8, 66.2]    & 57.10 
\notsotiny  [53.9, 60.2]    & 39.02 
\notsotiny  [35.7, 42.2]    & 58.19 
\notsotiny  [54.8, 61.4]    & 57.87 
\notsotiny  [54.7, 60.9]    & 57.85 
\notsotiny  [54.7, 61.1]    & 56.08 
\notsotiny  [52.8, 59.4]    & 54.29 
\notsotiny  [51.0, 57.3]    \\
& MSE(${y}, {A} {\hat{x}}_{0}^{u \mid t_0}$)           & 83.95 
\notsotiny  [81.8, 86.3]    & 67.32 
\notsotiny  [64.1, 70.4]    & 55.40 
\notsotiny  [52.3, 58.6]    & 35.58 
\notsotiny  [32.3, 38.7]    & 70.61 
\notsotiny  [67.5, 73.2]    & 65.36 
\notsotiny  [62.4, 68.3]    & 58.59 
\notsotiny  [55.5, 61.8]    & 56.55 
\notsotiny  [53.3, 59.8]    & 52.46 
\notsotiny  [49.1, 55.7]    \\
& MSE(${R}^{\dagger} {y}, {R}^{\dagger} {A} {\hat{x}}_{0}^{u \mid t_0}$) & 91.20 
\notsotiny  [89.6, 92.7]    & \textbf{\red{74.49}} 
\notsotiny  [71.7, 77.2]    & 63.00 
\notsotiny  [59.8, 66.0]    & \textbf{\red{47.54}} 
\notsotiny  [44.2, 50.5]    & 82.83 
\notsotiny  [80.5, 85.0]    & 76.48 
\notsotiny  [73.9, 79.0]    & 70.10 
\notsotiny  [67.1, 73.3]    & 71.17 
\notsotiny  [68.3, 74.0]    & 65.16 
\notsotiny  [62.1, 68.3]    \\
\cline{2-11}
& MSE(${R}^{\dagger} {y}, {\hat{x}}_{0}^{c \mid t_0}$)       & 63.17 
\notsotiny  [59.9, 66.3]    & 56.76 
\notsotiny  [53.5, 60.0]    & 54.98 
\notsotiny  [51.8, 58.1]    & 43.59 
\notsotiny  [40.3, 46.7]    & 47.67 
\notsotiny  [44.2, 51.2]    & 50.99 
\notsotiny  [47.8, 54.1]    & 57.40 
\notsotiny  [54.4, 60.4]    & 57.18 
\notsotiny  [53.8, 60.7]    & 56.33 
\notsotiny  [53.1, 59.5]    \\
& MSE(${y}, {A} {\hat{x}}_{0}^{c \mid t_0}$)           & \textbf{\red{97.01}} 
\notsotiny  [96.2, 97.8]    & 73.43 
\notsotiny  [70.2, 76.3]    & \textbf{\red{89.56}} 
\notsotiny  [87.9, 91.3]    & 39.29 
\notsotiny  [36.4, 42.5]    & \textbf{\red{94.65}} 
\notsotiny  [93.4, 95.8]    & \textbf{\red{95.73}} 
\notsotiny  [94.6, 96.7]    & \textbf{\red{99.03}} 
\notsotiny  [98.6, 99.4]    & \textbf{\red{98.61}} 
\notsotiny  [98.0, 99.1]    & \textbf{\red{92.17}} 
\notsotiny  [90.7, 93.6]    \\
& MSE(${R}^{\dagger} {y}, {R}^{\dagger} {A} {\hat{x}}_{0}^{c \mid t_0}$) & 94.28 
\notsotiny  [92.9, 95.5]    & 64.61 
\notsotiny  [61.4, 67.7]    & 82.73 
\notsotiny  [80.4, 85.1]    & 40.96 
\notsotiny  [37.6, 44.4]    & 92.09 
\notsotiny  [90.6, 93.4]    & 93.52 
\notsotiny  [92.2, 94.7]    & 95.53 
\notsotiny  [94.4, 96.5]    & 95.28 
\notsotiny  [94.1, 96.3]    & 88.10 
\notsotiny  [86.1, 90.0]    \\ 
\hline \hline 
\multirow{6}{*}{\rotatebox[origin=c]{90}{\parbox{3cm}{\centering {High Sparsity}}}} & MSE(${R}^{\dagger} {y}, {\hat{x}}_{0}^{u \mid  t_0}$)          & 80.42 \notsotiny [77.7, 82.8]     & 63.96 \notsotiny [60.7, 66.9]     & 59.37 \notsotiny [56.2, 62.6]     & 48.21 \notsotiny [44.9, 51.4]     & 40.98 \notsotiny [37.8, 44.1]     & 48.82 \notsotiny [45.5, 52.3]     & 59.80 \notsotiny [56.6, 62.9]     & 49.11 \notsotiny [45.9, 52.2]     & 42.95 \notsotiny [39.6, 46.2]     \\
& MSE(${y}, {A} {\hat{x}}_{0}^{u \mid  t_0}$)               & 81.20 \notsotiny [78.8, 83.4]     & 68.96 \notsotiny [66.0, 72.0]     & 55.46 \notsotiny [52.4, 58.7]     & 38.21 \notsotiny [34.8, 41.6]     & 63.67 \notsotiny [60.4, 66.8]     & 59.34 \notsotiny [56.2, 62.7]     & 57.28 \notsotiny [53.9, 60.4]     & 51.90 \notsotiny [48.6, 55.2]     & 48.12 \notsotiny [44.9, 51.4]     \\
& MSE(${R}^{\dagger} {y}, {R}^{\dagger}  {A} {\hat{x}}_{0}^{u \mid t_0}$) & 69.16 \notsotiny [66.4, 72.0]     & 64.93 \notsotiny [61.8, 68.0]     & 56.88 \notsotiny [53.8, 60.1]     & \textbf{\red{64.21}} \notsotiny [60.9, 67.5] & 64.03 \notsotiny [60.8, 67.0]     & 58.25 \notsotiny [55.1, 61.7]     & 57.48 \notsotiny [54.2, 61.0]     & 59.06 \notsotiny [56.0, 62.0]     & 57.96 \notsotiny [54.7, 60.9]     \\
\cline{2-11}
& MSE(${R}^{\dagger} {y}, {\hat{x}}_{0}^{c  \mid t_0}$)          & 78.07 \notsotiny [75.4, 80.6]     & 64.05 \notsotiny [60.8, 67.2]     & 59.56 \notsotiny [56.1, 63.0]     & 50.33 \notsotiny [47.2, 53.6]     & 33.93 \notsotiny [30.9, 37.0]     & 43.85 \notsotiny [40.7, 47.3]     & 62.09 \notsotiny [58.8, 65.0]     & 51.37 \notsotiny [48.2, 54.8]     & 45.28 \notsotiny [42.1, 48.7]     \\
& MSE(${y}, {A} {\hat{x}}_{0}^{c \mid  t_0}$)               & \textbf{\red{95.08}} \notsotiny [93.9, 96.2] & \textbf{\red{89.62}} \notsotiny [87.9, 91.2] & \textbf{\red{75.81}} \notsotiny [73.0, 78.3] & 49.28 \notsotiny [46.0, 52.5]     & \textbf{\red{91.36}} \notsotiny [89.7, 92.9] & \textbf{\red{88.18}} \notsotiny [86.4, 90.1] & \textbf{\red{95.33}} \notsotiny [94.2, 96.4] & \textbf{\red{94.32}} \notsotiny [93.2, 95.5] & \textbf{\red{86.91}} \notsotiny [84.8, 88.8] \\
& MSE(${R}^{\dagger} {y}, {R}^{\dagger}  {A} {\hat{x}}_{0}^{c \mid t_0}$) & 89.00 \notsotiny [87.0, 90.7]     & 69.79 \notsotiny [67.0, 72.6]     & 70.42 \notsotiny [67.4, 73.3]     & 62.13 \notsotiny [58.9, 65.6]     & 77.01 \notsotiny [74.2, 79.6]     & 80.15 \notsotiny [77.6, 82.4]     & 81.22 \notsotiny [78.9, 83.5]     & 81.01 \notsotiny [78.5, 83.4]     & 72.60 \notsotiny [69.7, 75.5]     
\end{tabular}}
\label{tab:AUC}
\end{table}

In Table \ref{tab:AUC}, the overall  AUC results for the models and corresponding ID/OOD test sets (See Table \ref{tab:mnistIDandOODdigits}) are reported. In most cases examined, using the error from the sparse sinograms of the conditionally sampled images proves to be the most effective approach by a significant margin. However,  there are instances where performance is significantly diminished. The model trained on MNIST7 completely fails to distinguish MNIST1 images, calling into question the reliability of such an OOD detector. Moreover, the model trained on MNIST4 appears to struggle with MNIST7 and MNIST9. Another point to note is that in the Moderate Sparsity scenario, projection domain comparison clearly outperforms FBP domain comparison when using conditional sampling, whereas the reverse is true for unconditional sampling. Given these observations, we will conduct further experiments to gain a deeper understanding of underlying issues. 

 
\subsubsection*{RQ-2: What are the failure scenarios associated with conditional sampling in OOD detection?}
Previous analyses and discussions have shown that, despite the potential of multi-scale scoring using comparisons of sparse sinograms from conditional samples, the approach can also fail significantly. Several factors could contribute to this, such as the relative positions of the ID and OOD data manifolds, the geometry of the ID data manifold, and the presence of enough informative data, even with a reduced number of projections. In this subsection, we experimentally focus on the failure cases to enhance our understanding and give rise to the development of coping strategies. 

\begin{figure*}[t!]
 \centering
 \includegraphics[width=\textwidth,keepaspectratio=true]{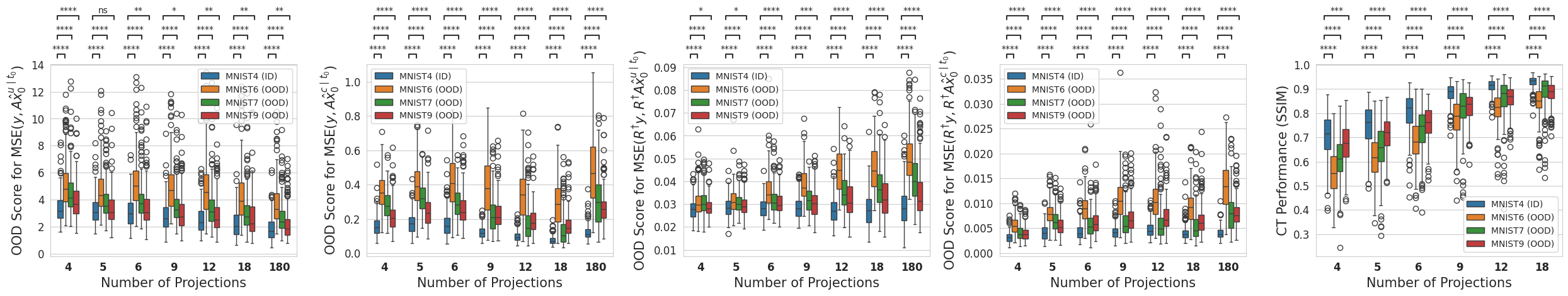}
 \caption{OOD score and SSIM distributions based on the model trained on MNIST4. The first two plots from the left represent Z-scores derived from MSE errors in the sinogram domain, with the first plot using unconditional reconstructions and the second using conditional reconstructions. Similarly, the third and fourth plots focus on the FBP domain, again using unconditional and conditional reconstructions, respectively. The final plot illustrates SSIM distributions to assess CT reconstruction ($t_0 = 1000$) performance rather than OOD detection. The statistical significance annotations indicated by asterisks denote p-values from t-tests for independent samples comparing ID and OOD distributions (* for $p \leq 5.00 \times 10^{-2}$, ** for $p \leq 1.00 \times 10^{-2}$, *** for $p \leq 1.00 \times 10^{-3}$, **** for $p \leq 1.00 \times 10^{-4}$, and $ns$ for non-significant results). Results for models trained on MNIST7 and MNIST8 are available in Supplementary Figure S3 online.}
 \label{fig:rq1boxplotwrtnproj}
\end{figure*}

Fig. \ref{fig:rq1boxplotwrtnproj} presents an analysis of the model trained on MNIST4, focusing on the reconstruction errors depicted in Fig. \ref{fig:blockdiagrams}, excluding MSE(${\mathbf{R}}^{\dagger} {\mathbf{y}}, {\hat{\mathbf{x}}}_{0}^{u \mid t_0}$) and MSE(${\mathbf{R}}^{\dagger} {\mathbf{y}}, {\hat{\mathbf{x}}}_{0}^{c \mid t_0}$). The first four plots enable a comparison between the distributions of Z-scores obtained from the ID and the corresponding OOD test sets. They illustrate distributions relative to the number of projections, offering insight into how sparsity impacts the separability of distributions, even though our main objective is to detect OOD images regardless of the number of projections. The final plot focuses on the Structural Similarity Index (SSIM), shedding light on the relationship between OOD detection performance and CT reconstruction quality. From this plot, we observe that OOD images typically show lower average SSIM scores compared to ID images, though significant overlaps are observed, particularly for MNIST7 and MNIST9. This suggests, that some OOD images attain a CT reconstruction quality on par with ID images. This phenomenon could arise either because the data fidelity is robust enough to counteract the misleading prior information, and/or due to a close proximity between the in-distribution and OOD data manifolds. For instance, when the number of projections is 18, the average SSIM for MNIST7 is very close to that of the ID images, yet the quality gap gets larger as the sparsity increases. This rapid decline in reconstruction performance for MNIST7 images could highlight the role of data fidelity here. For MNIST9, however, the difference in average SSIM scores remains consistent across different numbers of projections, potentially indicating comparable priors, or, in other words, a close proximity between the MNIST4 and MNIST9 data manifolds. When examining the Z-score distributions, we observe that the trend for sinogram domain comparisons with conditional reconstructions (MSE(${\mathbf{y}}, \mathbf{A} {\hat{\mathbf{x}}}_{0}^{c \mid t_0}$)) somewhat mirrors the SSIM performance. The more the model struggles with CT reconstruction of a signal, the more likely the OOD detector is to classify it as OOD, and vice versa. For example, when using 18 projection angles, this approach tends to assign indistinguishably low OOD scores to MNIST7 images, which are CT-reconstructed with near-ID quality in terms of SSIM. Sinogram domain comparisons with unconditional reconstructions (MSE(${\mathbf{y}}, \mathbf{A} {\hat{\mathbf{x}}}_{0}^{u \mid t_0}$)) tend to obscure the distinction between ID and OOD distributions. The ID signal appears to be less clean for the sparse measurements. Upon closer examination, the pattern resembles the conditional case, except for MNIST7 at 12 and 18 projection angles. In this instance, its distribution diverges slightly more from the ID distribution compared to MNIST9, a contrast not seen in the conditional scenarios. This observation signals the validity of our intuition about the influence of data fidelity and the potentially misleading effects of conditional sampling in OOD detection. When it comes to the FBP domain comparisons, employing the conditional reconstructions (MSE($\mathbf{R}^{\dagger} \mathbf{y}, \mathbf{\hat{x}}_{0}^{c \mid t_0}$)) do not appear to offer any advantage over its sinogram domain counterpart. The challenges in distinguishing MNIST9 when the number of projections is very low (such as 4, 5, and 6) and MNIST7 when the projections are slightly higher (like 12 and 18) remain unresolved, and are in fact exacerbated. However, in the unconditional case (MSE($\mathbf{R}^{\dagger} \mathbf{y}, \mathbf{\hat{x}}_{0}^{u \mid t_0}$)), since the data fidelity is disregarded, MNIST7 images can clearly be identified as OOD for 12 and 18 projection angles. Nevertheless, we see significant overlaps for MNIST9 in all cases, and for the typically well-separated MNIST6 as the projection angles become increasingly low. Unlike in the sinogram domain, the OOD signal deteriorates more than the ID signal as the number of projections decreases. This is likely because sparser-view FBP reconstructions become harder to distinguish with the rise in artifacts. Overall, relying solely on unconditional sampling appears inadequate for distinguishing OOD images from ID, although it does seem to provide valuable information in scenarios where the OOD detector is misled by the data fidelity. 

\begin{figure*}[t!]
 \centering
 \begin{subfigure}[b]{0.8\textwidth}
  \centering
{\includegraphics[width=\textwidth,keepaspectratio=true]{ 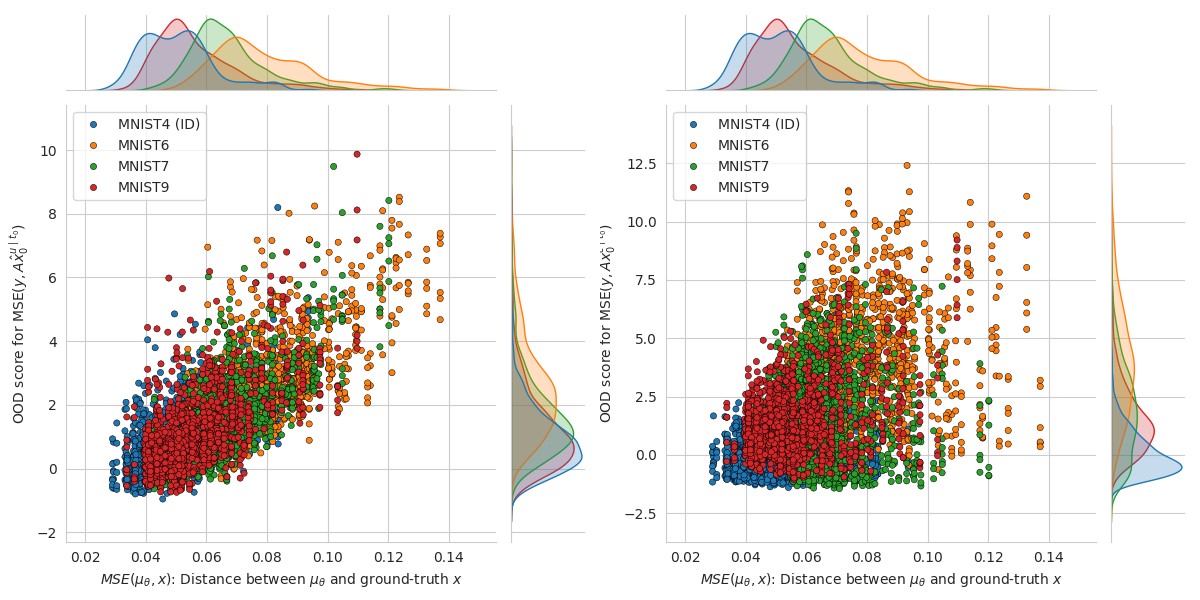}}
 \end{subfigure}
 \caption{Joint scatter plots illustrating how OOD scores for ID (MNIST4) and OOD images vary with increasing distance of the image from the mean of the prior distribution, in unconditional (left) and conditional (right) cases. For MSE(${y}, {A} {\hat{x}}_{0}^{u \mid   t_0}$), there is a strong positive correlation between distance and the OOD score (r = 0.71, p $<$ 0.0001).  However, with MSE(${y}, {A} {\hat{x}}_{0}^{c \mid   t_0}$), the correlation decreases to a moderate level but remains significant (r = 0.44, p $<$ 0.0001). Overall, conditioning reduces the correlation between the distance and the OOD score by incorporating additional information from the measurements. As a result, a substantial portion of MNIST7 images receive lower OOD scores, comparable to those of ID images. The complementary analyses for this figure can be found as Supplementary Figure S4 online.}
 \label{fig:rq2_uncondvscond}
\end{figure*}

Fig. \ref{fig:rq2_uncondvscond} shows the relationship between the OOD score and the MSE distance from the in-distribution mean to the ground truth, comparing reconstructions generated through unconditional sampling (left) and conditional sampling (right). In the unconditional case, there is a strong positive correlation. This is expected in reconstruction-based OOD detection, where the farther a point is from the in-distribution manifold, the higher the likelihood of it being classified as OOD. However, the correlation weakens when switching to conditional sampling for both ID and OOD datasets. This is due to the fact that the additional information from the measurements changes the marginal distributions of the reconstruction errors, thereby affecting the OOD scores. While distance from the in-distribution data manifold usually still leads to higher OOD scores, presence of overly informative measurements appears to cause some distant images to produce scores as low as those of ID images. A significant portion of MNIST7 seems to be affected by this phenomenon. Namely, while conditioning managed to concentrate ID images at low scores, the reconstruction errors of MNIST7 are also greatly condensed, resulting in overlaps with ID scores. In the same way that the sharply increased scores for ID images diminish the contrast between ID and OOD errors in the unconditional case, the drastically reduced scores for MNIST7 produce a similar outcome, resulting in poor OOD detection performance. A similar situation occurs with MNIST7 (ID) and MNIST1, this time leading to a dramatic failure. The analysis in Fig. \ref{fig:rq2_uncondvscond} also shows that MNIST9 images are in close proximity to ID images based on their distance from the in-distribution mean, which may account for the relatively lower OOD detection performance.

\subsubsection*{RQ-3: Can the contrast between ID and OOD scores be enhanced by weighting conditional and unconditional reconstruction errors?}
\begin{figure*}[t!]
 \centering
 \begin{subfigure}[b]{0.9\textwidth}
  \centering
{\includegraphics[width=\textwidth,keepaspectratio=true]{ 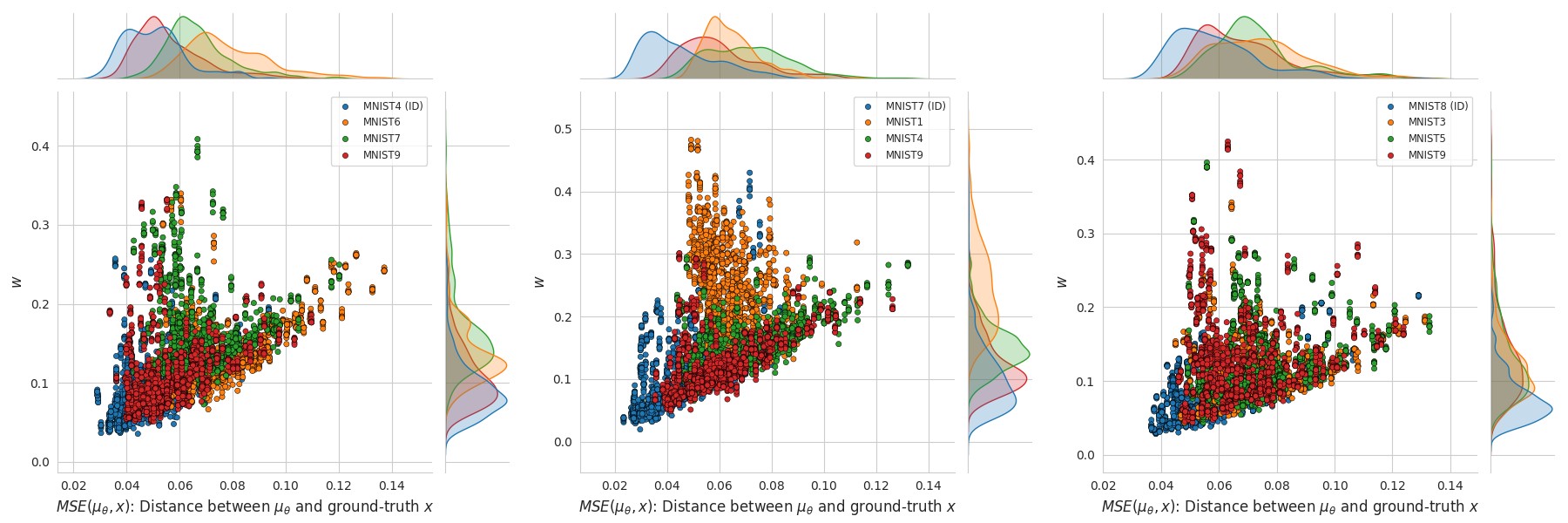}}
 \end{subfigure}
 \caption{Joint scatter plots illustrating how $w$ is related with the unknown MSE error between the in-distribution mean and the ground-truth, for all cases depicted in Table \ref{tab:mnistIDandOODdigits}. In the cases of MNIST4 (ID) vs. MNIST7 (first plot) and MNIST7 (ID) vs. MNIST1 (second plot), we observe noticeably elevated $w$ values that deviate from the general trend of positive correlation with the unknown image-to-manifold distance.}
 \label{fig:rq3_wvsdistance}
\end{figure*}

\begin{table}[t!]
\centering
 \caption{AUC scores for three models trained on full-view MNIST4, MNIST7, and MNIST8 images with the application of the weighting scheme. Averaging means the weight parameter $w = 0.5$. Red and blue texts highlight the highest and the second highest values in each column separately for two cases: Case-1: Moderate Sparsity, where the number of projection angles (\#proj) is \{18, 12, 9\}, and Case-2: High Sparsity, where \#proj = \{6, 5, 4\}. The yellow shaded columns show the cases with a sharp improvement. Confidence bounds are reported for all cases using 1000-fold bootstrapping.}
\resizebox{\textwidth}{!}{
\arrayrulecolor{black}
\begin{tabular}{p{0.3cm}|p{3cm}|P{5.2em}>{\columncolor{verylightyellow}}P{5.2em}P{5.2em}|>{\columncolor{verylightyellow}}P{5.2em}P{5.2em}P{5.2em}|P{5.2em}P{5.2em}P{5.2em}}
&\multicolumn{1}{l|}{ID Dataset:}  & \multicolumn{3}{c|}{MNIST4} & \multicolumn{3}{c|}{MNIST7} & \multicolumn{3}{c}{MNIST8} \\
\hline\hline
&\multicolumn{1}{l|}{OOD Dataset:} & \small MNIST6 & \small MNIST7 & \small MNIST9 & \small MNIST1 & \small MNIST4 & \small MNIST9 & \small MNIST3 & \small MNIST5 & \small MNIST9 \\
\hline
\multirow{5}{*}{\rotatebox[origin=c]{90}{\parbox{2cm}{\centering {\small Moderate Sp.}}}} & Top Results (Table \ref{tab:AUC})  & \textbf{\blue{97.01}} \notsotiny  [96.2, 97.8]    & 74.49 \notsotiny  [71.7, 77.2]    & \textbf{\red{89.56}}  \notsotiny  [87.9, 91.3]    & 47.54  \notsotiny  [44.2, 50.5]    & \textbf{\red{94.65}} \notsotiny  [93.4, 95.8]    & \textbf{\red{95.73}} \notsotiny  [94.6, 96.7]    & \textbf{\red{99.03}} \notsotiny  [98.6, 99.4]    & \textbf{\red{98.61}} \notsotiny  [98.0, 99.1]    & \textbf{\red{92.17}} \notsotiny  [90.7, 93.6]    \\
\cline{2-11}
&Averaging ($\Gamma = \mathbf{I}$)  & 86.62 \notsotiny [84.6, 88.5] & 69.34 \notsotiny [66.4, 72.3]  & 57.73 \notsotiny [54.6, 61.1] & 35.25 \notsotiny [32.3, 38.4]  & 73.07 \notsotiny [70.3, 75.8] & 68.69 \notsotiny [65.7, 71.5] & 63.11 \notsotiny [59.8, 66.2] & 60.79 \notsotiny [57.5, 64.2] & 54.69 \notsotiny [51.3, 57.9] \\
&Averaging ($\Gamma = \mathbf{R}^{\dagger}$)   & 95.75 \notsotiny [94.6, 96.7] & 78.09 \notsotiny [75.4, 80.6]  & 70.55 \notsotiny [67.7, 73.3] & 46.14 \notsotiny [43.0, 49.4]  & 89.12 \notsotiny [87.3, 91.0] & 84.37 \notsotiny [82.2, 86.5] & 81.65 \notsotiny [79.0, 84.0] & 82.17 \notsotiny [79.8, 84.5] & 73.79 \notsotiny [71.0, 76.8] \\
&Weighting ($\Gamma = \mathbf{I}$)  & 96.25 \notsotiny [95.2, 97.2]  & \red{\textbf{91.45}} \notsotiny [89.8, 93.1] & 76.51 \notsotiny [73.6, 79.3] & \blue{\textbf{78.33}} \notsotiny [75.6, 80.9]   & 88.06 \notsotiny [86.0, 89.9] & 83.95 \notsotiny [81.5, 86.1] & 87.79 \notsotiny [85.6, 89.9] & 87.72 \notsotiny [85.5, 89.9] & 81.78 \notsotiny [79.2, 84.4] \\
&Weighting ($\Gamma = \mathbf{R}^{\dagger}$) & \red{\textbf{97.98}} \notsotiny [97.3, 98.6] & \blue{\textbf{88.46}} \notsotiny [86.6, 90.3]  & \blue{\textbf{83.60}} \notsotiny [81.4, 85.9] & \red{\textbf{81.07}} \notsotiny [78.8, 83.3] & \blue{\textbf{94.20}} \notsotiny [92.9, 95.5] & \blue{\textbf{91.39}} \notsotiny [89.8, 93.0] & \blue{\textbf{94.58}} \notsotiny [93.3, 95.7] & \blue{\textbf{95.03}} \notsotiny [93.8, 96.2] & \blue{\textbf{90.34}} \notsotiny [88.6, 91.9]\\
\hline \hline
\multirow{5}{*}{\rotatebox[origin=c]{90}{\parbox{2cm}{\centering {\small High Sparsity}}}} & Top Results (Table \ref{tab:AUC})  & \textbf{\red{95.08}} \notsotiny [93.9, 96.2] & \textbf{\blue{89.62}} \notsotiny [87.9, 91.2] & \textbf{\red{75.81}} \notsotiny [73.0, 78.3] & 64.21 \notsotiny [60.9, 67.5]   & \textbf{\red{91.36}} \notsotiny [89.7, 92.9] & \textbf{\red{88.18}} \notsotiny [86.4, 90.1] & \textbf{\red{95.33}} \notsotiny [94.2, 96.4] & \textbf{\red{94.32}} \notsotiny [93.2, 95.5] & \textbf{\red{86.91}} \notsotiny [84.8, 88.8]    \\
\cline{2-11}
&Averaging ($\Gamma = \mathbf{I}$)  & 83.73 \notsotiny [81.3, 85.9]     & 71.57 \notsotiny [68.9, 74.5]     & 57.00 \notsotiny [53.8, 60.2] & 38.59 \notsotiny [35.4, 41.6]     & 65.85 \notsotiny [62.9, 68.8] & 61.64 \notsotiny [58.3, 64.8] & 60.96 \notsotiny [57.8, 64.0] & 55.71 \notsotiny [52.4, 59.0] & 50.69 \notsotiny [47.5, 54.0] \\
&Averaging ($\Gamma = \mathbf{R}^{\dagger}$)   &80.32 \notsotiny [77.8, 82.6]     & 70.12 \notsotiny [67.2, 73.1]     & 62.63 \notsotiny [59.6, 65.6] & 66.79 \notsotiny [63.7, 69.7]     & 71.30 \notsotiny [68.3, 74.1] & 67.18 \notsotiny [64.0, 70.2] & 68.41 \notsotiny [65.4, 71.3] & 68.66 \notsotiny [65.8, 71.6] & 65.57 \notsotiny [62.4, 68.7] \\
&Weighting ($\Gamma = \mathbf{I}$)  & \blue{\textbf{94.88}} \notsotiny [93.6, 96.1]     & \red{\textbf{93.60}} \notsotiny [92.2, 94.9] & 72.13 \notsotiny [69.2, 75.0] & 85.61 \notsotiny [83.2, 87.7]     & {85.28} \notsotiny [83.1, 87.6] & 77.20 \notsotiny [74.4, 80.0] & 81.61 \notsotiny [79.0, 84.1] & 85.66 \notsotiny [83.4, 88.0] & 82.57 \notsotiny  [80.1, 85.0] \\
&Weighting ($\Gamma = \mathbf{R}^{\dagger}$) & {93.38} \notsotiny [92.0, 94.6] & 88.94 \notsotiny [87.0, 90.8]     & \blue{\textbf{75.72}} \notsotiny [73.2, 78.3] & \red{\textbf{92.56}} \notsotiny [90.9, 94.0] & \blue{\textbf{87.20}} \notsotiny [85.1, 89.2] & \blue{\textbf{81.86}} \notsotiny [79.5, 84.2] & \blue{\textbf{85.63}} \notsotiny  [83.5, 87.7] & \blue{\textbf{86.83}} \notsotiny [84.9, 88.7] & \blue{\textbf{85.55}} \notsotiny  [83.6, 87.7]
\end{tabular}}
\label{tab:AUCafterweighting}
\end{table}

This subsection presents the experimental results demonstrating the extent to which applying Eq. \eqref{eq:weighting} enhances robustness in failure scenarios, as well as its inherent limitations. Our weighting approach can be viewed as a blend of the simplest form of distance-based OOD detection and reconstruction-based OOD detection methods. OODs in both frameworks are often points that are far from the in-distribution manifold. Although the parameter $w$ is defined as the distance between the measurement subspace and the in-distribution mean, it is actually affected by the unknown distance between the in-distribution mean and the ground-truth image.  Fig. \ref{fig:rq3_wvsdistance} illustrates how closely $w$ is associated with the unknown MSE error between the in-distribution mean and the ground-truth. Since these two measures are likely to exhibit a positive correlation, $w$ can be seen as an approximation of the unknown image-to-manifold distance, allowing this information to be incorporated into the OOD score. However, $w$ proves to be more than this. It effectively distinguishes MNIST7 in the first plot and MNIST1 in the second, even though their image-to-manifold distances appear to be lower than those of MNIST6 and MNIST4, respectively. These are two failure cases from Table \ref{tab:AUC}, where we attribute the OOD detector's difficulty in identifying OOD images to the presence of highly informative measurements. In Table \ref{tab:AUCafterweighting}, we compare the AUC scores obtained using the proposed weighting scheme to the best results from Table \ref{tab:AUC}. The columns highlighted in yellow illustrate how the detector's significant failures are improved in both moderate and high sparsity scenarios, although for 4 (ID) vs 7, the best results are achieved with $\Gamma = \mathbf{I}$, while for 7 (ID) vs 1, $\Gamma = \mathbf{R}^{\dagger}$ is the winner. The reason could be that the straight-line structures of the digits 7 and 1 provide little unique information in their projections. From Table \ref{tab:AUCafterweighting}, we see that 4 (ID) vs 6 shows neither improvement nor decline. Referring back to Fig. \ref{fig:rq3_wvsdistance}, we see that the parameter $w$ for MNIST6 images are well-correlated with the true distance. That means, the OOD scores of distant images will be elevated by the unconditional sampling errors, while scores for closer images are driven by conditional sampling. Since MNIST6 appears to lie on a distant manifold, even unconditional sampling proves effective (See ``Averaging" results in Table \ref{tab:AUCafterweighting}). Yet for the closer portion, as supported by the low $w$ values, conditioning does not help them improve reconstruction, thereby aiding effective OOD detection. A similar situation occurs in the case of 7 (ID) vs. 4. Table \ref{tab:AUCafterweighting} also reports several instances of performance decline caused by the weighting scheme. While some are minor (e.g., 8 (ID) vs 9), others are rather significant (e.g., 8 (ID) vs 3). One possible reason for the case of 7 (ID) vs 9 is the greater intra-class distance variability in MNIST7 (ID), as one can see in the second plot of Fig. \ref{fig:rq3_wvsdistance}. Such a skewed in-distribution could lead to a drop in performance for our weighting scheme by generating large errors that overshadow those of OOD images with low $w$ values. The other notable failure cases, 8 (ID) vs 3 and 8 (ID) vs 5, face a similar issue. These pairs are already well-separated when using conditional samples, but the weighting scheme increases the false positives (FP) by emphasizing errors from unconditional reconstructions of the distant ID images. Despite such drops in AUC scores, depending on the application, selecting a threshold that allows for some FPs can enable effective use of the weighting scheme, which ensures a True Positive Rate (TPR) at least as high as when using only conditional reconstruction errors. Supplementary Figure S4 and Supplementary Figure S5 (both available online) provide ROC curves to validate this situation. Furthermore, in nearly all cases, weighting the FBP domain errors ($\Gamma = \mathbf{R}^{\dagger}$) outperformed weighting the sinogram domain errors ($\Gamma = \mathbf{I}$). This is likely because FBP reconstruction errors are more distinctive in the unconditional scenario. Additionally, in most instances of performance loss, the declines were even more pronounced when the measurements were highly sparse, possibly because FBP reconstructions no longer maintained the disparities in the unconditional case. 

Supplementary Figure S7 online presents a sensitivity analysis of the impact of \( w \) on the overall OOD detection performance, considering different scales of \( w \). Moreover, Supplementary Figure S8 online presents AUC scores for different numbers of projections separately, offering a preliminary insight into the weighting scheme's stability across varying levels of sparsity. As another stability analysis, we provide Supplementary Figure S9 online, where AUC scores across three different noise levels (i.e., \{0.01, 0.03, 0.05\}) are reported for the number of 18 projections. These results highlight the limitations of FBP-domain comparison in the presence of noise, as artifacts introduced by FBP reconstructions from noisy projections distort the OOD scores more than the noise itself. In contrast, multi-scale reconstruction-based OOD detection using sinogram domain comparisons remains stable across conditional, unconditional, and weighted schemes. Note that, all three analyses were conducted using a model trained on MNIST4 and evaluated on MNIST6, MNIST7, and MNIST9 as OOD datasets. This selection provides a diverse set of OOD distributions, reflecting varying levels of similarity to and deviation from the in-distribution data.

\section*{Discussion}
 Our findings indicate that computing the reconstruction error in the projection domain with conditional samples is the most effective and straightforward approach for assessing OOD reconstruction quality. However, challenges arise because the standard practice of conditioning in CT reconstruction can also generalize to the reconstruction of OOD images, thereby complicating OOD detection. In other words, while conditioning prevents sparse-view in-distribution (ID) images from being falsely marked as OOD, it may lead more false negatives (FN) by enabling the accurate reconstruction of OOD images. In certain contexts, particularly when addressing localized anomalies, because a CT reconstruction framework generalizes well to anomalous images does not mean the image is free of anomalies. In this regard, we proposed a weighting approach that balances the errors from both conditional and unconditional reconstructions. It determines the weight of the unconditional errors to be incorporated into the OOD score based on the normalized distance of the measurement subspace from the in-distribution mean. The results demonstrate that this approach effectively prevents the detector from assigning low scores to OOD images with sufficient measurements, enhancing robustness against highly informative data fidelities in OOD images. However, this improvement comes with a performance trade-off; leading to more false positives (FP), especially when in-distribution images deviate notably from the in-distribution mean. Therefore, the decision whether to use the weighting scheme should be based on the costs associated with FPs and FNs in the specific scenario. For instance, in non-destructive testing, a defect detection system with many FPs leads to unnecessary further scans or rejection of defect-free items, which can increase production time and costs. On the other hand, an FN in this context means that a defected item passes through production, which may have serious implications. Similarly, in the medical domain, an FN could result in a missed diagnosis of a serious condition, leading to delayed treatment and potentially severe health outcomes. To further improve the weighting scheme, alternative distance metrics that account for the geometry of the in-distribution data manifold could be explored. As a reference example, Supplementary Table S1 online presents results using the normalized \( L^2 \) distance averaged across all training set projections, inspired by clustering literature, which favors average linkage over single-point distances to better capture global structure. Note that, this alternative distance metric is included as an demonstrative example but is neither effective nor practical due to its high computational cost. 

 In this study, we utilized FBP reconstructions of sparse measurements as input, relying on the diffusion models' ability to map off-manifold data onto the learned data manifold. The choice of FBP emerged as a practical and efficient option, with its effectiveness demonstrated through experiments. The results from unconditional sampling indicate that FBP reconstructions alone are insufficient to generate a clear contrast between ID and OOD scores. At higher sparsity levels, an unconditional detector struggles to produce low errors for ID images. Yet, in a non-negligible number of cases, the retaining structures help prevent images from straying off the generative trajectory toward the learned data manifold. This effect, unattainable through unconditional generation, underscores the potential of integrating unconditional reconstruction from noisy FBP images into the OOD detection process. On the other hand, in the conditional setting, FBP images serve as complementary data alongside the measurements. This additional information does not appear to negatively impact the ID signal while enhancing the OOD signal by facilitating poorer reconstructions of OOD images. This outcome wouldn't always occur when starting directly from noise, as distinctively informative OOD projections would be mapped closer to their original manifolds. Acquiring OOD errors across multiple scales offers shorter pathways to the learned data manifold, reducing the interventions imposed by measurement consistency constraints. These observations highlight the need for further investigation to identify the optimal balance between unconditional and conditional reconstruction-based OOD detection approaches. This could entail using more advanced weighting techniques in the proposed scheme, as previously discussed, or even implementing adaptive weighting mechanisms. Additionally, further research may involve adjusting the frequency of the measurement consistency step during conditional reconstruction. Yet another intriguing direction could be adopting the concept of multi-scale reconstructions to inverse problems by utilizing a range of regularization parameters $\lambda$ during the conditional generation. This setup eliminates the need for partial diffusions and FBP images as inputs, while maintaining the realism-faithfulness trade-off by varying $\lambda$ from low to high, respectively. However, it may increase the computational demand as it necessitates multiple passes through the full reconstruction chain. Further exploratory work by diving into the latent-space of the diffusion model to demonstrate the proximity of the FBP reconstructions to the learned data manifold could also be beneficial. 

Before concluding the discussion, it is important to note that the concept of multiple reconstructions introduces a computationally intensive approach to OOD detection. In this work, we performed 12 reconstructions to calculate a multi-scale score, requiring 642 model evaluations in total. Still, it remains lower than the 1000 evaluations required for a single CT reconstruction without the PLMS sampler (\(T=1000\)) \cite{liu2022pseudo}, ensuring that the overall computational cost remains less than twice that of the standard CT process without OOD detection when integrated before the actual CT module. Meanwhile, the weighting scheme demands twice the model evaluations, as it relies on both conditional and unconditional samples, leading to additional computational overhead. As a result, this approach does not seem well-suited for applications requiring real-time predictions. That said, as noted in \cite{graham2023denoising}, conditional and unconditional reconstructions starting from different points can be parallelized, helping to alleviate some of the computational burden.

\section*{Conclusion}
Through a comprehensive exploration, we have shown that a diffusion model trained on full-view target images for CT reconstruction can be effectively multi-tasked for out-of-distribution (OOD) detection. Our work builds on a recent OOD detection method that leverages reconstruction errors from multiple forward and reverse partial diffusion processes, where the input image is exposed to varying noise levels. By redefining the notions of ``input'' and ``reconstruction error'', we have demonstrated the potential of this approach for sparse-view computed tomography. We explored different approaches for comparisons in both the sinogram and image domains, and found that the sinogram domain was generally more effective for OOD scoring. The experiments also reveal that while a model trained on full-view images struggles to generalize effectively to the distribution shift introduced by sparse-view filtered backprojection reconstructions, conditional reconstruction counteracts this limitation by generating a cleaner in-distribution signal. However, conditioning on the measurements introduces another challenge: in some cases, it blurs OOD signal. Reliance on highly informative measurements minimize the impact of the misleading prior and cause inadvertently well-reconstructed OOD images. To address this, we proposed a simple weighting scheme that reduces the influence of conditioning when the projection data noticeably deviates from the forward-projected mean of in-distribution images. This weighting scheme proves to be effective in enhancing the OOD detector's robustness against false negatives (i.e., misclassifying an OOD image as in-distribution), though it may increase false positives (i.e., flagging an in-distribution image as OOD). The study offers several opportunities for improvement, including the use of more sophisticated distance metrics, adaptive weighting mechanisms, and conditional sampling with less frequent measurement consistency step. Additionally, leveraging semi-supervised approaches that integrate a small number of OOD images into the training set could also enhance the performance. Further exploratory work is necessary to gain a better insight on the effectiveness and applicability of both multi-scale reconstruction-based out-of-distribution detection approach and the proposed weighting scheme in handling the complexities of real-world CT challenges. Evaluating the proposed approach on real clinical or industrial CT data with noisy measurements also remains an important direction for future work.

\section*{Code availability}
The software package TorchRadon \cite{torch_radon} has been used for solving computed tomography (CT) reconstruction problem, while the codebase from \cite{graham2023denoising, graham2023unsupervised} was adapted for out-of-distribution detection in the sparse-view CT setting. This adapted version will be publicly available upon acceptance.

\section*{Data availability}
The datasets analyzed in this study are part of the MNIST database \cite{lecun1998mnist,deng2012mnist}, which is publicly accessible at \url{https://yann.lecun.com/exdb/mnist/}. The train and test splits will also be made publicly available together with the code upon acceptance to ensure reproducibility.
\bibliography{sample}

\section*{Acknowledgements}
This work was carried out during the tenure of an ERCIM ‘Alain Bensoussan’ Fellowship Programme. 

\section*{Author contributions statement}
E.D.T. conceptualized the study, conceived and conducted the experiments, wrote the original draft of the manuscript. E.D.T. analysed the results together with F.L. and T.v.L., who also edited the manuscript. All authors reviewed the manuscript. 

\section*{Competing interests}
The author(s) declare no competing interests.


\end{document}


\font\myfont=cmr12 at 14pt
\font\myfontt=cmr12 at 10pt
\title{\myfont Exploring Out-of-distribution Detection for Sparse-view Computed Tomography with Diffusion Models: Supplementary Experiments}


\author{\myfontt Ezgi Demircan-Tureyen\textsuperscript{*}, Felix Lucka, Tristan van Leeuwen}
\date{} 
\maketitle
\vspace{-3.5em} 
\begin{center}
\textsuperscript{*}{\myfontt edt@cwi.nl}
\end{center}

\begin{figure*}[h!]
    \centering
   \begin{subfigure}[b]{0.24\textwidth}
        \centering
        \includegraphics[width=\textwidth,keepaspectratio=true]{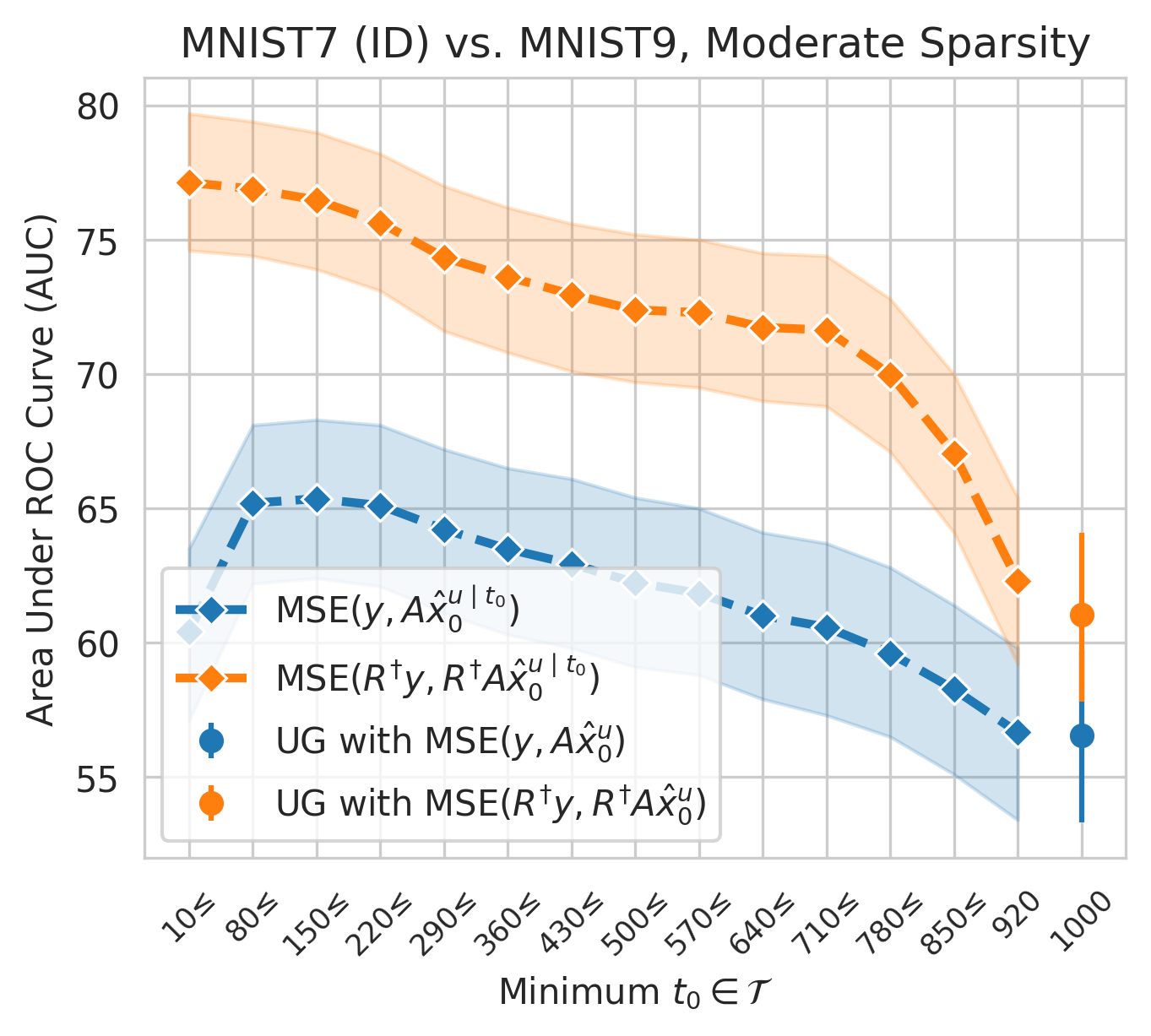}
    \end{subfigure}
        \begin{subfigure}[b]{0.24\textwidth}
        \centering
        \includegraphics[width=\textwidth,keepaspectratio=true]{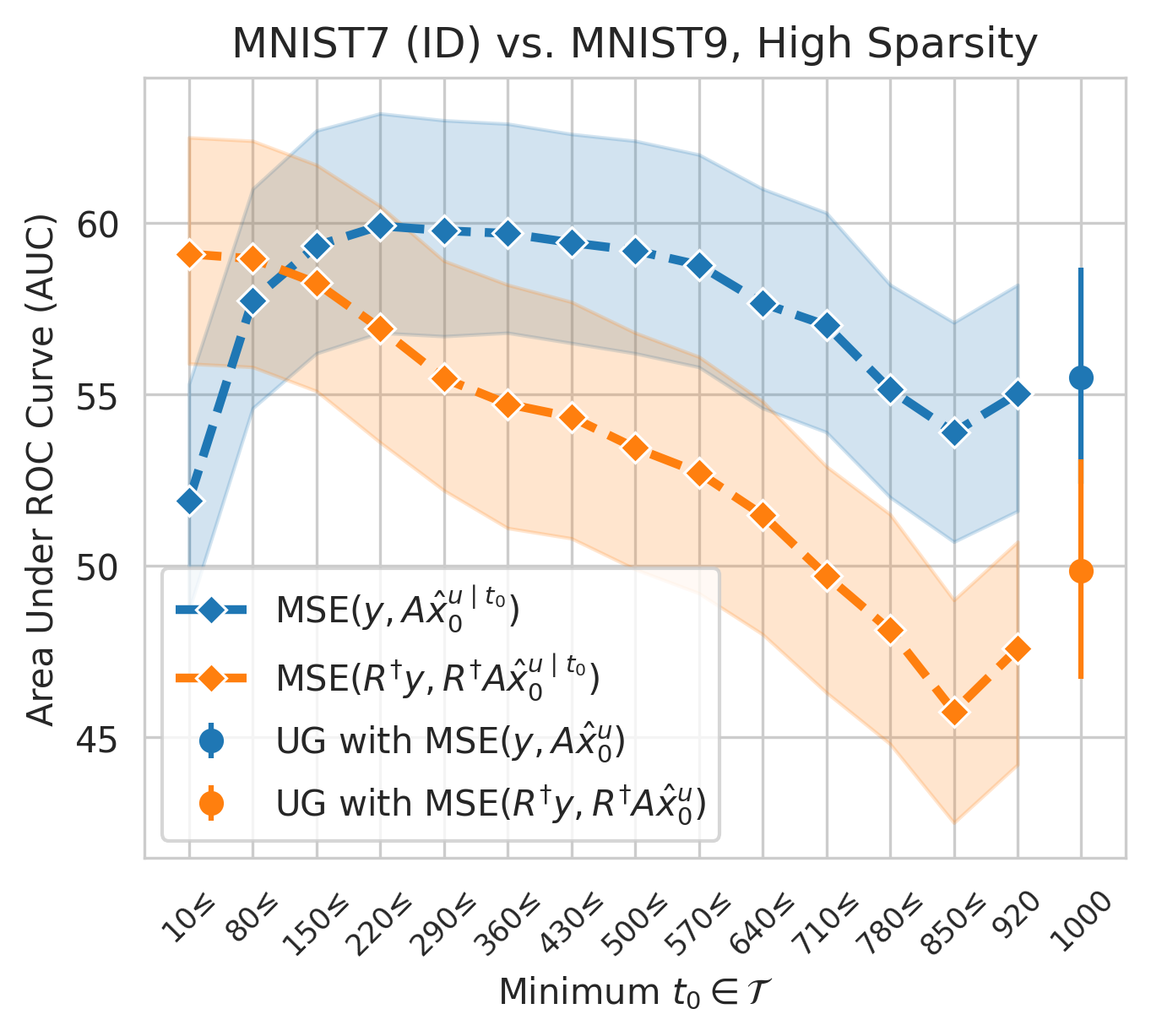}
    \end{subfigure}
   \begin{subfigure}[b]{0.24\textwidth}
        \centering
        \includegraphics[width=\textwidth,keepaspectratio=true]{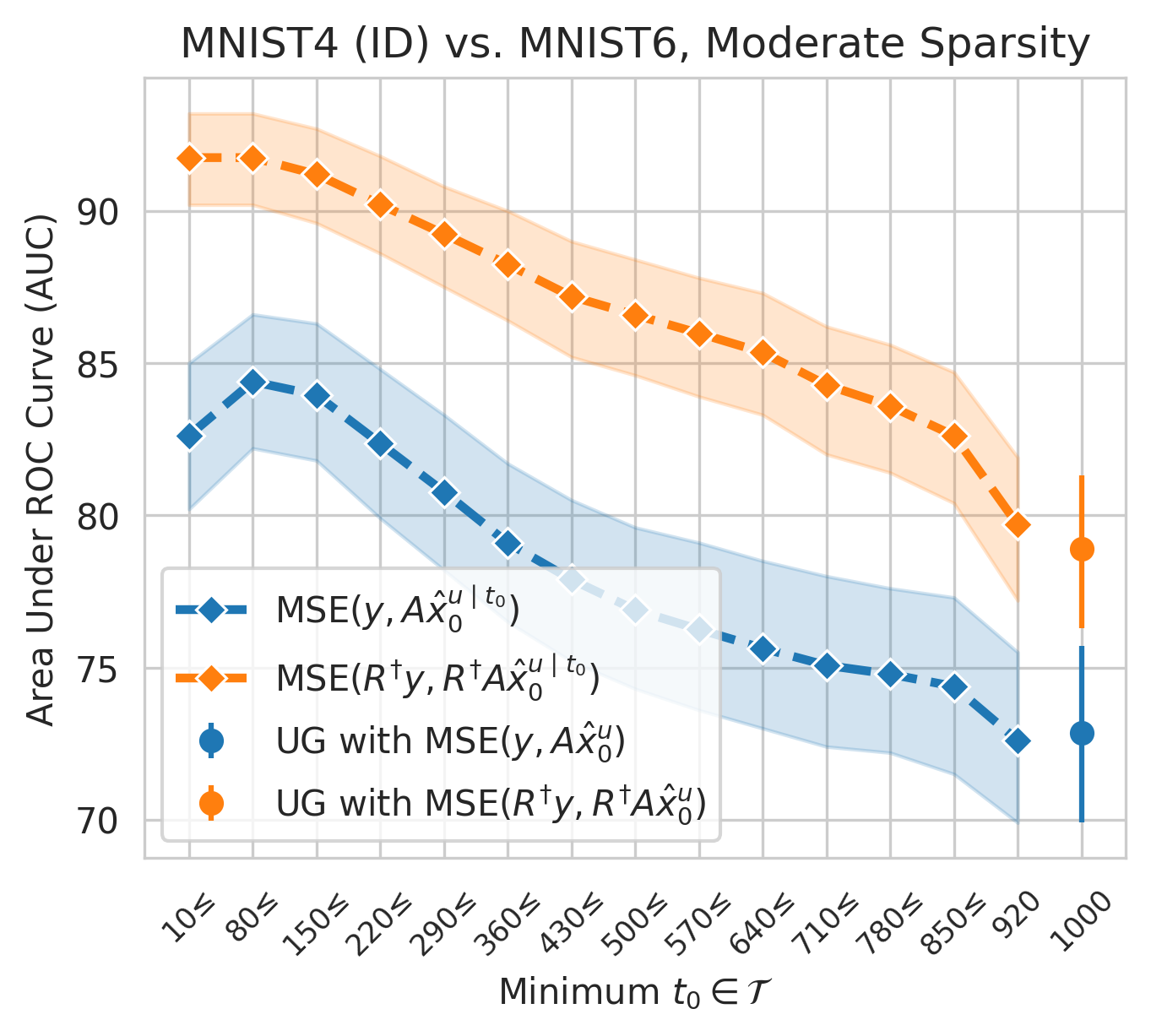}
    \end{subfigure}
        \begin{subfigure}[b]{0.24\textwidth}
        \centering
        \includegraphics[width=\textwidth,keepaspectratio=true]{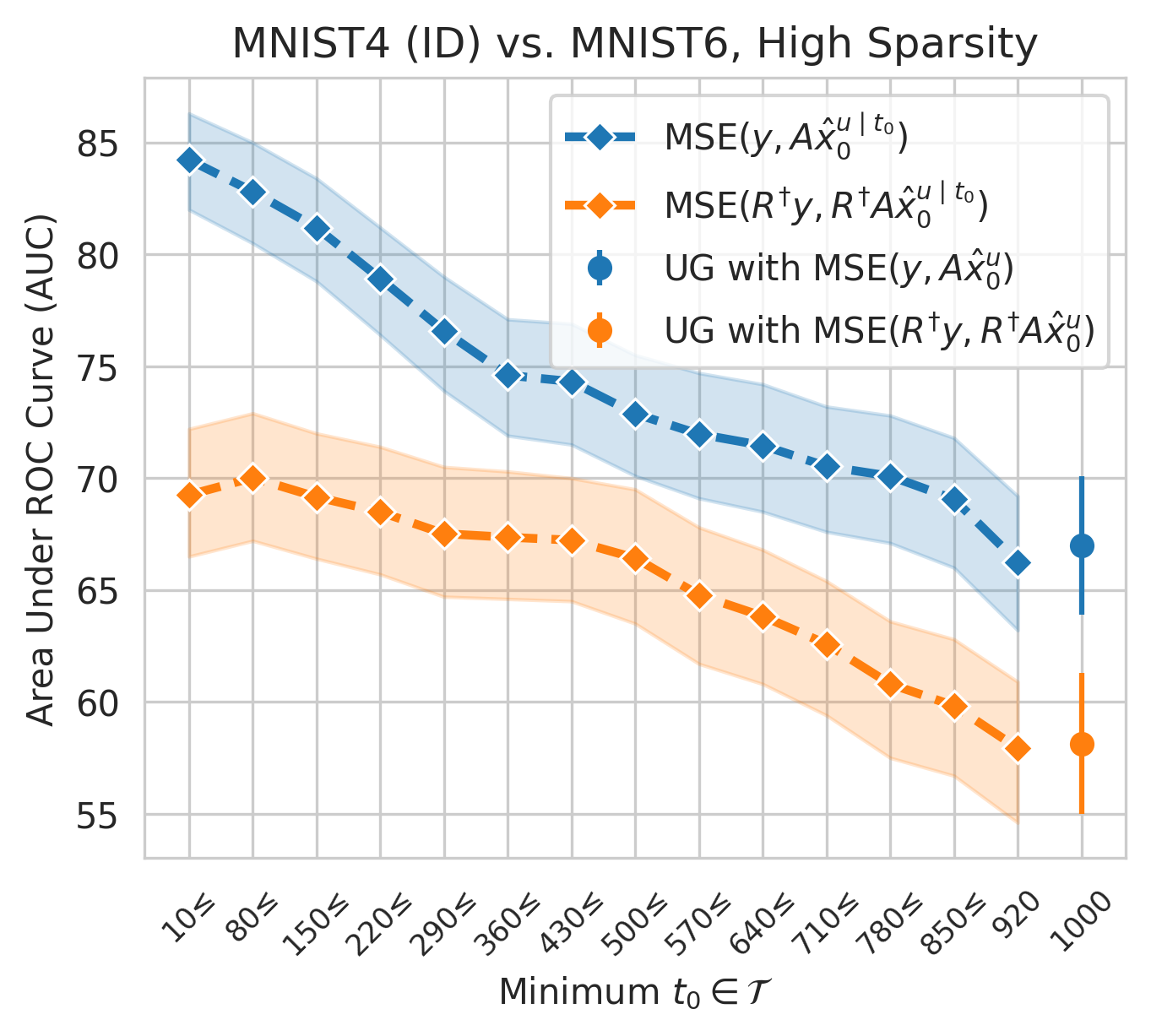}
    \end{subfigure}    
    \\
       \begin{subfigure}[b]{0.24\textwidth}
        \centering
        \includegraphics[width=\textwidth,keepaspectratio=true]{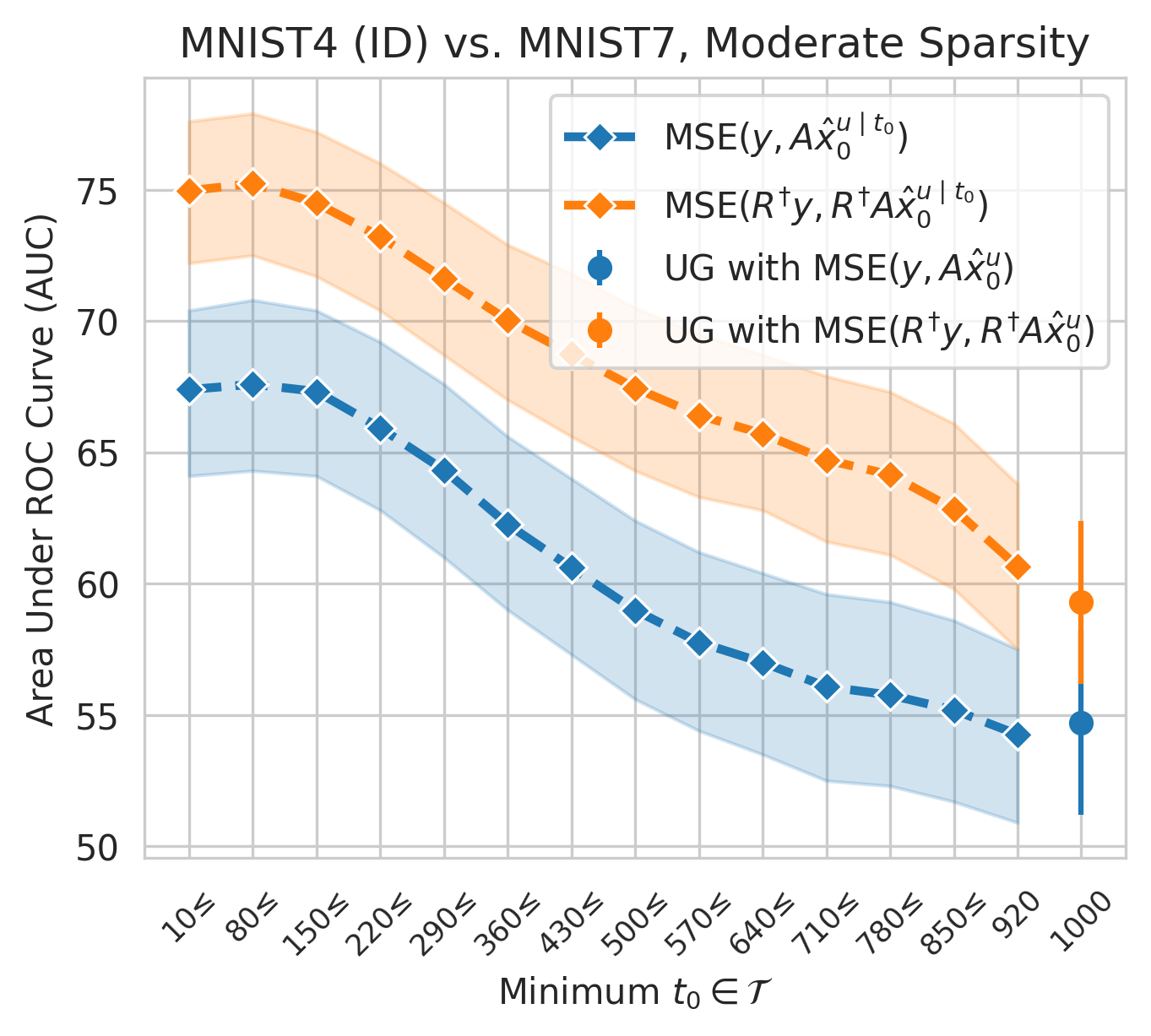}
    \end{subfigure}
        \begin{subfigure}[b]{0.24\textwidth}
        \centering
        \includegraphics[width=\textwidth,keepaspectratio=true]{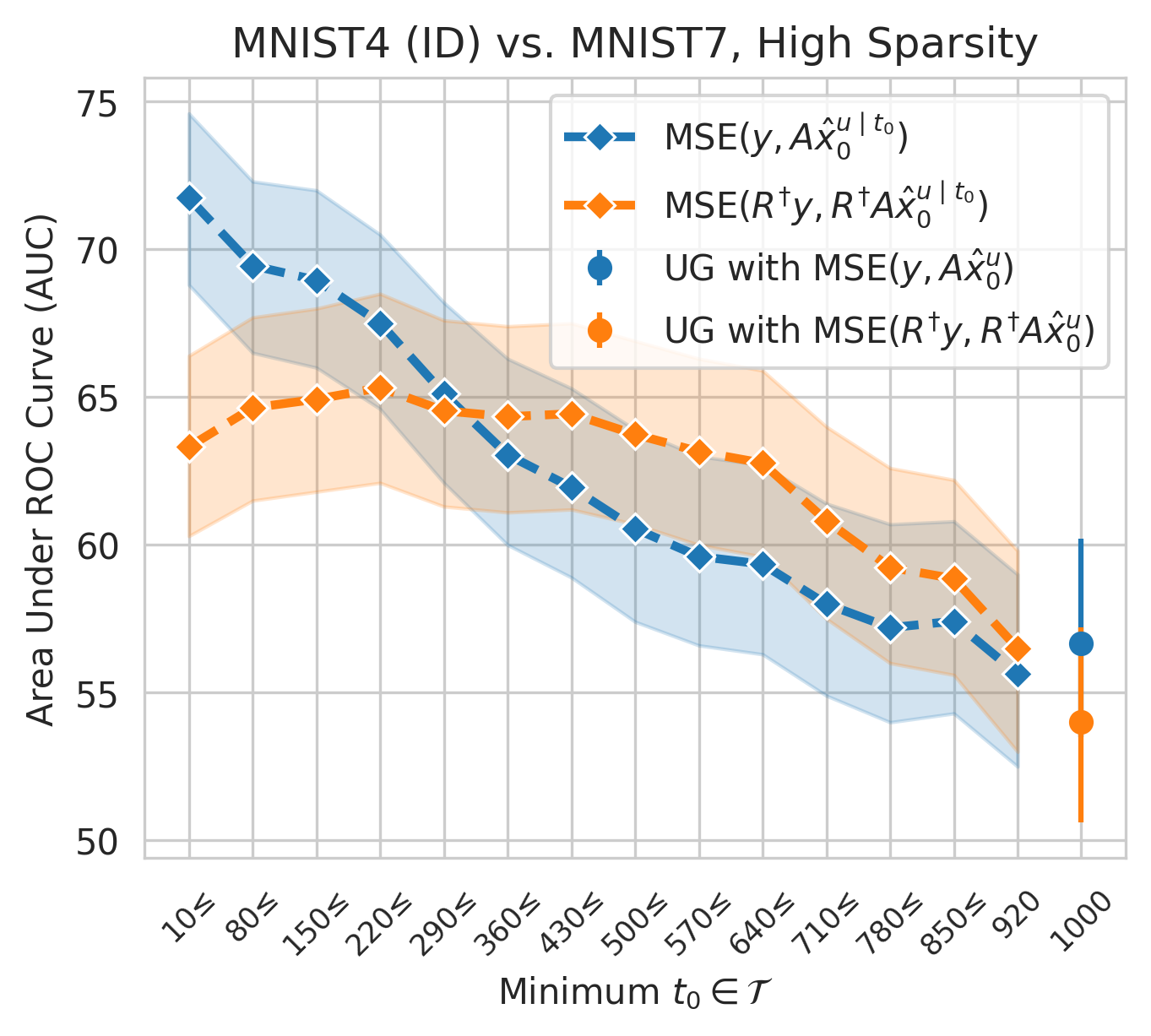}
    \end{subfigure}
   \begin{subfigure}[b]{0.24\textwidth}
        \centering
        \includegraphics[width=\textwidth,keepaspectratio=true]{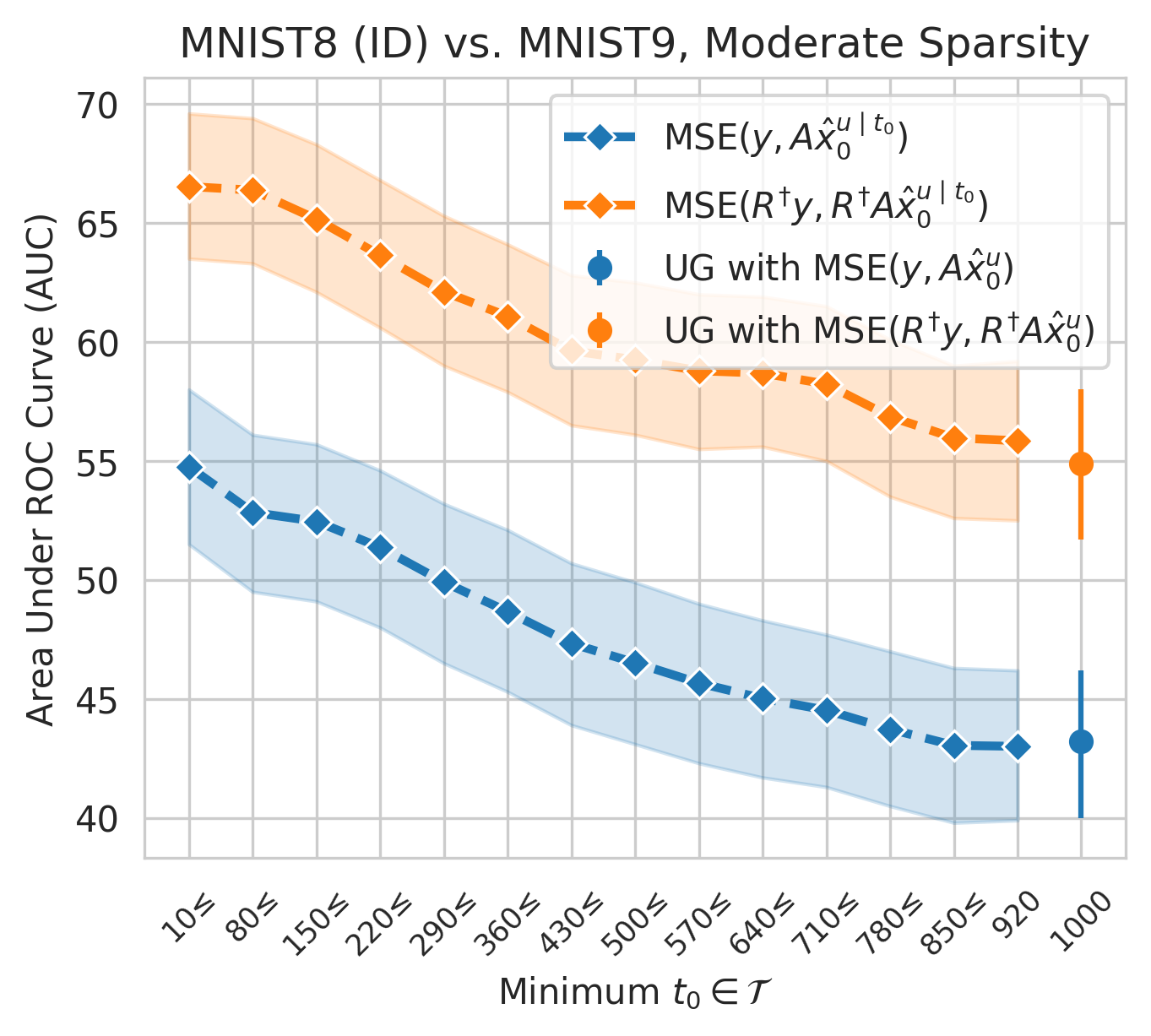}
    \end{subfigure}
        \begin{subfigure}[b]{0.24\textwidth}
        \centering
        \includegraphics[width=\textwidth,keepaspectratio=true]{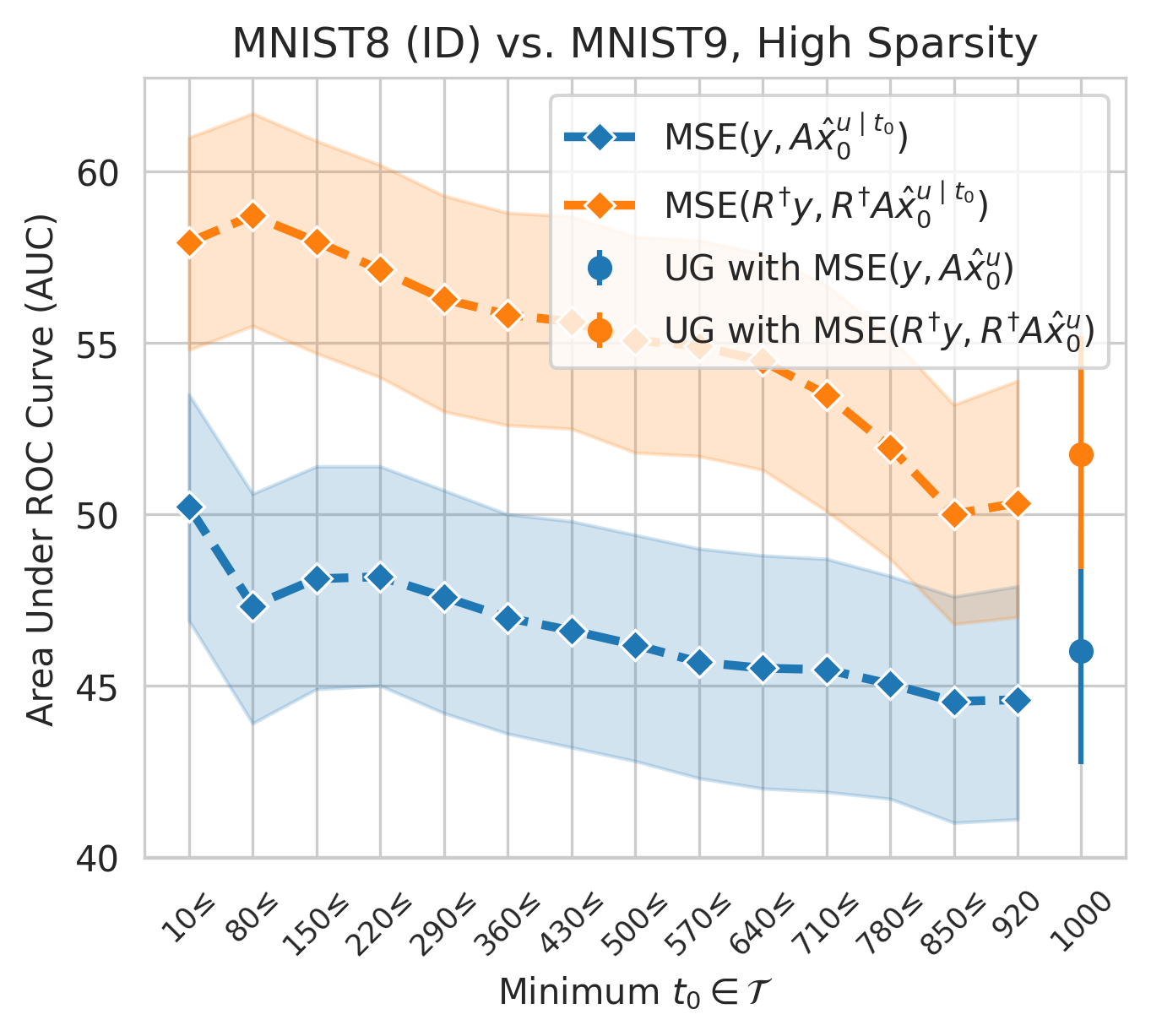}
    \end{subfigure}    
    \\
    \begin{subfigure}[b]{0.24\textwidth}
        \centering
        \includegraphics[width=\textwidth,keepaspectratio=true]{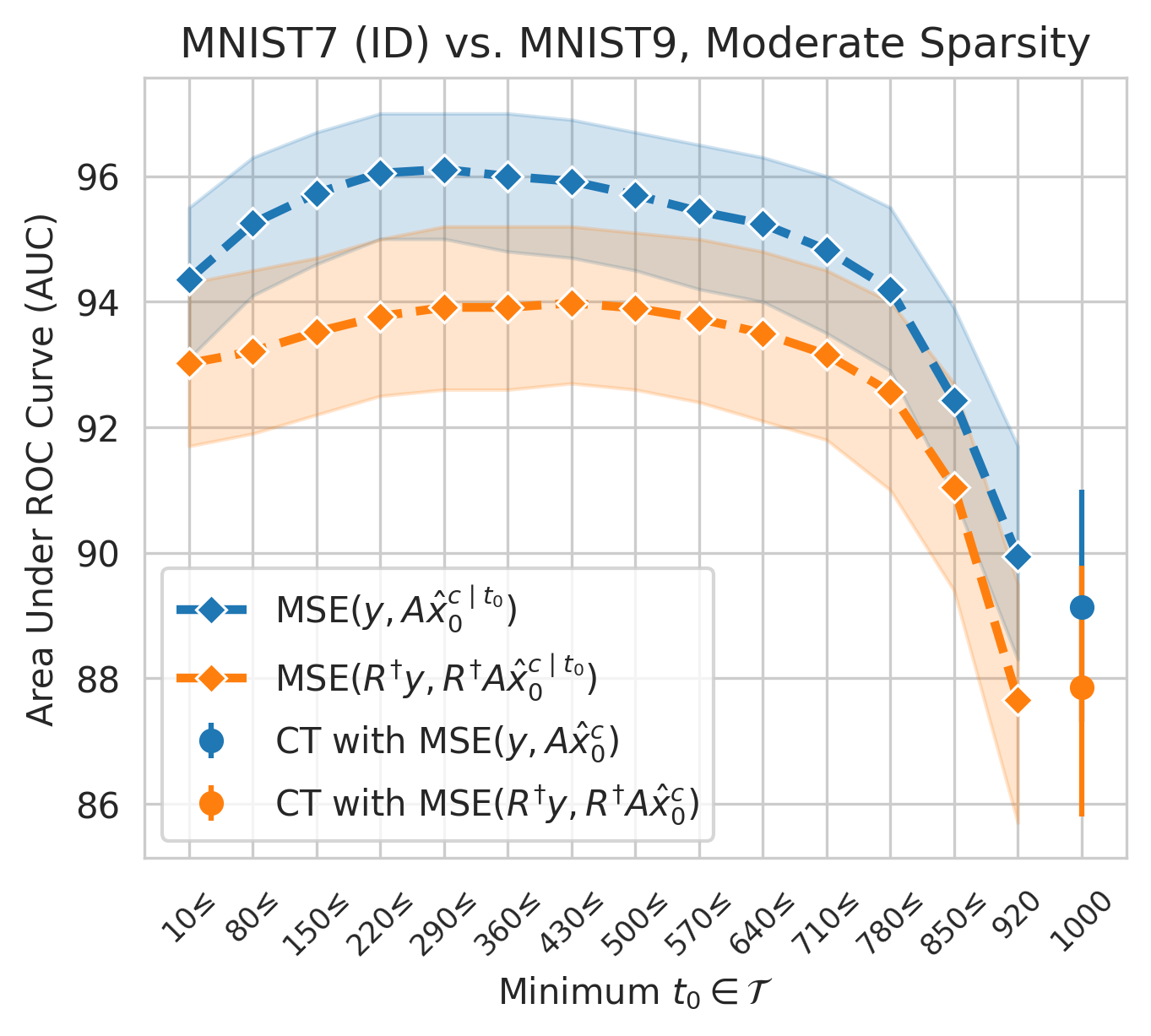}
    \end{subfigure}
    \begin{subfigure}[b]{0.24\textwidth}
        \centering
        \includegraphics[width=\textwidth,keepaspectratio=true]{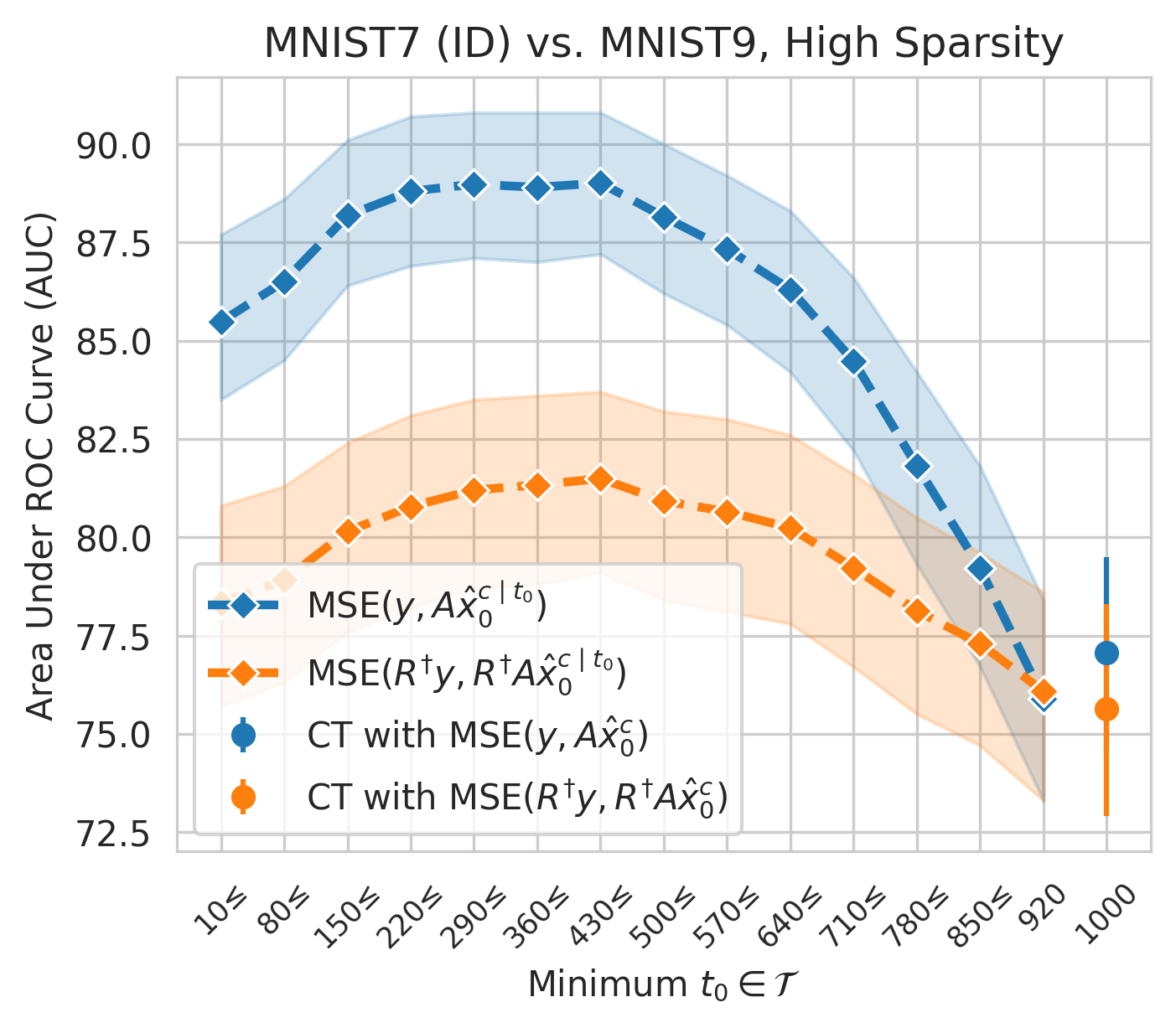}
    \end{subfigure}    
    \begin{subfigure}[b]{0.24\textwidth}
        \centering
        \includegraphics[width=\textwidth,keepaspectratio=true]{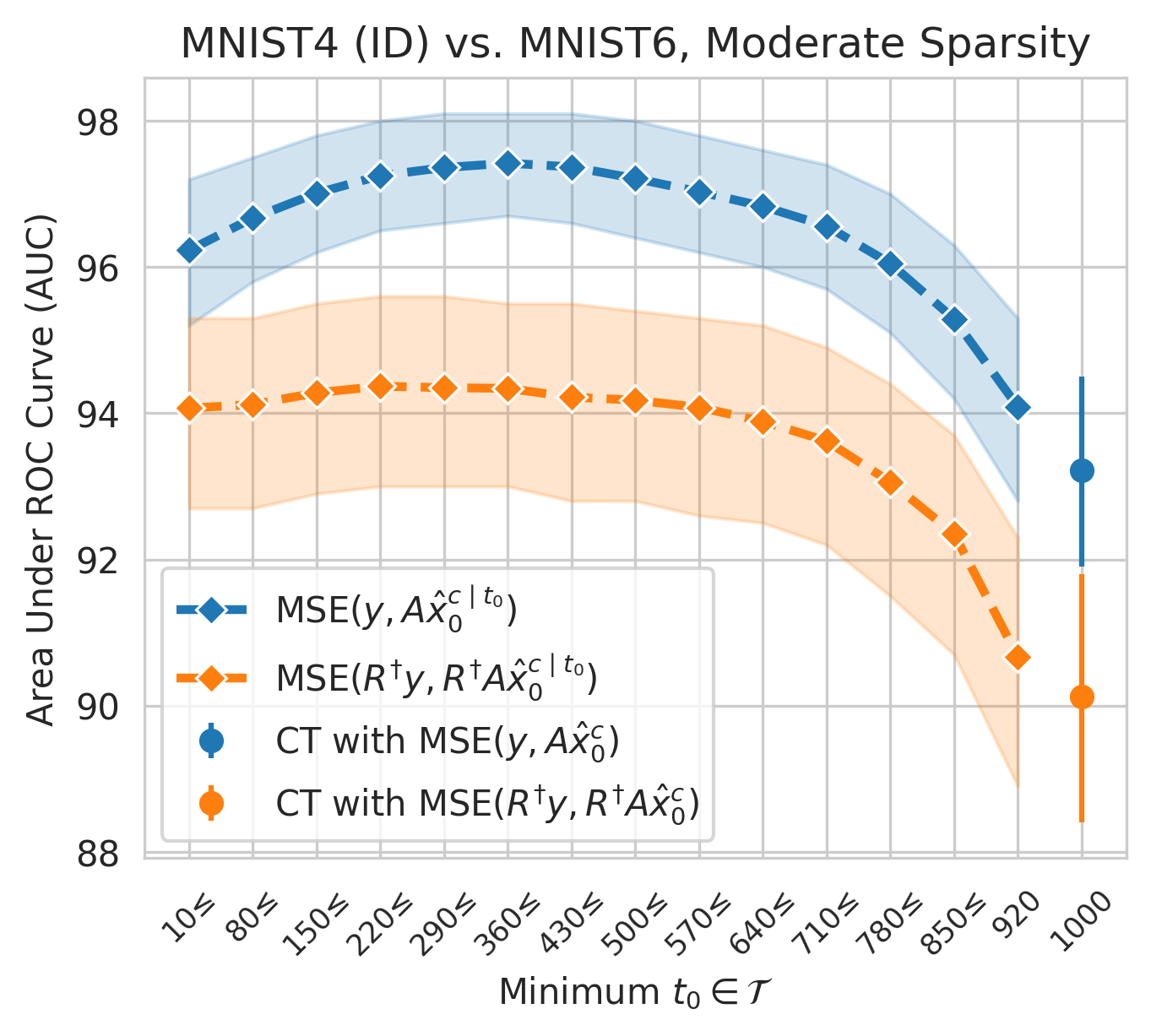}
    \end{subfigure}
    \begin{subfigure}[b]{0.24\textwidth}
        \centering
        \includegraphics[width=\textwidth,keepaspectratio=true]{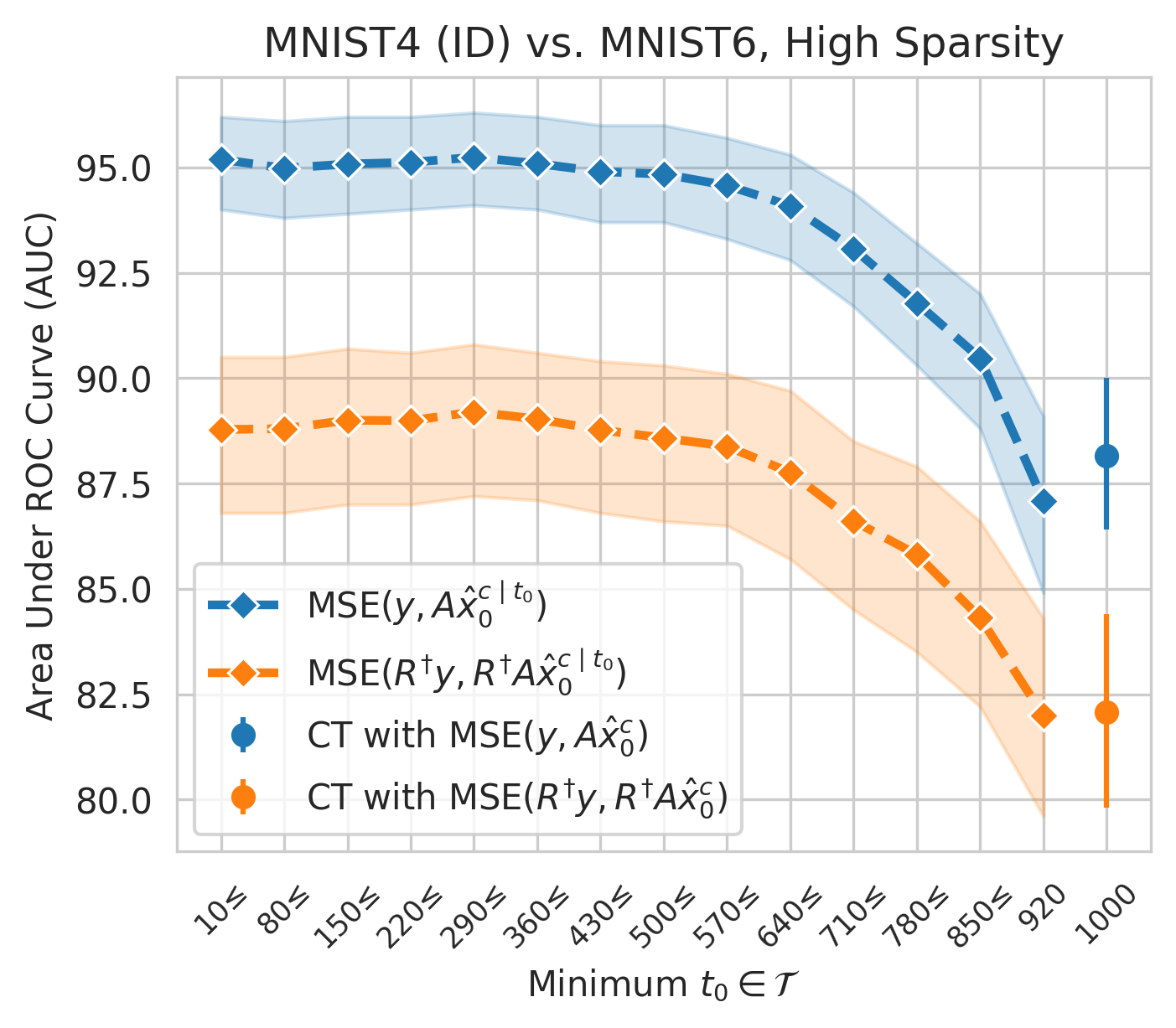}
    \end{subfigure} 
        \\
    \begin{subfigure}[b]{0.24\textwidth}
        \centering
        \includegraphics[width=\textwidth,keepaspectratio=true]{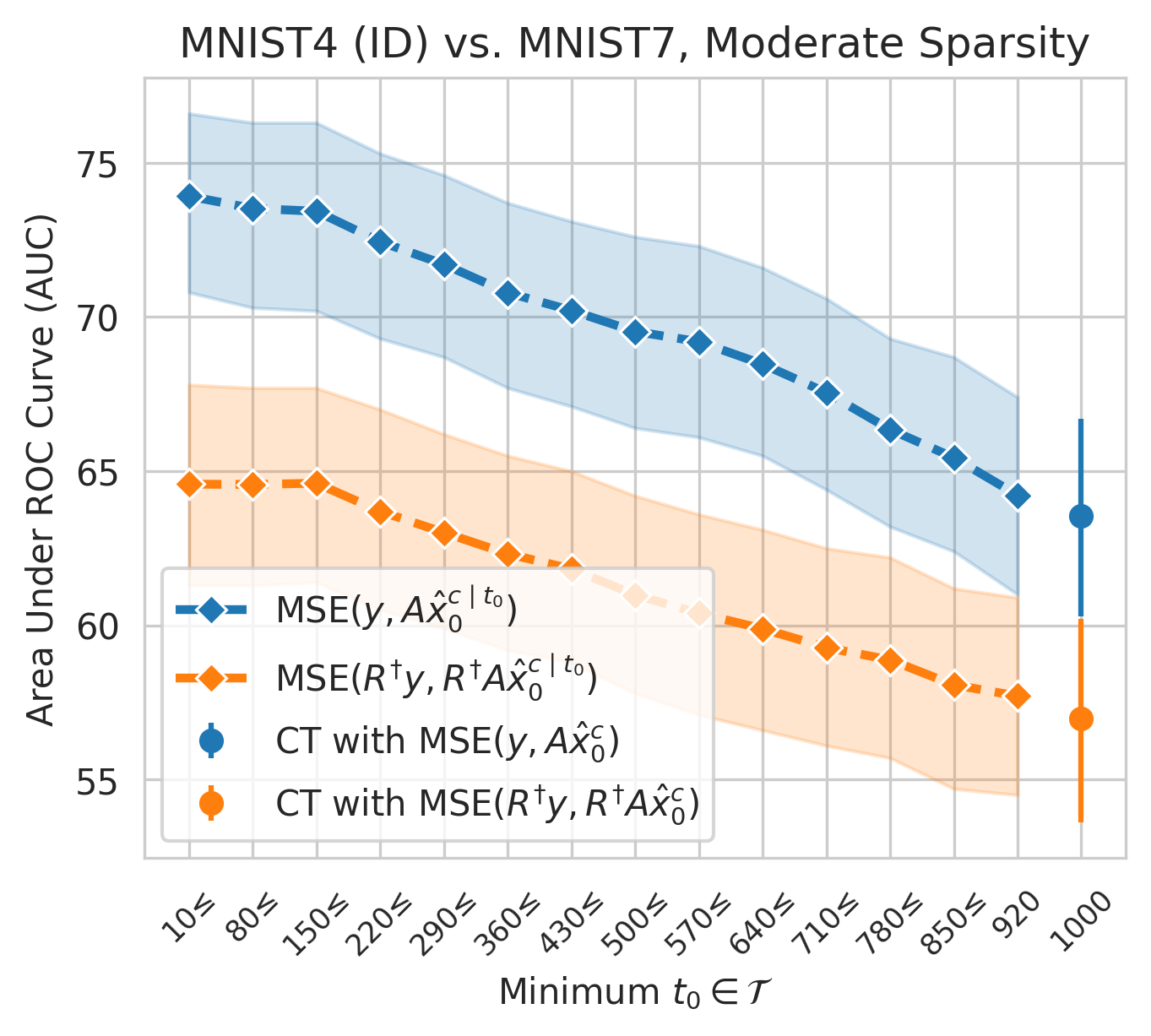}
    \end{subfigure}
    \begin{subfigure}[b]{0.24\textwidth}
        \centering
        \includegraphics[width=\textwidth,keepaspectratio=true]{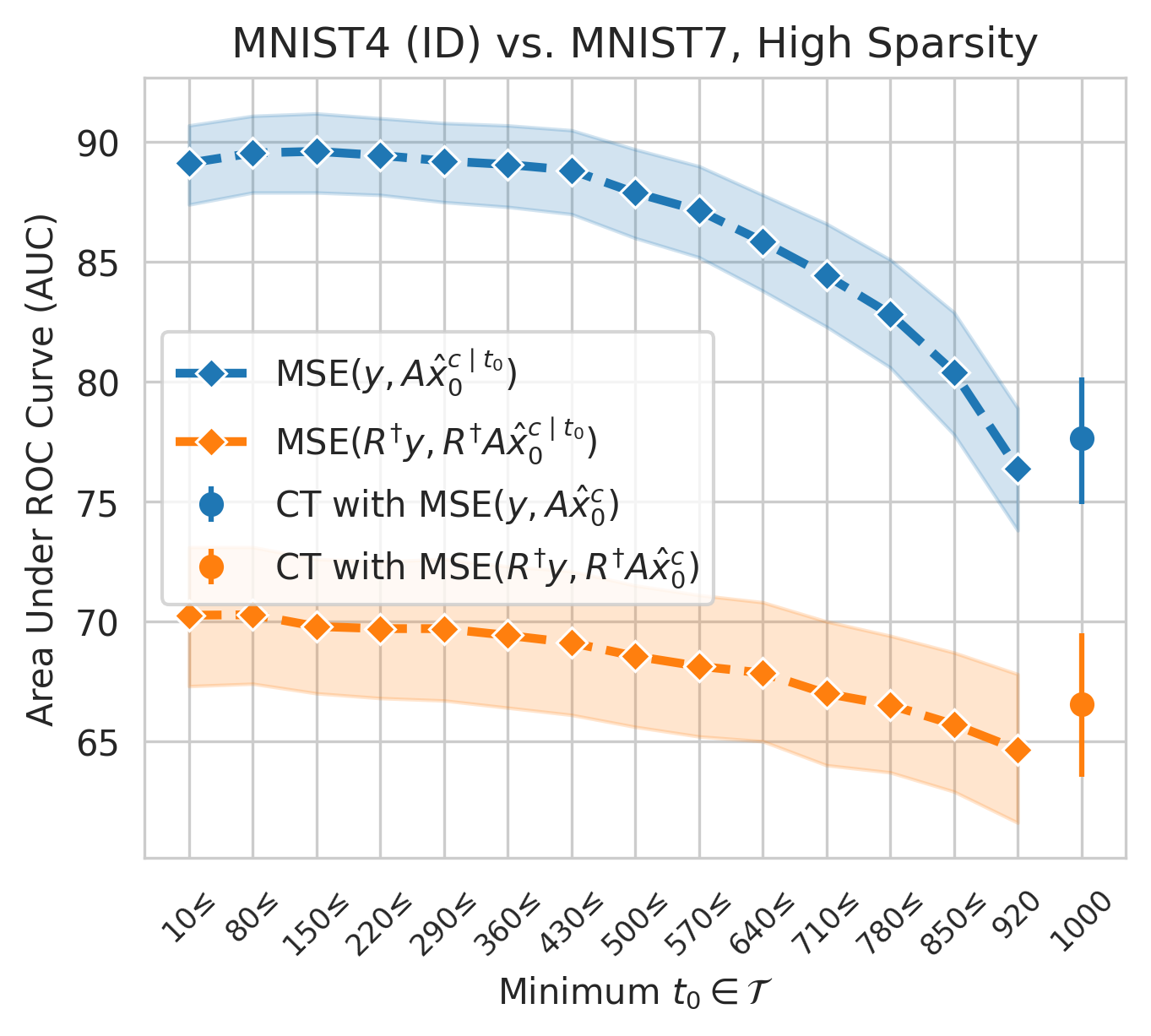}
    \end{subfigure}    
    \begin{subfigure}[b]{0.24\textwidth}
        \centering
        \includegraphics[width=\textwidth,keepaspectratio=true]{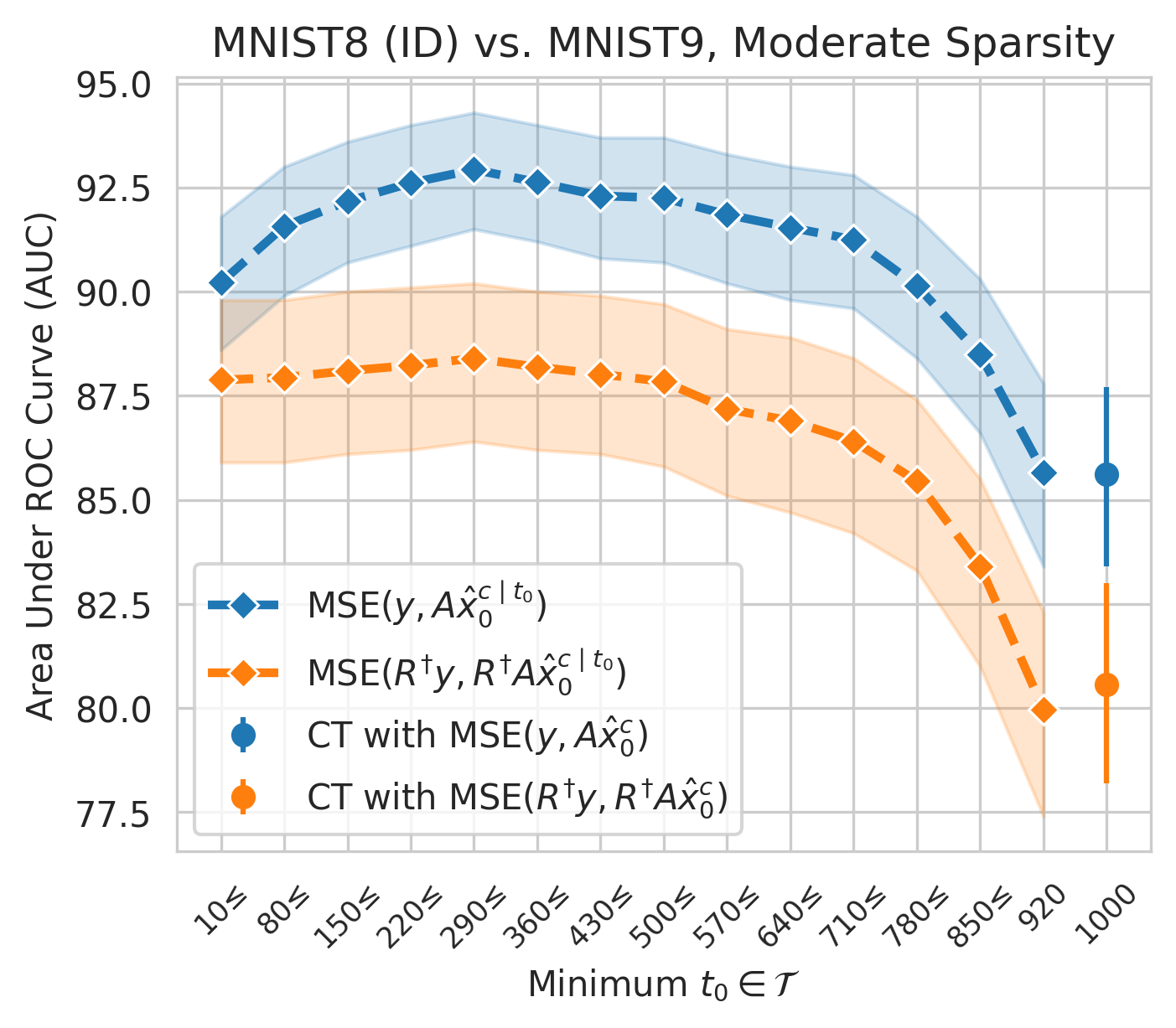}
    \end{subfigure}
    \begin{subfigure}[b]{0.24\textwidth}
        \centering
        \includegraphics[width=\textwidth,keepaspectratio=true]{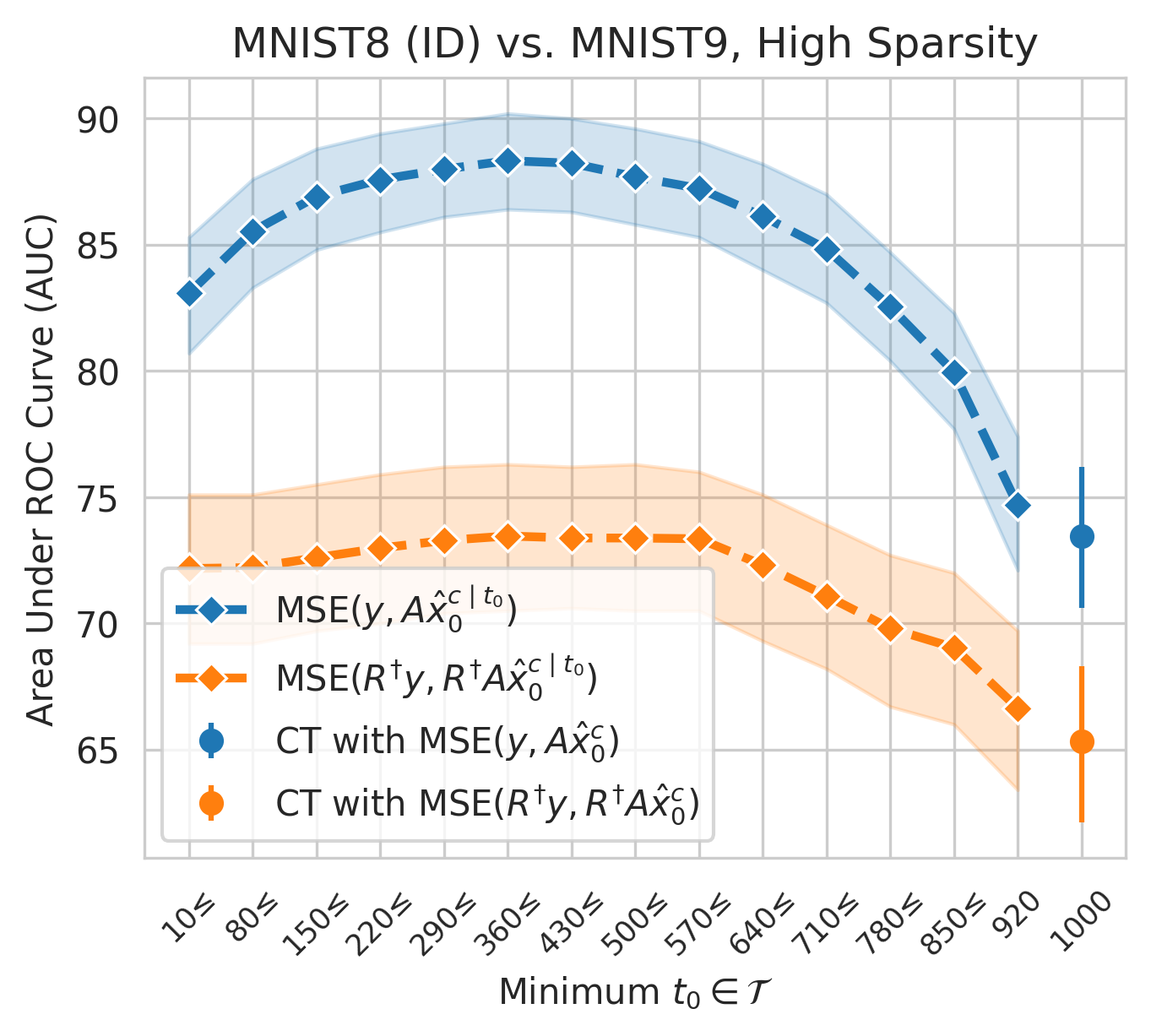}
    \end{subfigure}
 \caption{AUC scores with respect to different minimum values of $t_0$, shown for unconditional sampling \textbf{(top two rows)} and conditional sampling \textbf{(bottom two rows)}, across various ``ID vs. OOD, Moderate/High Sparsity'' scenarios, indicated in the individual plot titles, with \{18, 12, 9\} and \{6, 5, 4\} number of projections, respectively. As the minimum $t_0$ increases from 10 to 920, the number of reconstructions for OOD detection decreases from 14 to 1. The result shown at $t_0 = 1000$ represents what would be obtained without applying multi-scale scoring with FBP inputs. This corresponds to unconditional generation (denoted as ``UG'') and CT reconstruction (denoted as ``CT'') in unconditional \textbf{(top two rows)} and conditional cases \textbf{(bottom two rows)}, respectively. The shaded areas around the AUC values indicate confidence bounds, reported for all cases based on 1000-fold bootstrapping.}
\label{fig:rq1mint0_sup}  
\end{figure*}

\begin{figure*}[h!]
    \centering
   \begin{subfigure}[b]{0.24\textwidth}
        \centering
        \includegraphics[width=\textwidth,keepaspectratio=true]{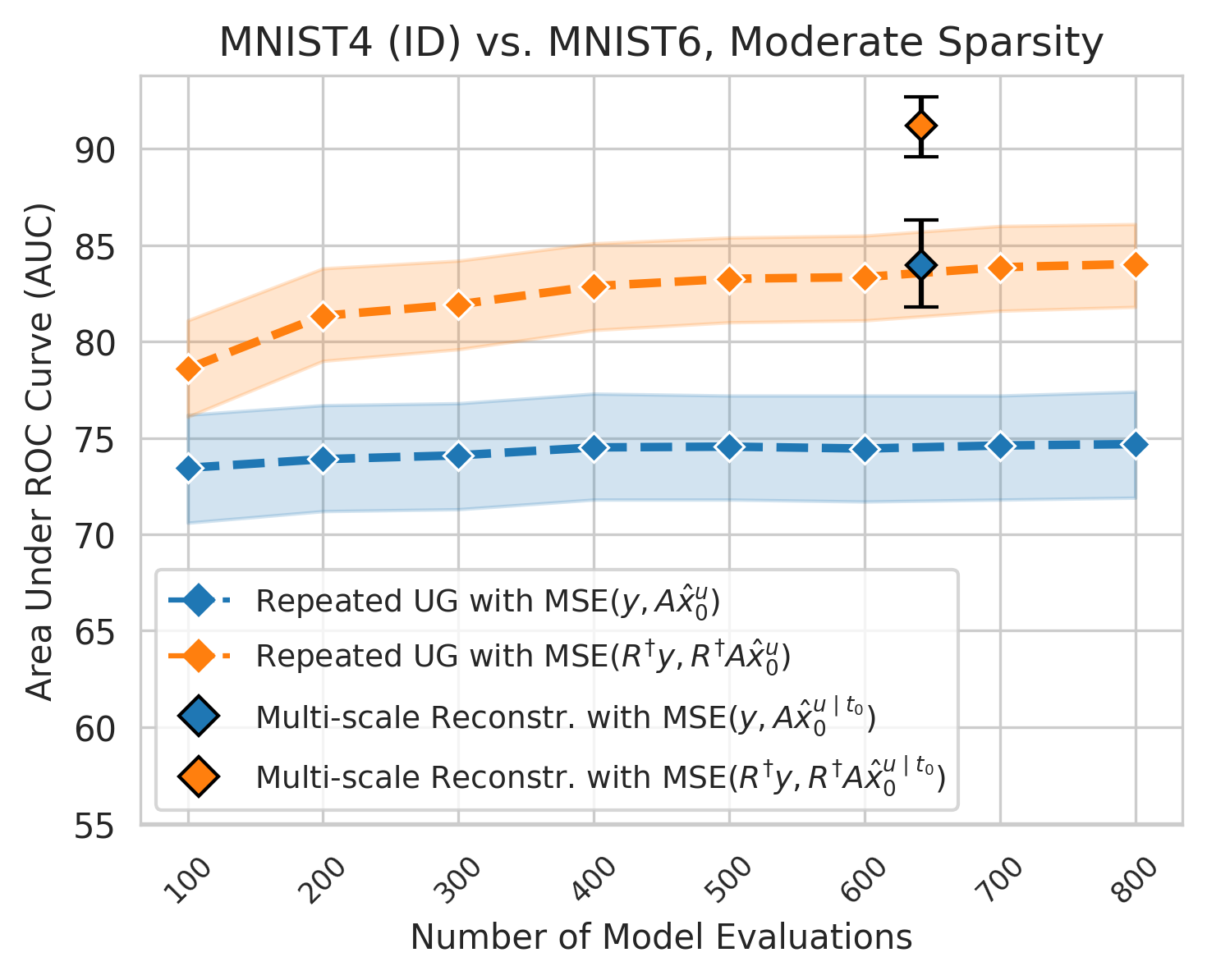}
    \end{subfigure}
       \begin{subfigure}[b]{0.24\textwidth}
        \centering
        \includegraphics[width=\textwidth,keepaspectratio=true]{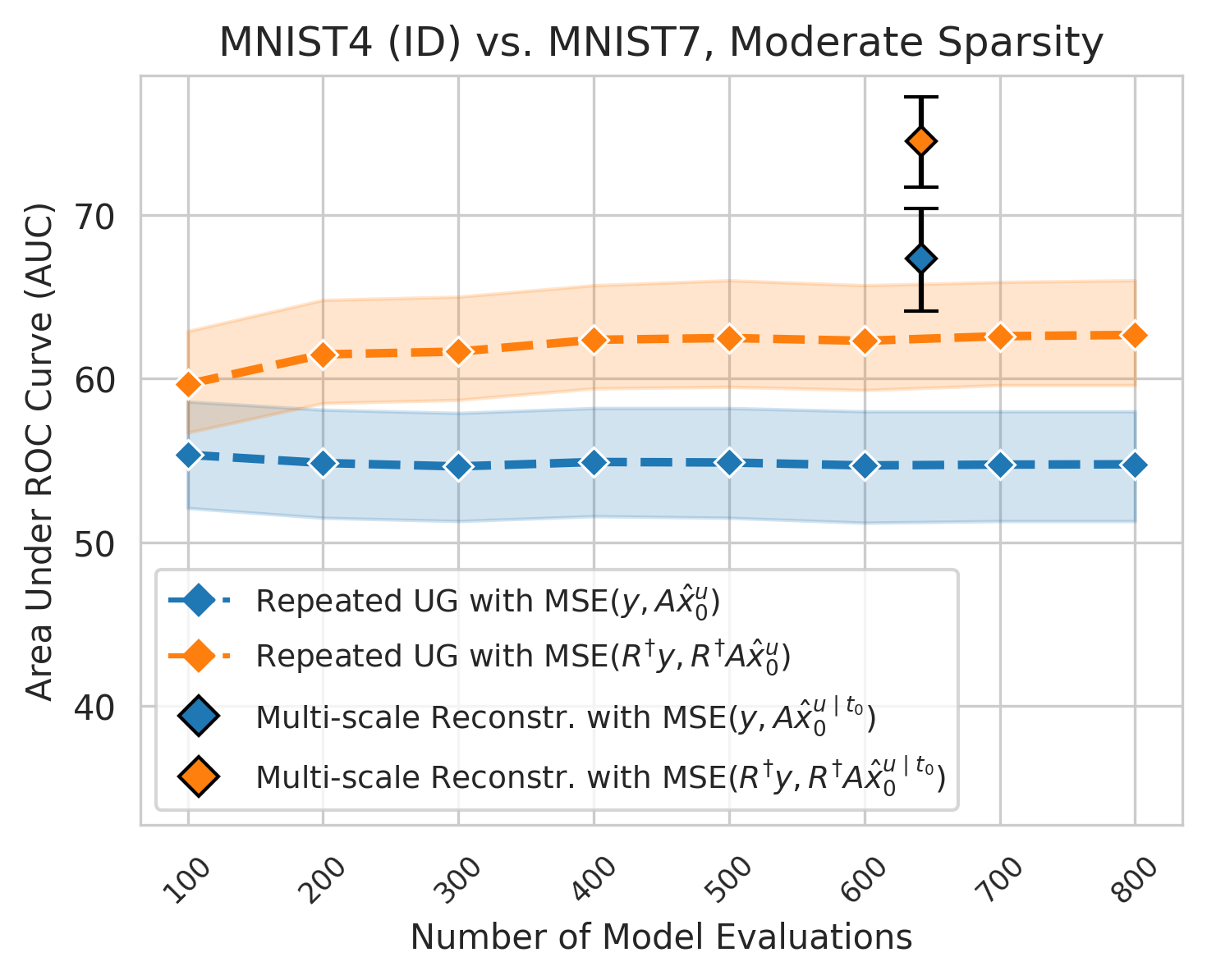}
    \end{subfigure}
            \begin{subfigure}[b]{0.24\textwidth}
        \centering
        \includegraphics[width=\textwidth,keepaspectratio=true]{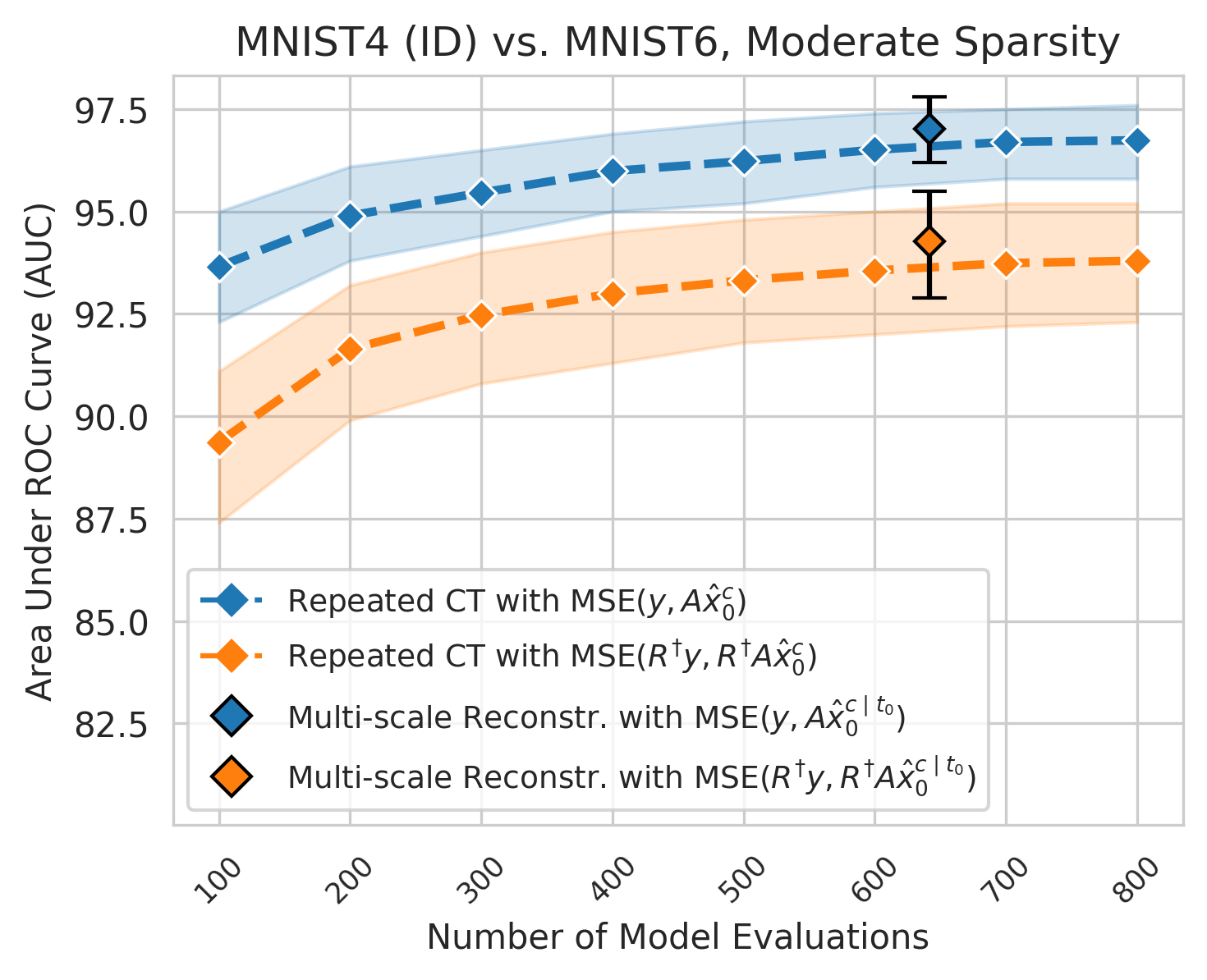}
    \end{subfigure}
      \begin{subfigure}[b]{0.24\textwidth}
        \centering
        \includegraphics[width=\textwidth,keepaspectratio=true]{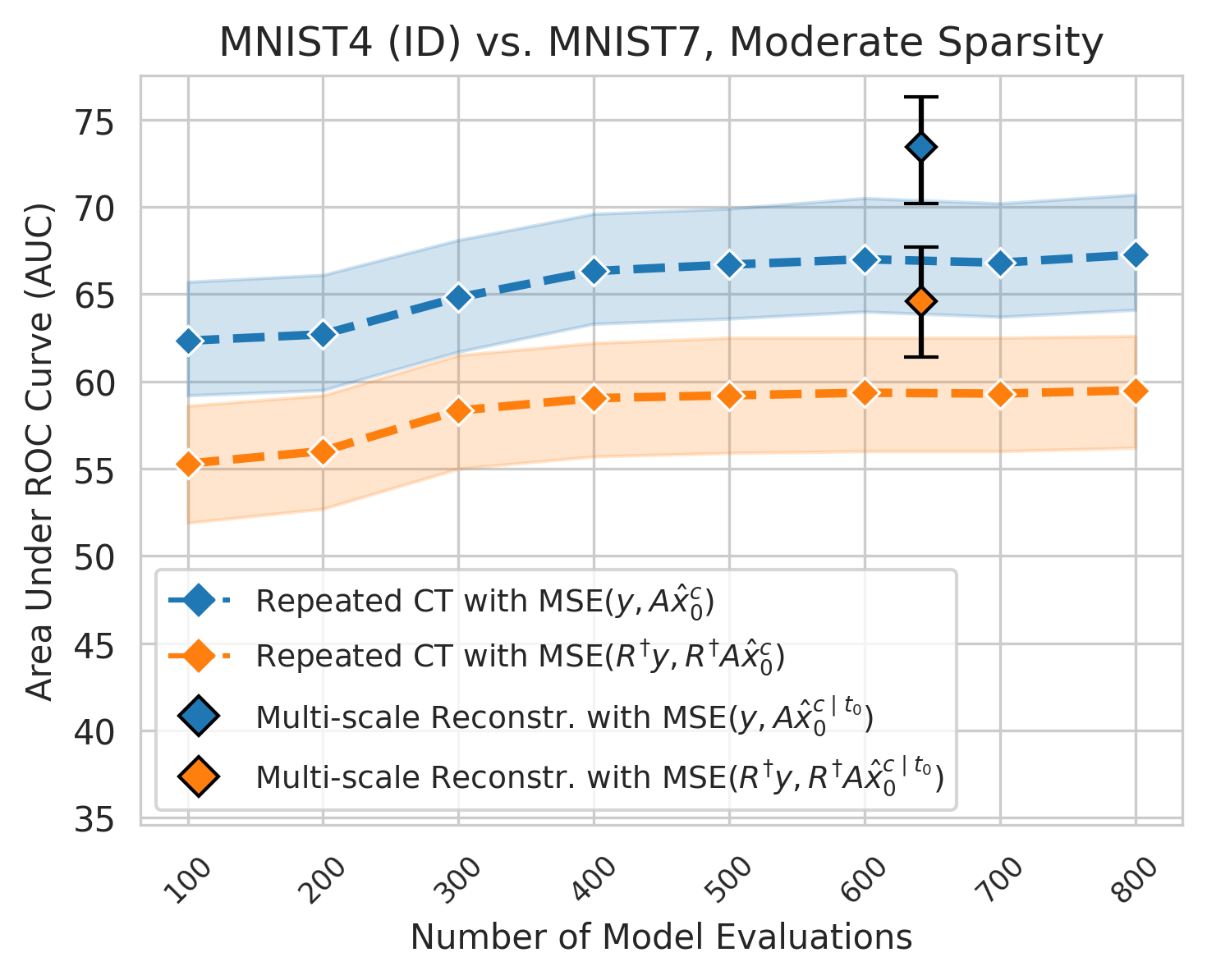}
    \end{subfigure}
    \\
        \begin{subfigure}[b]{0.24\textwidth}
        \centering
        \includegraphics[width=\textwidth,keepaspectratio=true]{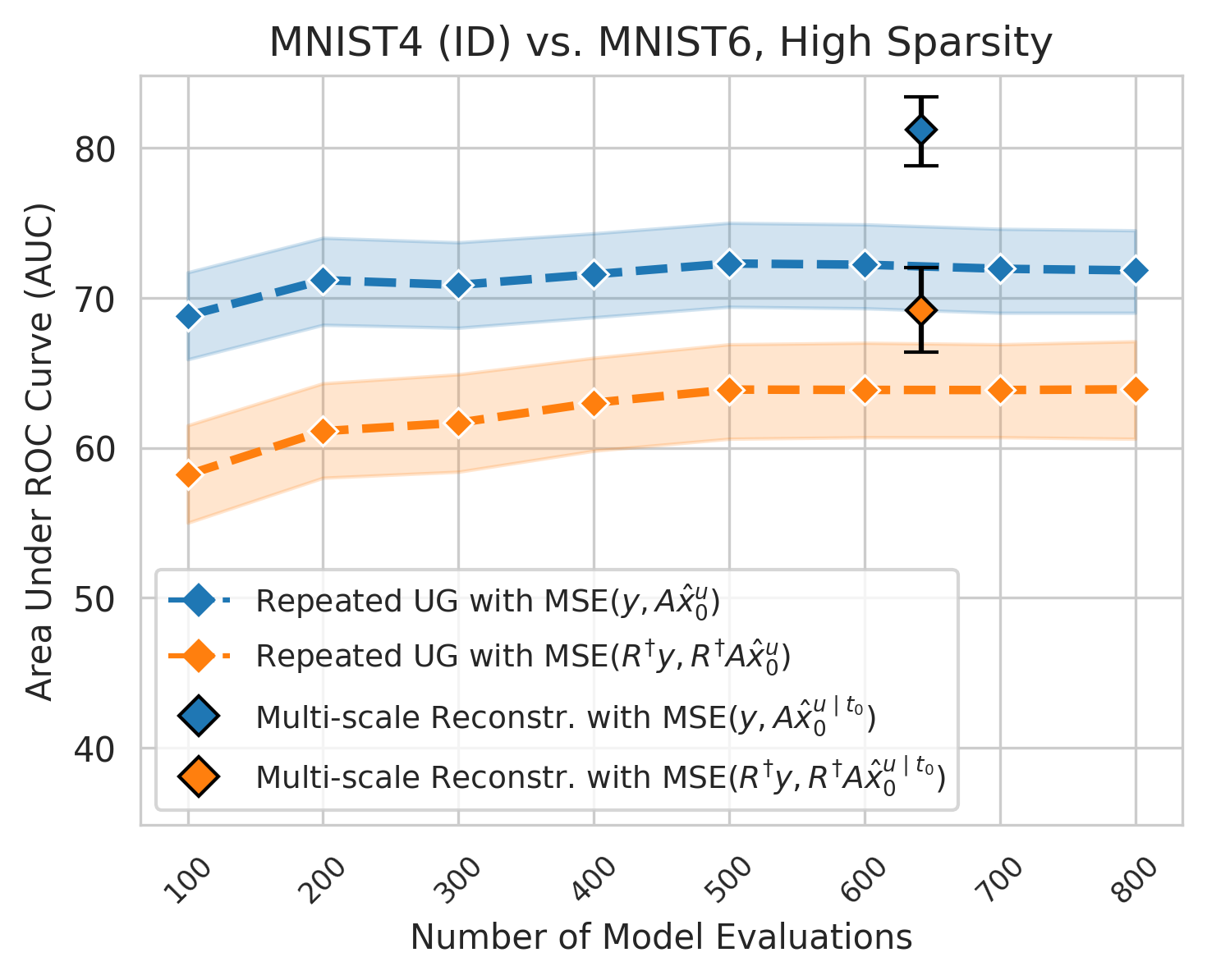}
    \end{subfigure}    
        \begin{subfigure}[b]{0.24\textwidth}
        \centering
        \includegraphics[width=\textwidth,keepaspectratio=true]{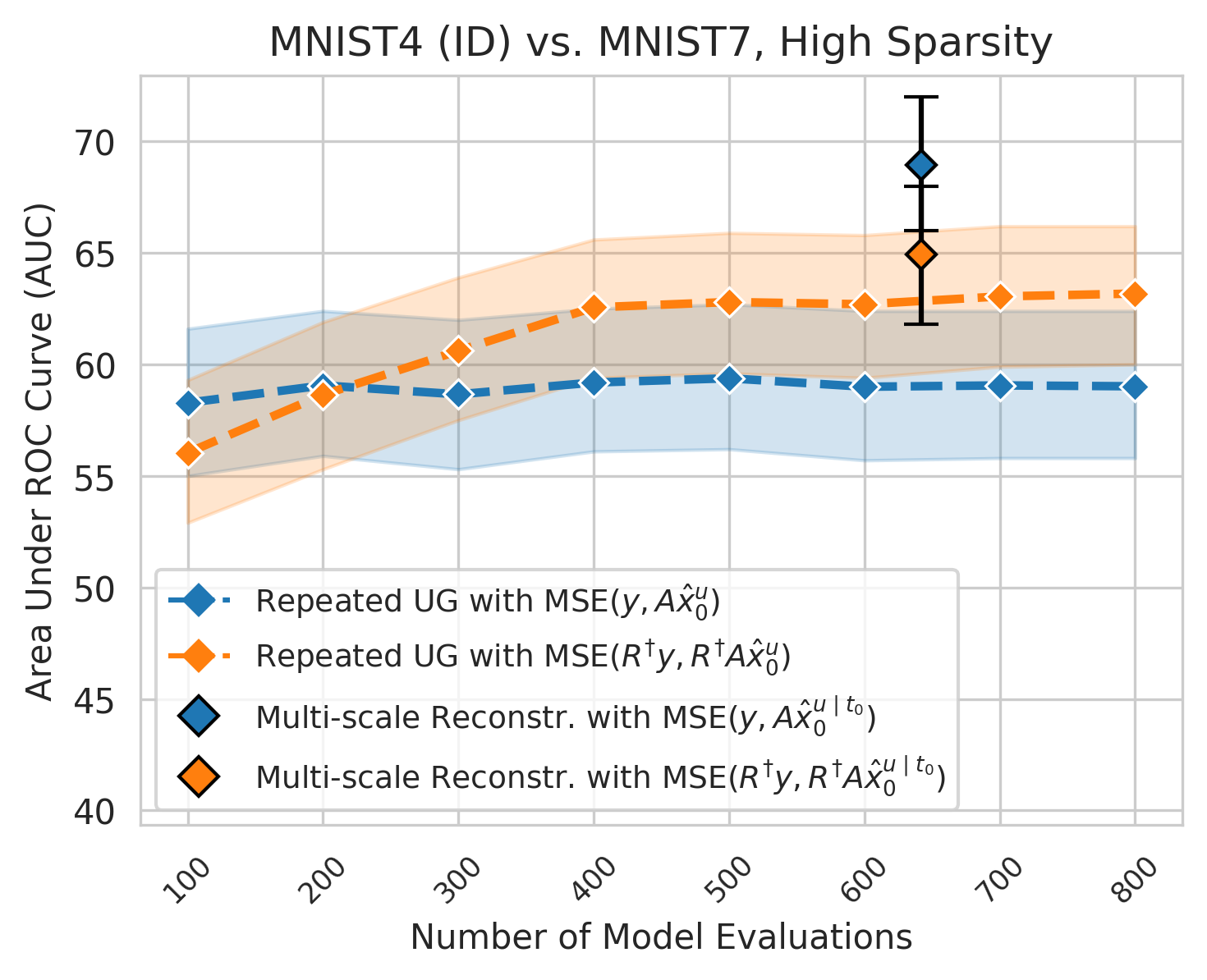}
    \end{subfigure} 
        \begin{subfigure}[b]{0.24\textwidth}
        \centering
        \includegraphics[width=\textwidth,keepaspectratio=true]{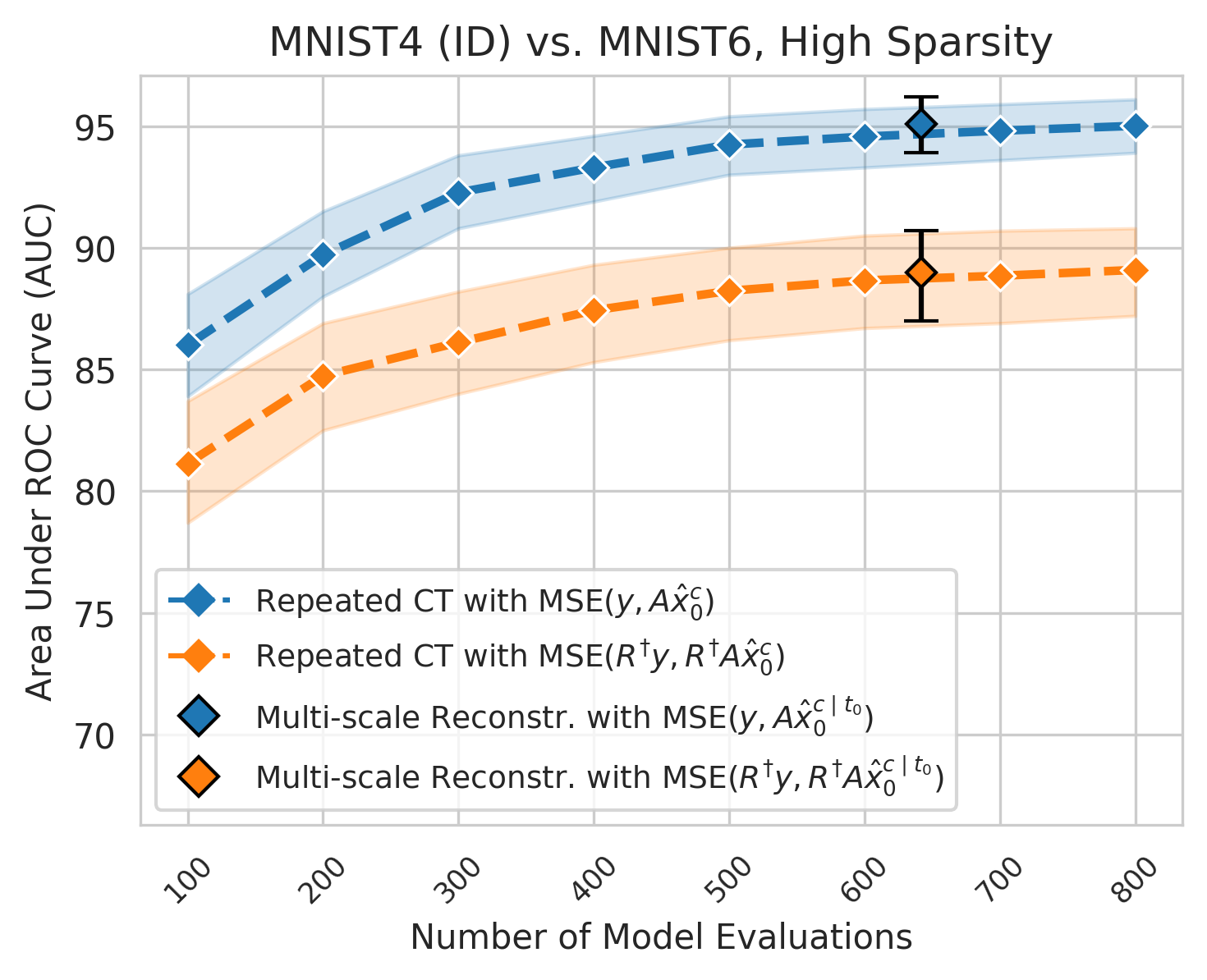}
    \end{subfigure} 
    \begin{subfigure}[b]{0.24\textwidth}
        \centering
        \includegraphics[width=\textwidth,keepaspectratio=true]{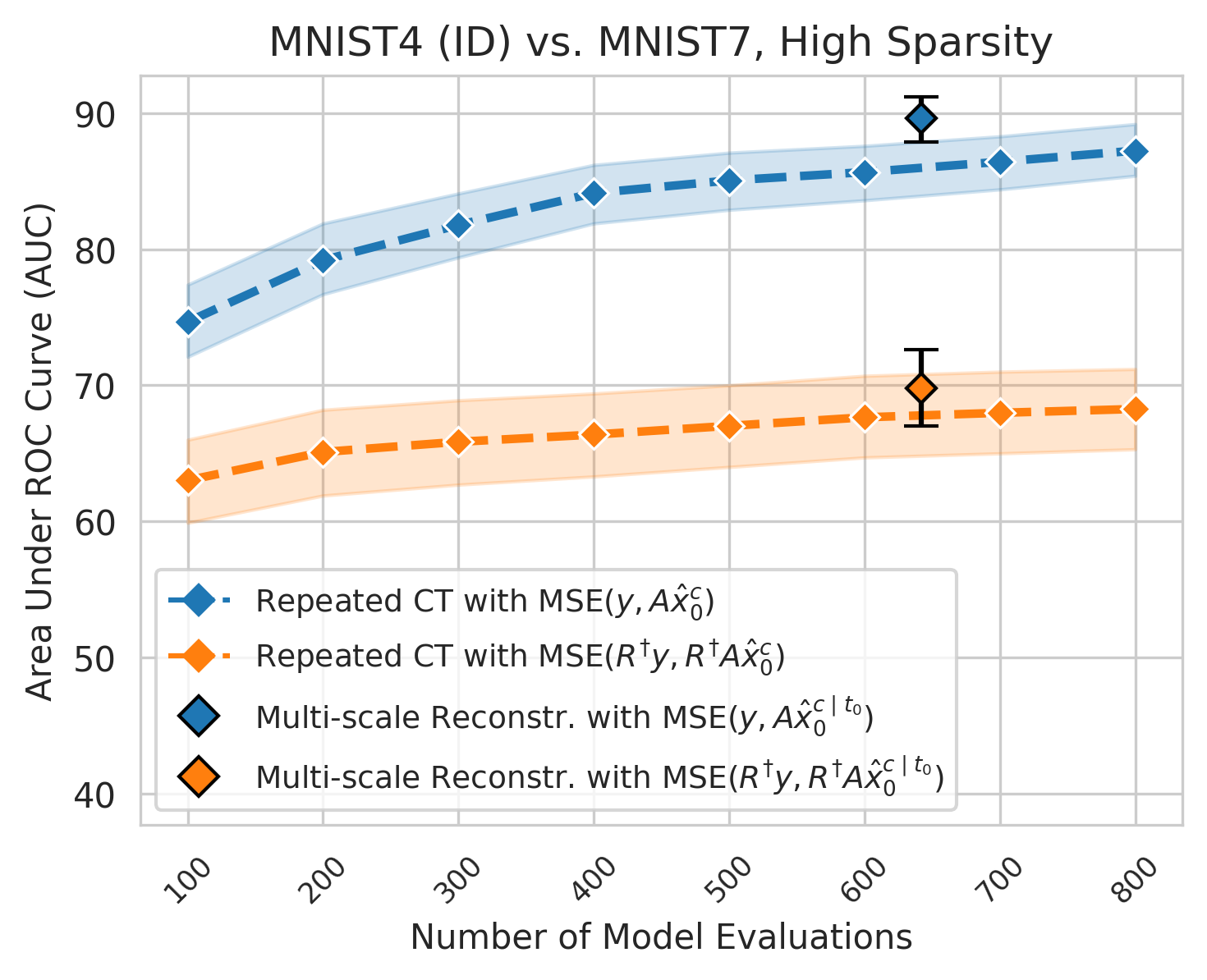}
    \end{subfigure}    
 \caption{AUC scores as a function of the total number of model evaluations, where each evaluation count of \(100 \times n\) represents running the generation process \(n\) times to mitigate randomness and ensure a fair comparison. The evolutions of both unconditional (``UG'') and conditional (``CT'', since it corresponds to CT reconstruction) generations are compared against the AUC scores obtained from multi-scale reconstructions by using noisy FBP inputs, depicted by individual error bars at $642$ model evaluations. The analysis considers the model trained on MNIST4, with MNIST6 and MNIST7 treated as OOD to complement the visual results in Fig. 5. The \textbf{top row} presents results for Moderate Sparsity, while the \textbf{bottom row} corresponds to High Sparsity. Shaded regions around AUC values indicate confidence bounds, derived from 1000-fold bootstrapping. The use of 100 model evaluations per generation is due to the PLMS sampler.}
\label{fig:rq1repeatedct_sup}  
\end{figure*}

\clearpage

\begin{figure*}[t!]
 \centering
 \begin{subfigure}[b]{\textwidth}
 \includegraphics[width=\textwidth,keepaspectratio=true]{ 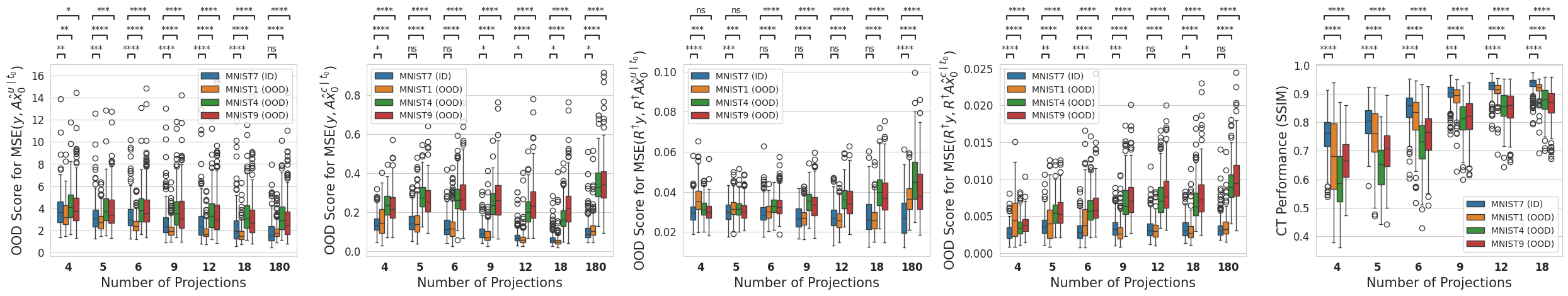}
  \caption{OOD score and SSIM distributions based on the model trained on MNIST7.}
 \end{subfigure}  
 \vspace{2pt}
 \begin{subfigure}[b]{\textwidth}
  \includegraphics[width=\textwidth,keepaspectratio=true]{ 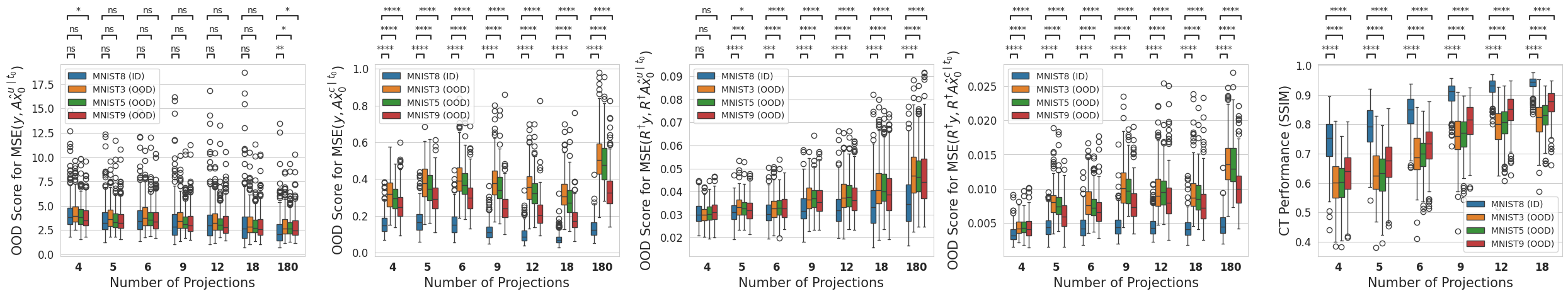}
  \caption{OOD score and SSIM distributions based on the model trained on MNIST8.}
 \end{subfigure} 
 \caption{This figure presents additional analyses complementary to Fig. 7. In both (a) and (b), the first two plots from the left represent Z-scores derived from MSE errors in the sinogram domain, with the first plot using unconditional reconstructions and the second using conditional reconstructions. Similarly, the third and fourth plots focus on the FBP domain, again using unconditional and conditional reconstructions, respectively. The final plot in (a) illustrates SSIM distributions to assess CT reconstruction ($t_0 = 1000$) performance rather than OOD detection. The statistical significance annotations indicated by asterisks denote p-values from t-tests for independent samples comparing ID and OOD distributions (* for $p \leq 5.00 \times 10^{-2}$, ** for $p \leq 1.00 \times 10^{-2}$, *** for $p \leq 1.00 \times 10^{-3}$, **** for $p \leq 1.00 \times 10^{-4}$, and $ns$ for non-significant results).}
 \label{fig:rq1boxplotwrtnproj_sup}
\end{figure*}

\clearpage

\begin{figure*}[t!]
 \centering
 \begin{subfigure}[b]{0.4\textwidth}
  \centering
{\includegraphics[width=\textwidth,keepaspectratio=true]{ 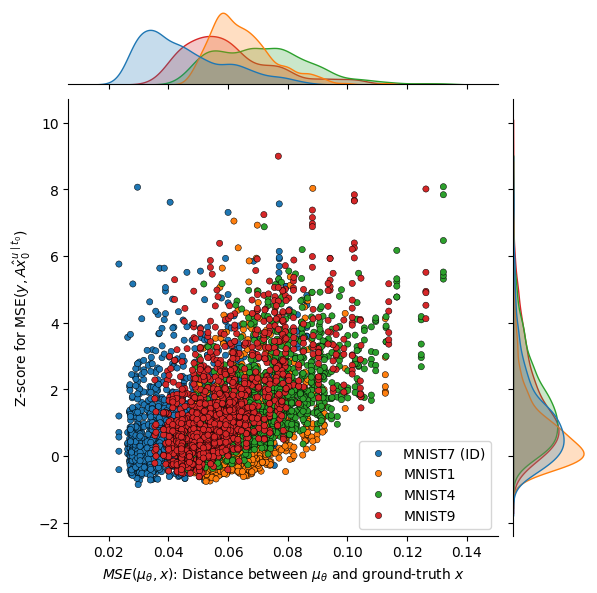}}
 \end{subfigure}
   \begin{subfigure}[b]{0.4\textwidth}
  \centering
{\includegraphics[width=\textwidth,keepaspectratio=true]{ 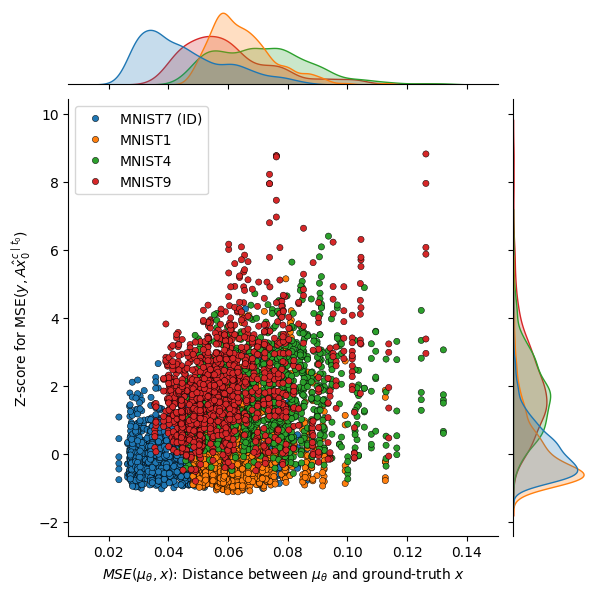}}
 \end{subfigure}
  \begin{subfigure}[b]{0.4\textwidth}
  \centering
{\includegraphics[width=\textwidth,keepaspectratio=true]{ 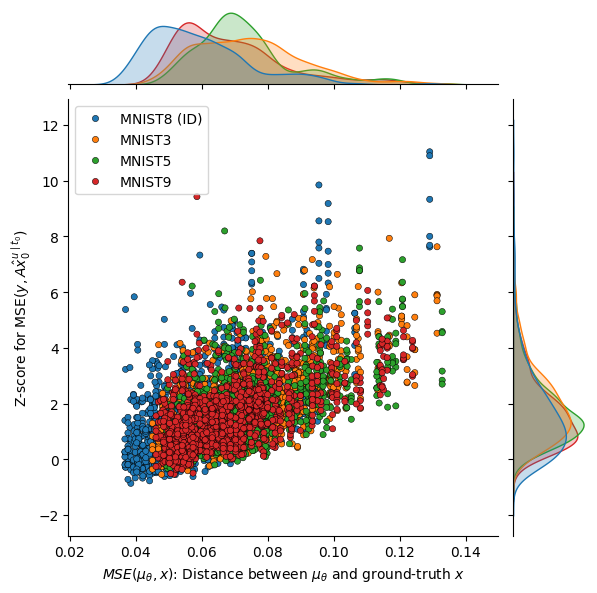}}
 \end{subfigure}
   \begin{subfigure}[b]{0.4\textwidth}
  \centering
{\includegraphics[width=\textwidth,keepaspectratio=true]{ 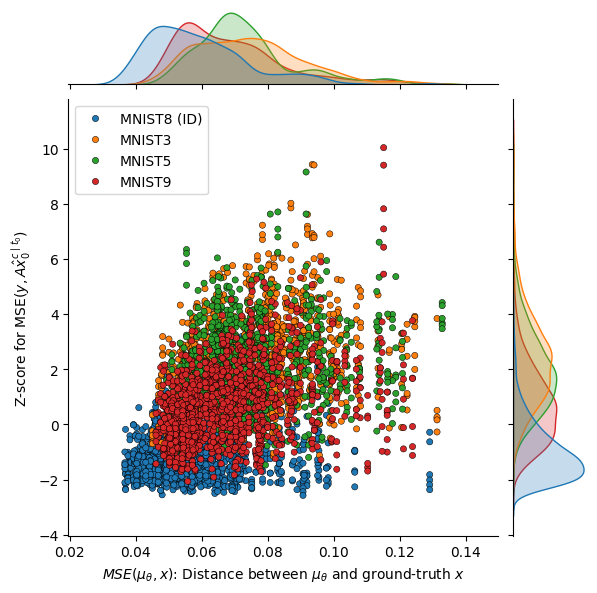}}
 \end{subfigure}
 \caption{This figure presents additional analyses complementary to Fig. 8. Joint scatter plots illustrating how OOD scores vary with increasing distance of the image from the mean of the prior distribution, in unconditional (left column) and conditional (right column) cases, for the models trained on MNIST7 (top row) and MNIST8 (bottom row). For both cases, the positive correlation between the distance and the OOD score, initially moderate with r = 0.47 and r = 0.58 (p $<$ 0.0001) when using MSE(${y}, {A} {\hat{x}}_{0}^{u \mid t_0}$), weakens to r = 0.31 and r = 0.41 (p $<$ 0.0001) with MSE(${y}, {A} {\hat{x}}_{0}^{c \mid t_0}$). For the model trained on MNIST7, conditioning results in a substantial portion of MNIST1 images receiving lower OOD scores, similar to those of ID images. For the model trained on MNIST8, however, this effect is not observed. Conditioning compresses the ID scores to distinctly lower levels than those of all OOD sets, making them more distinguishable.}
 \label{fig:rq2_uncondvscond_sup}
\end{figure*}

\begin{figure*}[t!]
 \centering
 \begin{subfigure}[b]{0.4\textwidth}
  \centering
{\includegraphics[width=\textwidth,keepaspectratio=true]{ 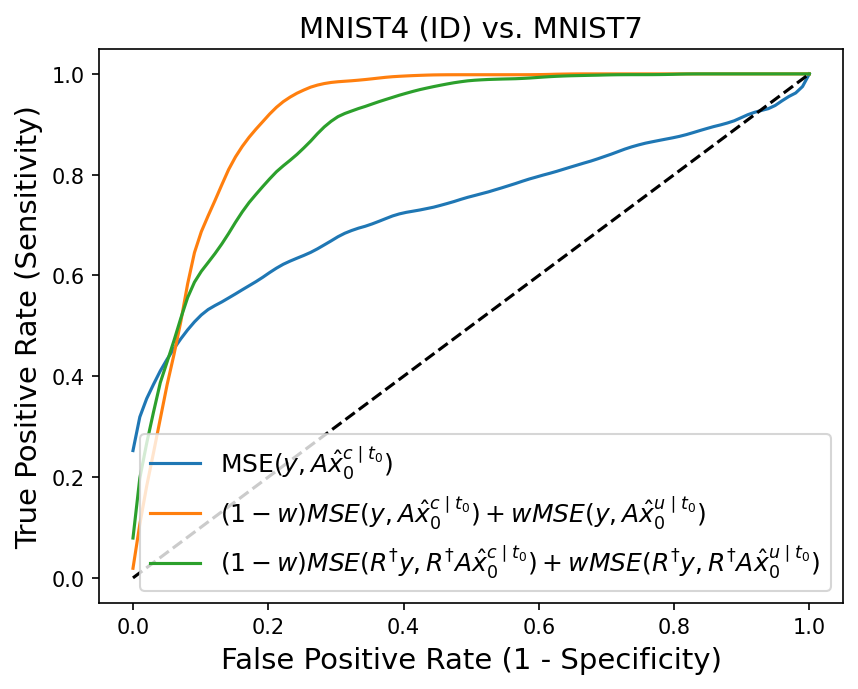}}
 \end{subfigure}
   \begin{subfigure}[b]{0.4\textwidth}
  \centering
{\includegraphics[width=\textwidth,keepaspectratio=true]{ 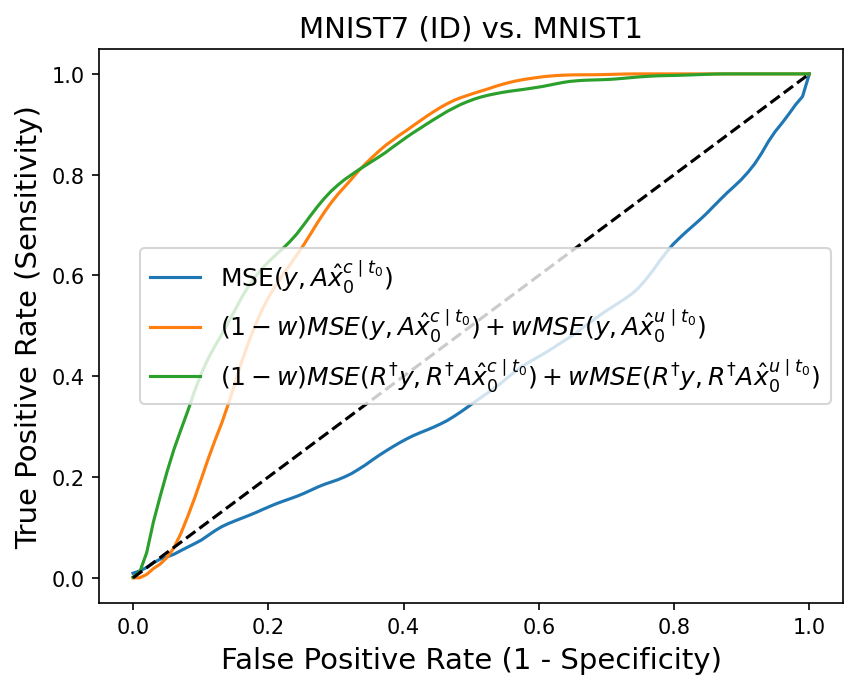}}
 \end{subfigure}
 \begin{subfigure}[b]{0.4\textwidth}
  \centering
{\includegraphics[width=\textwidth,keepaspectratio=true]{ 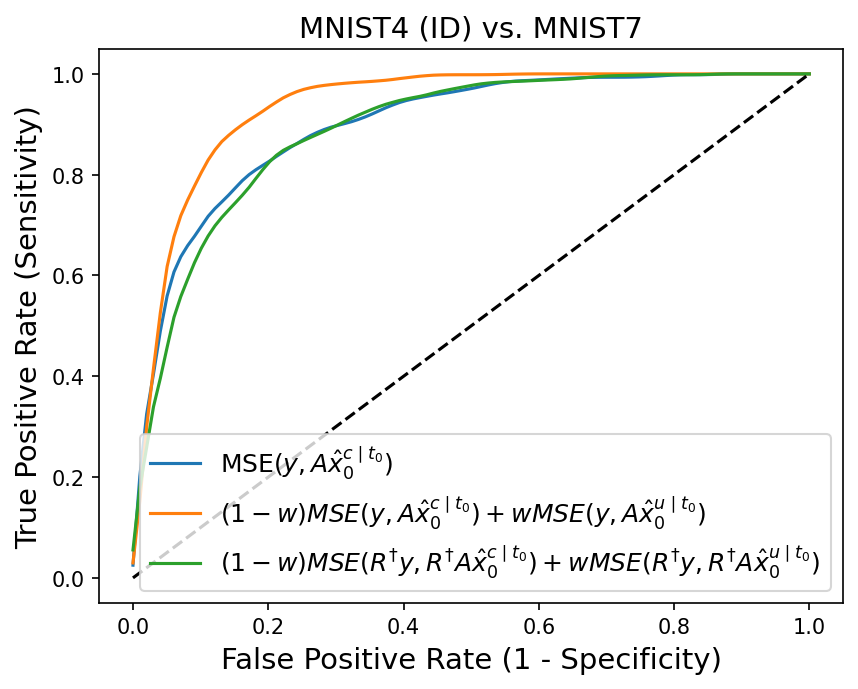}}
 \end{subfigure}
   \begin{subfigure}[b]{0.4\textwidth}
  \centering
{\includegraphics[width=\textwidth,keepaspectratio=true]{ 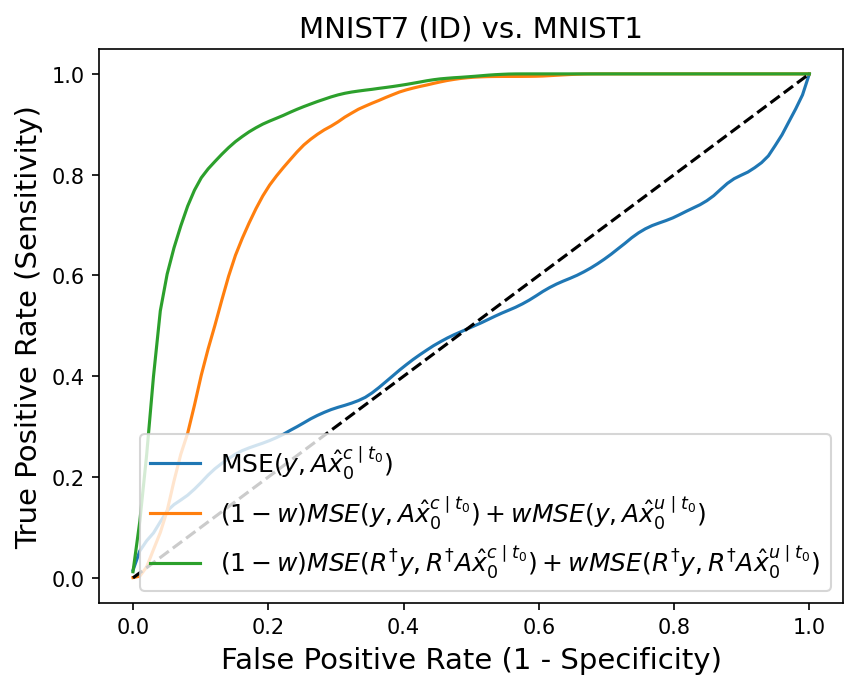}}
 \end{subfigure}
 \caption{The ROC curves demonstrate the trade-off between TPR (sensitivity) and \(1 - \text{FPR}\) (specificity) in cases where the proposed weighting scheme achieved a \textbf{higher} AUC than using reconstruction error in the projection domain with conditional samples. The top row plots the results for the Moderate Sparsity scenario (with \{18, 12, 9\} projections), whereas the bottom row considers the High Sparsity scenario (with \{6, 5, 4\} projections).}
 \label{fig:rq3_roccurves_sup}
\end{figure*}

\begin{figure*}[t!]
 \centering
 \begin{subfigure}[b]{0.4\textwidth}
  \centering
{\includegraphics[width=\textwidth,keepaspectratio=true]{ 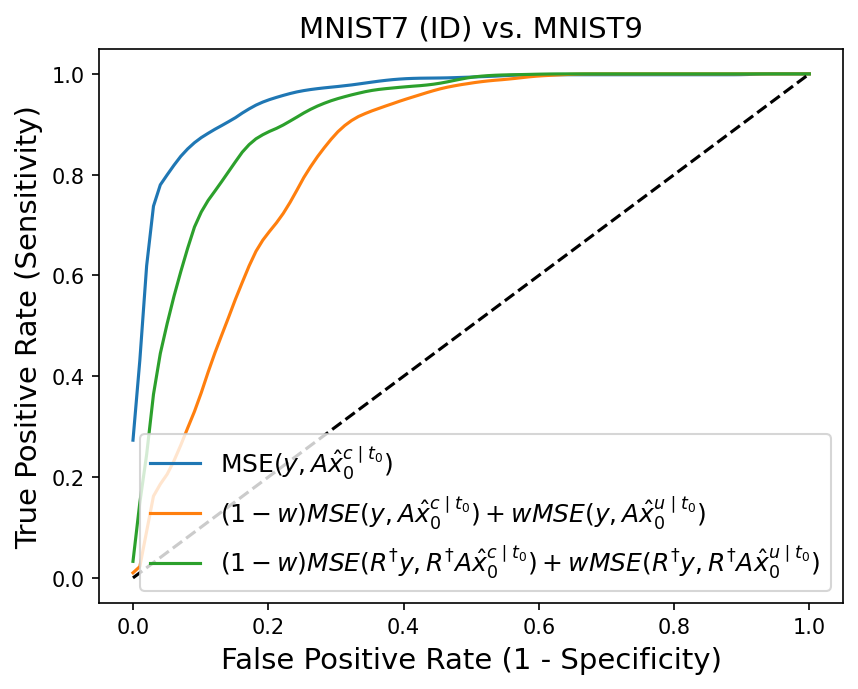}}
 \end{subfigure}
   \begin{subfigure}[b]{0.4\textwidth}
  \centering
{\includegraphics[width=\textwidth,keepaspectratio=true]{ 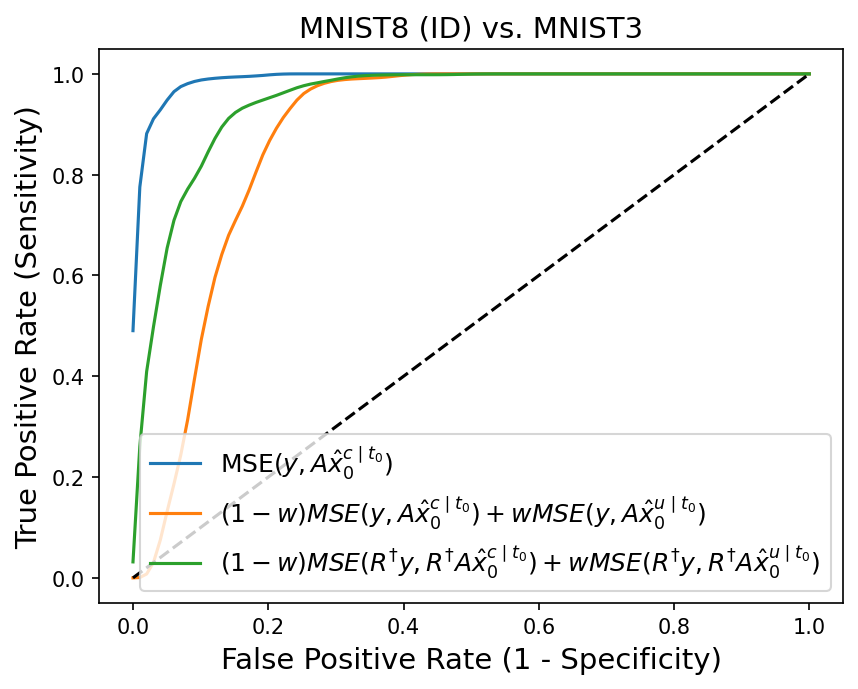}}
 \end{subfigure}
 \begin{subfigure}[b]{0.4\textwidth}
  \centering
{\includegraphics[width=\textwidth,keepaspectratio=true]{ 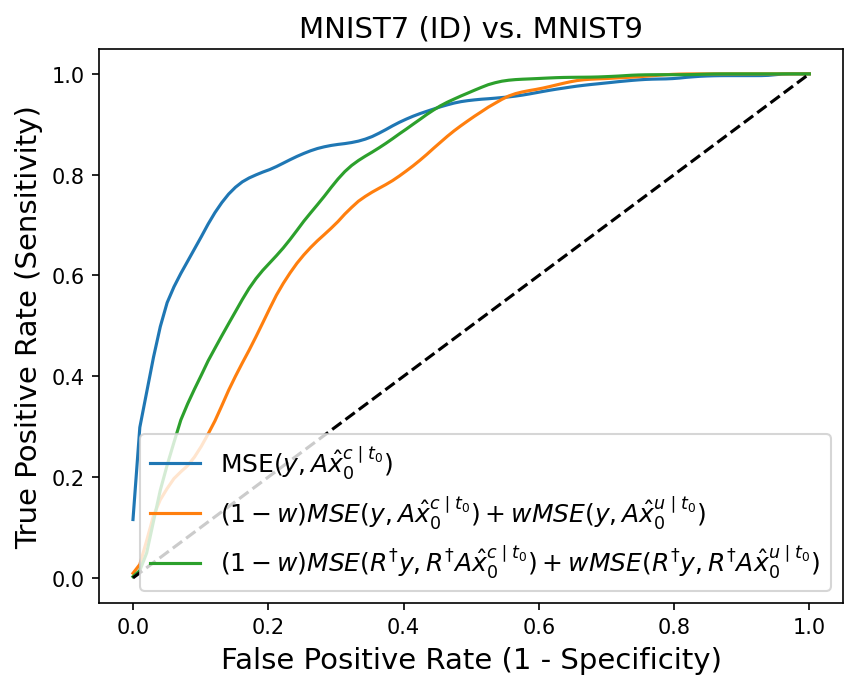}}
 \end{subfigure}
   \begin{subfigure}[b]{0.4\textwidth}
  \centering
{\includegraphics[width=\textwidth,keepaspectratio=true]{ 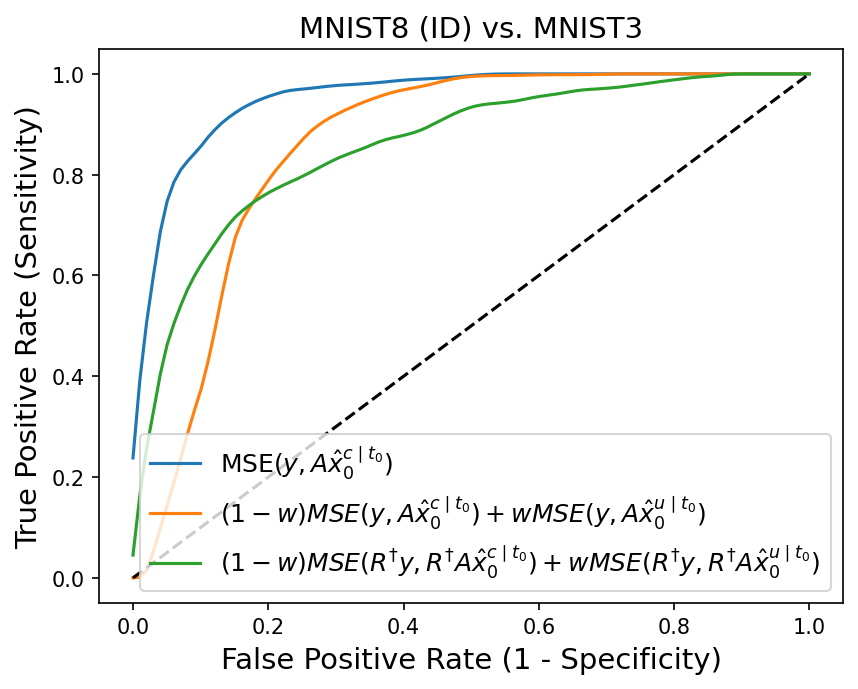}}
 \end{subfigure}
 \caption{The ROC curves demonstrate the trade-off between TPR (sensitivity) and \(1 - \text{FPR}\) (specificity) in cases where the proposed weighting scheme resulted in a \textbf{lower} AUC compared to using reconstruction error in the projection domain with conditional samples. The top row plots the results for the Moderate Sparsity scenario (with \{18, 12, 9\} projections), whereas the bottom row considers the High Sparsity scenario (with \{6, 5, 4\} projections).}
 \label{fig:rq3_roccurvesii_sup}
\end{figure*}

\begin{figure*}[t!]
  \centering
  \begin{subfigure}[b]{0.8\textwidth}
    \centering
    \includegraphics[width=\textwidth, keepaspectratio=true]{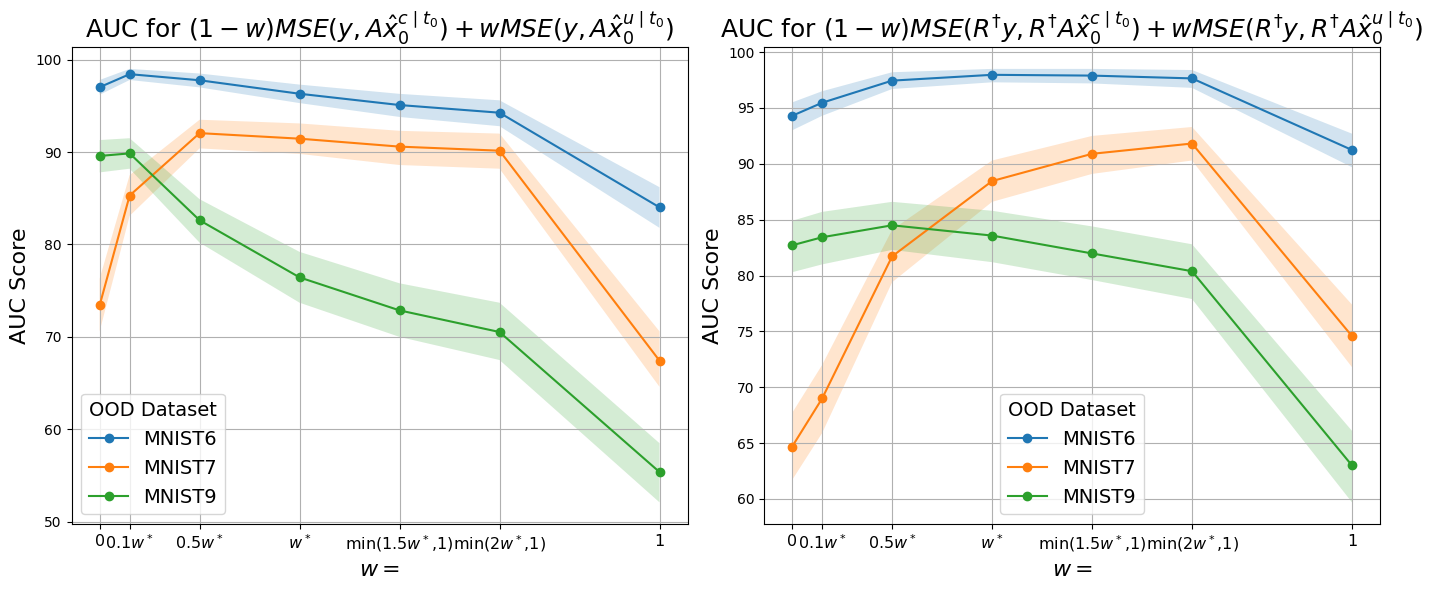}
    \caption{Moderate Sparsity with \#proj = \{18, 12, 9\}.}  
    \label{fig:subfig_a}
  \end{subfigure}\\
  \begin{subfigure}[b]{0.8\textwidth}
    \centering
    \includegraphics[width=\textwidth, keepaspectratio=true]{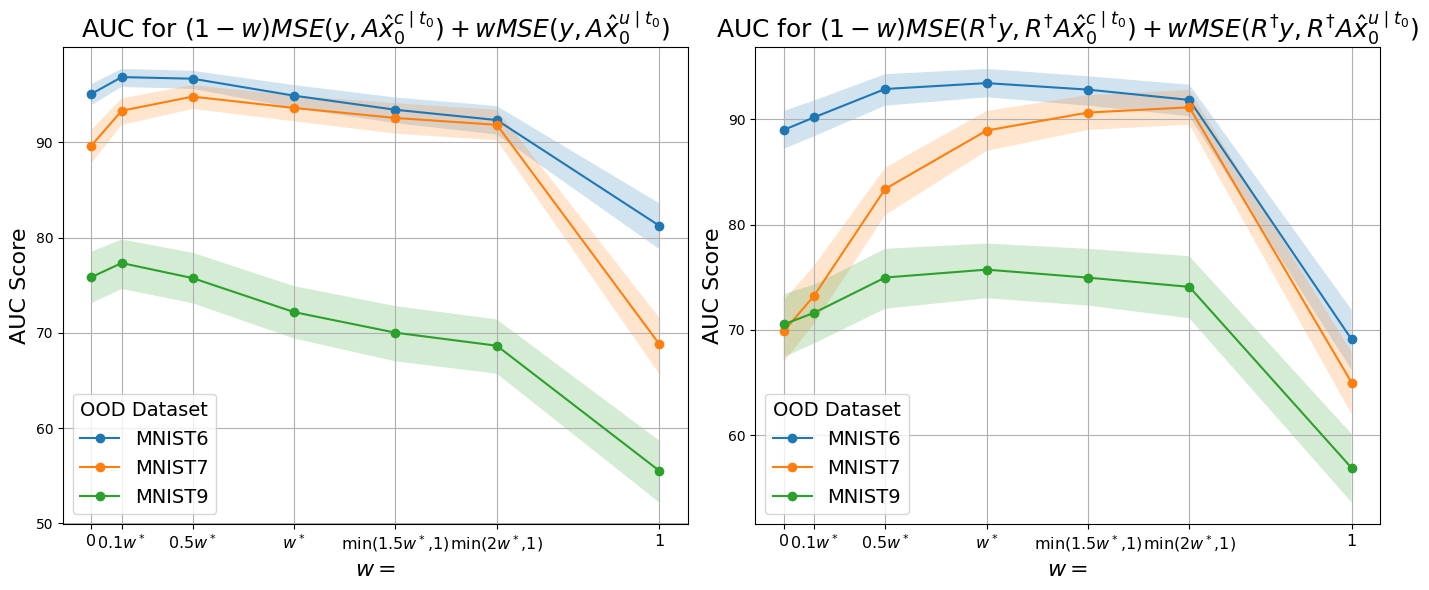}
    \caption{High Sparsity with \#proj = \{6, 5, 4\}.}  
    \label{fig:subfig_b}
  \end{subfigure}\\
 \caption{AUC scores obtained using the proposed weighting scheme, plotted against different scales of \( w^* \), where \( w^* \) represents the normalized \( L^2 \) distance between the measurement and the forward-projected mean of the training set. When scaled up, values exceeding 1 are capped at 1. The cases $\mathbf{w} = \textbf{0}$ and $\mathbf{w} = \textbf{0}$ correspond to conditional and unconditional reconstructions, respectively. This experiment is performed using the model trained on MNIST4, with MNIST6, MNIST7, and MNIST9 considered as OOD datasets. Shaded regions around AUC values indicate confidence bounds, derived from 1000-fold bootstrapping.}
 \label{fig:sensitivityanalysis_sup}
\end{figure*}

\begin{figure*}[t!]
  \centering
  \begin{subfigure}[b]{0.7\textwidth}
    \centering
    \includegraphics[width=\textwidth, keepaspectratio=true]{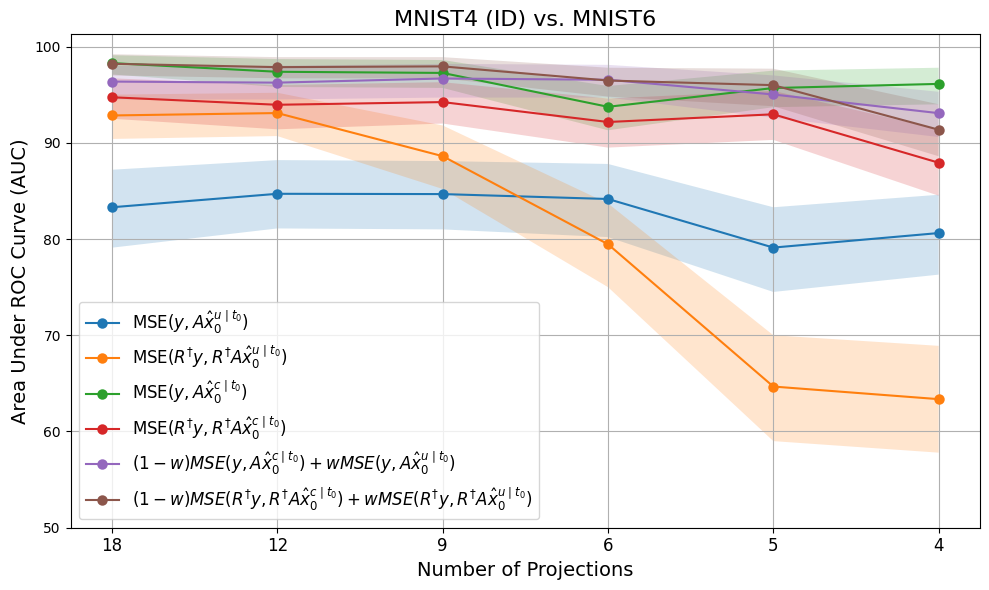}
  \end{subfigure}\\
  \begin{subfigure}[b]{0.7\textwidth}
    \centering
    \includegraphics[width=\textwidth, keepaspectratio=true]{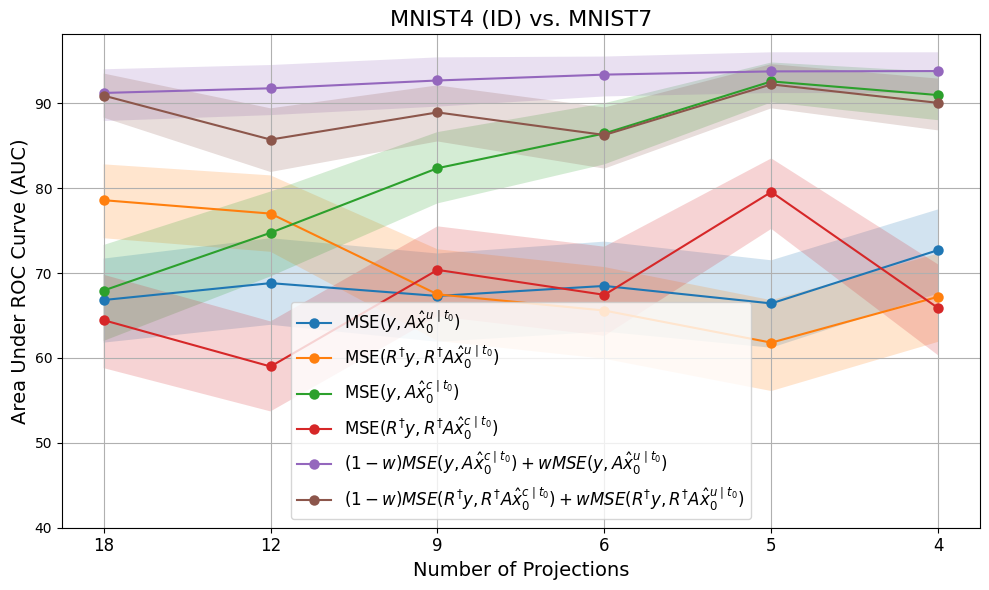}
  \end{subfigure}\\
  \begin{subfigure}[b]{0.7\textwidth}
    \centering
    \includegraphics[width=\textwidth, keepaspectratio=true]{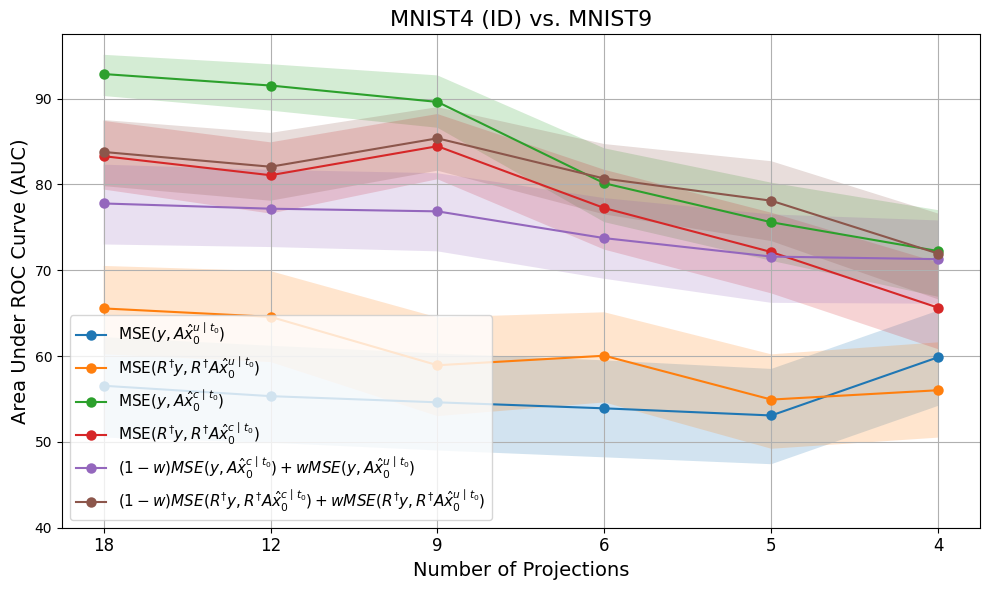}
  \end{subfigure}\\  
 \caption{AUC scores across different number of projections, considering six different reconstruction errors discussed in the paper. This experiment is performed using the model trained on MNIST4, with MNIST6, MNIST7, and MNIST9 being the OOD datasets. Shaded regions around AUC values indicate confidence bounds, derived from 1000-fold bootstrapping. }
 \label{fig:stabilityanalysis_nproj_sup}
\end{figure*}

\begin{figure*}[t!]
  \centering
  \begin{subfigure}[b]{0.7\textwidth}
    \centering
    \includegraphics[width=\textwidth, keepaspectratio=true]{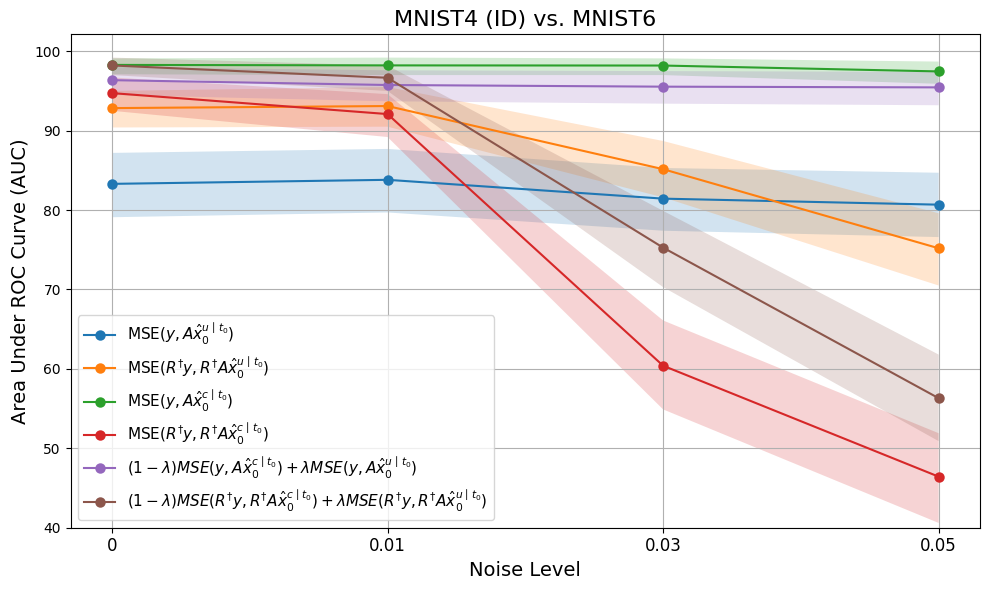}
  \end{subfigure}\\
  \begin{subfigure}[b]{0.7\textwidth}
    \centering
    \includegraphics[width=\textwidth, keepaspectratio=true]{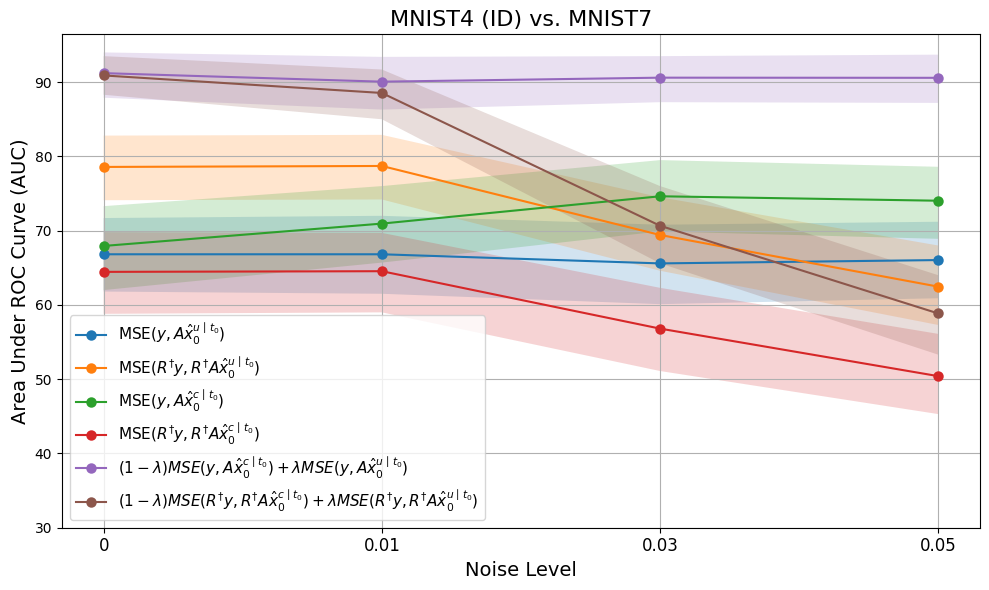}
  \end{subfigure}\\
  \begin{subfigure}[b]{0.7\textwidth}
    \centering
    \includegraphics[width=\textwidth, keepaspectratio=true]{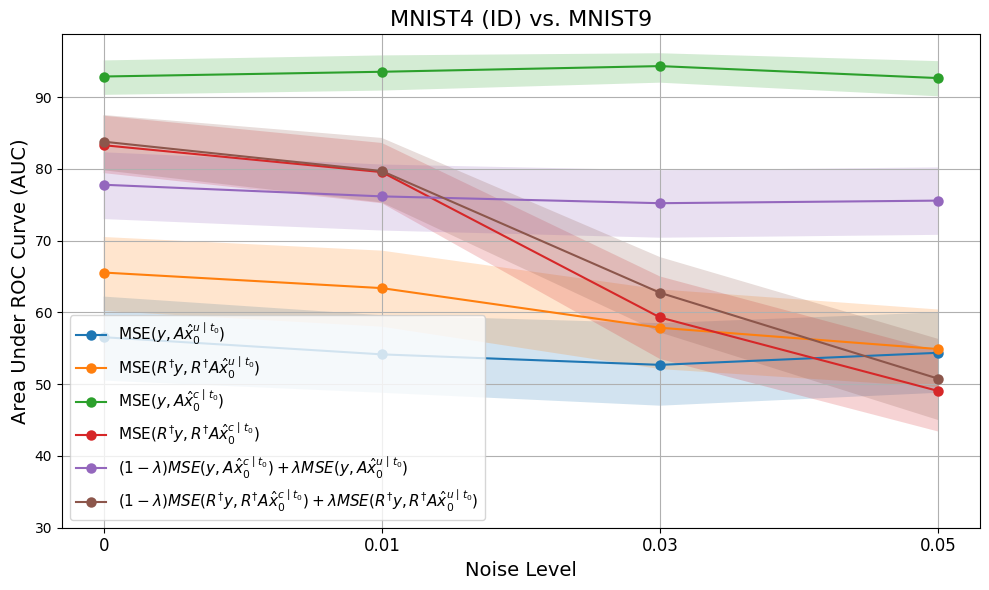}
  \end{subfigure}\\  
 \caption{AUC scores across different levels of noise, considering six different reconstruction errors discussed in the paper. This experiment is performed using the model trained on MNIST4, with MNIST6, MNIST7, and MNIST9 being the OOD datasets. The number of projections is kept fixed at $18$. Shaded regions around AUC values indicate confidence bounds, derived from 1000-fold bootstrapping. }
 \label{fig:stabilityanalysis_noise_sup}
\end{figure*}

\newcommand{\blue}[1]{\textcolor{blue}{#1}}
\newcommand{\red}[1]{\textcolor{red}{#1}}

\begin{table}[h!]
\centering
 \caption{AUC scores for three models trained on full-view MNIST4, MNIST7, and MNIST8 images, evaluated using two different distance metrics for the weighting scheme. Here, $w_{\mu_\theta}$ represents the normalized \( L^2 \) distance between the measurement and the forward-projected mean of the training set, while $w_{avg}$ corresponds to the normalized \( L^2 \) distance averaged across all training set projections. (i.e., $w = \frac{\frac{1}{|\mathcal{D}|} \sum_{x \in \mathcal{D}} \left\| \mathbf{y} - \mathbf{A} x \right\|^2}{\left\| \mathbf{y} \right\|^2 + \frac{1}{|\mathcal{D}|} \sum_{x \in \mathcal{D}} \left\| \mathbf{A} x \right\|^2}$). Red texts highlight the highest values in each column separately for two cases: Moderate Sparsity, where the number of projection angles (\#proj) is \{18, 12, 9\}, and High Sparsity, where \#proj = \{6, 5, 4\}. Confidence bounds are reported for all cases using 1000-fold bootstrapping. This alternative distance metric is included as an illustrative example but is neither effective nor practical due to its high computational cost.}
\resizebox{\textwidth}{!}{
\begin{tabular}{p{0.3cm}|p{3cm}|P{5.2em}P{5.2em}P{5.2em}|P{5.2em}P{5.2em}P{5.2em}|P{5.2em}P{5.2em}P{5.2em}}
&\multicolumn{1}{l|}{ID Dataset:}  & \multicolumn{3}{c|}{MNIST4} & \multicolumn{3}{c|}{MNIST7} & \multicolumn{3}{c}{MNIST8} \\
\hline\hline
&\multicolumn{1}{l|}{OOD Dataset:} & \small MNIST6 & \small MNIST7 & \small MNIST9 & \small MNIST1 & \small MNIST4 & \small MNIST9 & \small MNIST3 & \small MNIST5 & \small MNIST9 \\
\hline
\multirow{5}{*}{\rotatebox[origin=c]{90}{\parbox{2cm}{\centering {\small Moderate}}}} 
&$w_{\mu_\theta}$ ($\Gamma = \mathbf{I}$)  & 96.25 \notsotiny [95.2, 97.2]  & \red{\textbf{91.45}} \notsotiny [89.8, 93.1] & 76.51 \notsotiny [73.6, 79.3] & 78.33 \notsotiny [75.6, 80.9]   & 88.06 \notsotiny [86.0, 89.9] & 83.95 \notsotiny [81.5, 86.1] & 87.79 \notsotiny [85.6, 89.9] & 87.72 \notsotiny [85.5, 89.9] & 81.78 \notsotiny [79.2, 84.4] \\
&$w_{\mu_\theta}$ ($\Gamma = \mathbf{R}^{\dagger}$) & \red{\textbf{97.98}} \notsotiny [97.3, 98.6] & 88.46 \notsotiny [86.6, 90.3]  & \red{\textbf{83.60}} \notsotiny [81.4, 85.9] & \red{\textbf{81.07}} \notsotiny [78.8, 83.3] & \red{\textbf{94.20}} \notsotiny [92.9, 95.5] & \red{\textbf{91.39}} \notsotiny [89.8, 93.0] & \red{\textbf{94.58}} \notsotiny [93.3, 95.7] & \red{\textbf{95.03}} \notsotiny [93.8, 96.2] & \red{\textbf{90.34}} \notsotiny [88.6, 91.9]\\
 \cline{2-11}
&$w_{avg}$ ($\Gamma = \mathbf{I}$)  & 95.52 \notsotiny [94.4, 96.6] & 89.33 \notsotiny [87.3, 91.1]  & 71.02 \notsotiny [68.2, 73.9] & 69.34 \notsotiny [66.3, 72.3] & 85.57 \notsotiny [83.5, 87.5] & 80.34 \notsotiny [77.8, 82.8] & 84.59 \notsotiny [82.2, 86.8] & 84.50 \notsotiny [81.8, 86.9] & 78.29 \notsotiny [75.5, 81.0] \\
&$w_{avg}$ ($\Gamma = \mathbf{R}^{\dagger}$)   & 97.90 \notsotiny [97.2, 98.5] & 89.40 \notsotiny [87.5, 91.2] & 81.75 \notsotiny [79.4, 84.0]
 & 79.23 \notsotiny [76.8, 81.7] & 94.06 \notsotiny [92.6, 95.4] & 90.64 \notsotiny [89.0, 92.3] & 93.92 \notsotiny [92.4, 95.2] & 94.65 \notsotiny [93.4, 96.0] & 90.17 \notsotiny [88.5, 91.8] \\
\hline \hline
\multirow{5}{*}{\rotatebox[origin=c]{90}{\parbox{2cm}{\centering {\small High}}}} 
 &$w_{\mu_\theta}$ ($\Gamma = \mathbf{I}$)  & \red{\textbf{94.88}} \notsotiny [93.6, 96.1]     & \red{\textbf{93.60}} \notsotiny [92.2, 94.9] & 72.13 \notsotiny [69.2, 75.0] & 85.61 \notsotiny [83.2, 87.7]     & {85.28} \notsotiny [83.1, 87.6] & 77.20 \notsotiny [74.4, 80.0] & 81.61 \notsotiny [79.0, 84.1] & 85.66 \notsotiny [83.4, 88.0] & 82.57 \notsotiny  [80.1, 85.0] \\
&$w_{\mu_\theta}$ ($\Gamma = \mathbf{R}^{\dagger}$) & {93.38} \notsotiny [92.0, 94.6] & 88.94 \notsotiny [87.0, 90.8]     & \red{\textbf{75.72}} \notsotiny [73.2, 78.3] & \red{\textbf{92.56}} \notsotiny [90.9, 94.0] & \red{\textbf{87.20}} \notsotiny [85.1, 89.2] & \red{\textbf{81.86}} \notsotiny [79.5, 84.2] & \red{\textbf{85.63}} \notsotiny  [83.5, 87.7] & \red{\textbf{86.83}} \notsotiny [84.9, 88.7] & 85.55 \notsotiny  [83.6, 87.7]\\
\cline{2-11}
&$w_{avg}$ ($\Gamma = \mathbf{I}$)  & 94.08 \notsotiny [92.7, 95.3] & 92.13 \notsotiny [90.5, 93.6] & 68.60 \notsotiny [65.7, 71.4]
 & 78.57 \notsotiny [75.9, 81.2] & 81.71 \notsotiny [79.3, 84.0] & 73.19 \notsotiny [70.3, 76.0] & 82.91 \notsotiny [80.6, 85.3] & 82.96 \notsotiny [80.4, 85.4] & 79.32 \notsotiny [76.5, 81.9] \\
&$w_{avg}$ ($\Gamma = \mathbf{R}^{\dagger}$)  & 91.55 \notsotiny [90.0, 93.1] & 87.98 \notsotiny [86.0, 89.8] & 74.09 \notsotiny [71.0, 76.9] &
91.78 \notsotiny [90.2, 93.3] & 85.57 \notsotiny [83.3, 87.8] & 79.29 \notsotiny [76.6, 81.8] & 84.91 \notsotiny [82.7, 87.1] & 86.61 \notsotiny [84.5, 88.5] & \red{\textbf{85.91}} \notsotiny [83.8, 87.9]
\end{tabular}}
\label{tab:AUCafterweighting}
\end{table}